\documentclass{aa}

\usepackage{adjustbox} % Auto-adjust table size
\usepackage{afterpage}
\usepackage{amsfonts} % More math fonts
\usepackage{amsmath} % Facilitate writing of formulas
\usepackage{amssymb} % More math symbols
\usepackage{booktabs}% Better tables
\usepackage{fontawesome} % Plenty of nice symbols
\usepackage{lipsum}
\usepackage{lmodern} % Fix font size warnings
\usepackage{multirow} % Allow line merging in tables
\usepackage{siunitx} % Facilitate writing of units and numbers 
\usepackage{soul} % Highlights
\usepackage{subcaption} % Allow to put subfigures into figures
\usepackage{threeparttable} % Allow footnotes in tables
\usepackage{wasysym} % Astrological symbols and more
\usepackage{xcolor} % Color management

\usepackage[varg]{txfonts}
\usepackage{graphicx}
\usepackage{natbib,twoopt}
\usepackage{hyperref} %% to avoid \citeads line fills
\bibpunct{(}{)}{;}{a}{}{,}             %% natbib format for A&A and ApJ

\hypersetup{
colorlinks=true,% Set to false to disable coloring links
citecolor=blue,% The color of citations
linkcolor=blue,% The color of references to document elements (sections, figures, etc)
urlcolor=blue,% The color of hyperlinks (URLs)
}

\makeatletter
  \newcommandtwoopt{\citeads}[3][][]{\href{http://adsabs.harvard.edu/abs/#3}%
    {\def\hyper@linkstart##1##2{}%
     \let\hyper@linkend\@empty\citealp[#1][#2]{#3}}}
  \newcommandtwoopt{\citepads}[3][][]{\href{http://adsabs.harvard.edu/abs/#3}%
    {\def\hyper@linkstart##1##2{}%
     \let\hyper@linkend\@empty\citep[#1][#2]{#3}}}
  \newcommandtwoopt{\citetads}[3][][]{\href{http://adsabs.harvard.edu/abs/#3}%
    {\def\hyper@linkstart##1##2{}%
     \let\hyper@linkend\@empty\citet[#1][#2]{#3}}}
  \newcommandtwoopt{\citeyearads}[3][][]%
    {\href{http://adsabs.harvard.edu/abs/#3}
    {\def\hyper@linkstart##1##2{}%
     \let\hyper@linkend\@empty\citeyear[#1][#2]{#3}}}
\makeatother

\begin{document}

% Preamble
\title{1D atmospheric study of the temperate sub-Neptune K2-18b}

\author{D. Blain \inst{1} \and B. Charnay \inst{1} \and B. Bézard \inst{1}}

\institute{LESIA, Observatoire de Paris, PSL Research University, CNRS, Sorbonne Université, Université de Paris, 92195, Meudon, France\\
\email doriann.blain@obspm.fr
}

\date{Received July 30, 2020~/ Accepted November 13, 2020}

\abstract
% context heading (optional)
{The atmospheric composition of exoplanets with masses between 2 and 10 M$_\oplus$ is poorly understood. In that regard, the sub-Neptune K2-18b, which is subject to Earth-like stellar irradiation, offers a valuable opportunity for the characterisation of such atmospheres. Previous analyses of its transmission spectrum from the Kepler, Hubble (HST), and Spitzer space telescopes data using both retrieval algorithms and forward-modelling suggest the presence of H$_2$O and an H$_2$--He atmosphere, but have not detected other gases, such as CH$_4$.}
% aims heading (mandatory)
{We present simulations of the atmosphere of K2-18 b using Exo-REM, our self-consistent 1D radiative-equilibrium model, using a large grid of atmospheric parameters to infer constraints on its chemical composition.}
% methods heading (mandatory)
{We compared the transmission spectra computed by our model with the above-mentioned data (0.4 to 5 $\mu$m), assuming an H$_2$--He dominated atmosphere. We investigated the effects of irradiation, eddy diffusion coefficient, internal temperature, clouds, C/O ratio, and metallicity on the atmospheric structure and transit spectrum.}
% results heading (mandatory)
{We show that our simulations favour atmospheric metallicities between 40 and 500 times solar and indicate, in some cases, the formation of H$_2$O-ice clouds, but not liquid H$_2$O clouds. We also confirm the findings of our previous study, which showed that CH$_4$ absorption features nominally dominate the transmission spectrum in the HST spectral range. We compare our results with results from retrieval algorithms and find that the H$_2$O-dominated spectrum interpretation is either due to the omission of CH$_4$ absorptions or a strong overfitting of the data. Finally, we investigated different scenarios that would allow for a CH$_4$-depleted atmosphere. We were able to fit the data to those scenarios, finding, however, that it is very unlikely for \object{K2-18b} to have a high internal temperature. A low C/O ratio ($\approx$ 0.01--0.1) allows for H$_2$O to dominate the transmission spectrum and can fit the data but so far, this set-up lacks a physical explanation. Simulations with a C/O ratio $<$ 0.01 are not able to  fit the data satisfactorily.}
% conclusions heading (optional)
{}


\maketitle

% Main
\section{Introduction}

\begin{table*}[pt]
\centering
\caption{\label{tab:general_parameters} Parameters of K2-18b and its star}
\begin{tabular}{@{}lccc@{}} 
\hline
\hline
Parameter                                               & Value                                                         &                                                 & References                            \\ \hline
\textbf{Host star:} \\
Spectral type                                   & M2.5 V                                                        &                                                 & 1     \\
$M_\ast$ (kg)   & 9.8445 $\pm$ 0.086 $\times 10^{29}$                           & (0.50 M$_\odot$)                & 1     \\
$R_\ast$ (Mkm)                                  & 0.3092 $\pm$ 0.0102                           & (0.44 R$_\odot$)                & 2     \\
$T_{\ast,\,\text{eff}}$ (K)     & 3457 $\pm$ 39                                         &                                                 & 3     \\
$g_\ast$ (m$\cdot$s$^{-2}$)& 690$^{+ 48}_{- 44}$                                                                 &                                                               & Derived                                 \\
$[$Fe/H$]$                                              & 0.123 $\pm$ 0.157                                 &                                               & 3       \\
$t_{\ast}$ (Gyr)                        & $2.4 \pm 0.6$                                         &                                                 & 4
\\
\textbf{Planet:} \\
$a_p$ (Mkm)                                     & 23.801$^{+0.069}_{-0.070}$         & (0.16 au)                     & 2     \\
$e$                                                     & 0.09$^{+0.12}_{-0.09}$                                         &                                               & 1       \\
$M_p$ (kg)                                              & 5.15 $\pm$ 0.08 $\times 10^{25}$        & (8.63 M$_\oplus$)     & 1     \\
$R_p$ (km)                                              & 16$\,$640 $\pm$ 550\tablefootmark{$\ast$}                                       & (2.61 R$_\oplus$)         & 2     \\
$T_{p,\,\text{int}}$ (K)                & $83^{+8}_{-6}$                                        &                                                 & 5 (model)     \\
$E_{p,\,\text{e}}$ (W$\cdot$m$^{-2}$)           & 1368$^{+114}_{-107}$  & (1.005 $E_{\oplus,\,\text{e}}$)         & 2     \\ \hline
\end{tabular}
\tablefoot{
\tablefoottext{$\ast$}{We actually use a $10^5$-Pa radius of 16400 km in our simulations. See text. The radius from \citep{Benneke2019} corresponds to a pressure of $10^3$ Pa.
}
\tablebib{
(1)~\citet{Cloutier2019}; (2) \citet{Benneke2019}; (3) \citet{Benneke2017}; (4) \citet{Guinan2019}; (5) \citet{Rogers2010}.
}
}
\end{table*}

While fairly common among the thousands of exoplanets discovered to date\footnote{\href{exoplanet.eu}{exoplanet.eu}} (see \autoref{fig:exoplanet_distribution}), the atmospheres of super-Earths and sub-Neptunes -- with masses between 2 and 10 M$_\oplus$ -- are poorly understood. By essentially using the mass and radius of the planets, it has been well-established that planets with low masses ($<$ 2 M$_\oplus$) must primarily be made up of iron and silicates, and generally with a thin atmosphere. On the other hand, planets with high masses ($> 10$ M$_\oplus$) must retain a thick atmosphere composed mainly of H$_2$ and He, representing a significant portion of the planet mass \citep{Chen2016}. Within the transition between the two populations, however, studies with models struggle to give a clear answer \citep{Valencia2013, Fulton2017, Zeng2019, Otegi2020}, exhibiting a degeneracy between massive rocky planets, ocean planets, and small gaseous planets. Thus, spectral observations are crucial for characterising the atmosphere of these objects and to better constrain their internal composition.

The recently discovered, transiting exoplanet K2-18b (see \autoref{tab:general_parameters}) offers not only the opportunity to retrieve spectroscopic data on the atmosphere of a sub-Neptune, but also to study such atmospheres under nearly Earth-like conditions. Indeed, the stellar irradiance received by K2-18b ($E_{p,\,\text{e}} = 1368^{+114}_{-107}$ W$\cdot$m$^{-2}$) is very close to that of the Earth. Nine transits were acquired using the Wide Field Camera 3 on the Hubble Space Telescope (HST/WFC3), as well as data from Kepler K2 and Spitzer IRAC channels 1 and 2 \citep{Benneke2019}. These data have already been analysed by several teams \citep{Benneke2019, Tsiaras2019, Madhusudhan2020, Scheucher2020, Bezard2020}. Also, \citet{Scheucher2020} and \citet{Bezard2020} used self-consistent models (respectively, '1D-TERRA' and 'Exo-REM') to analyse the data , while the other used free retrieval algorithms. The first five investigations concluded that there is a presence of H$_2$O as well as a significant amount of H$_2$-He. They also derived upper limits for the abundance of CH$_4$, which in all cases was lower than $\approx$ 3.5$\%$ at a 99$\%$ confidence level. In contrast, \citet{Bezard2020} found that the HST/WFC3 spectrum is dominated by CH$_4$ absorption and found abundances of CH$_4$ and H$_2$O, respectively, between 3\%\ and 10$\%$ and 5\%\ and 11$\%$ at a 1-$\sigma$ confidence level. They assumed an H$_2$-dominated atmosphere and varied the atmospheric metallicity, but their simulations did not include H$_2$O clouds and they did not simulate non-solar C/O ratios.

Given the irradiance of the planet and the presence of H$_2$O in the atmosphere, the question of the existence of liquid H$_2$O is naturally posed. \citet{Benneke2019} found that a cloud layer was needed to reproduce the data. They then used a self-consistent model and found that liquid H$_2$O could condense at the right pressure to explain this cloud layer. Contrary to \citet{Benneke2019}, \citet{Madhusudhan2020} did not find compelling  evidence for clouds or hazes in the atmosphere. Using an interior model, \citet{Madhusudhan2020} indicated that if the planet had a small rocky core and a thin H$_2$/He atmosphere, an ocean of liquid H$_2$O could exist. However, \citet{Scheucher2020} ruled out this possibility, arguing that an H$_2$O ocean would partially evaporate in the atmosphere, giving a spectrum that would be incompatible with the data.

In the present work, we use Exo-REM, our self-consistent one-dimensional (1D) atmospheric model adapted for transiting exoplanets to study K2-18b atmospheric composition assuming a thick H$_2$--He atmosphere. We first present the extension of Exo-REM \citep{Baudino2015, Baudino2017, Charnay2018} to irradiated planets and detail the calculations of the absorption cross-sections. We then expand on the work of \citet{Bezard2020} by including clouds and studying their formation, and by investigating the effects of irradiation, the eddy diffusion coefficient, internal temperature, and metallicity on the atmosphere properties.

\afterpage{
\begin{figure}[pt]
\centering
\includegraphics[width=1.0\linewidth]{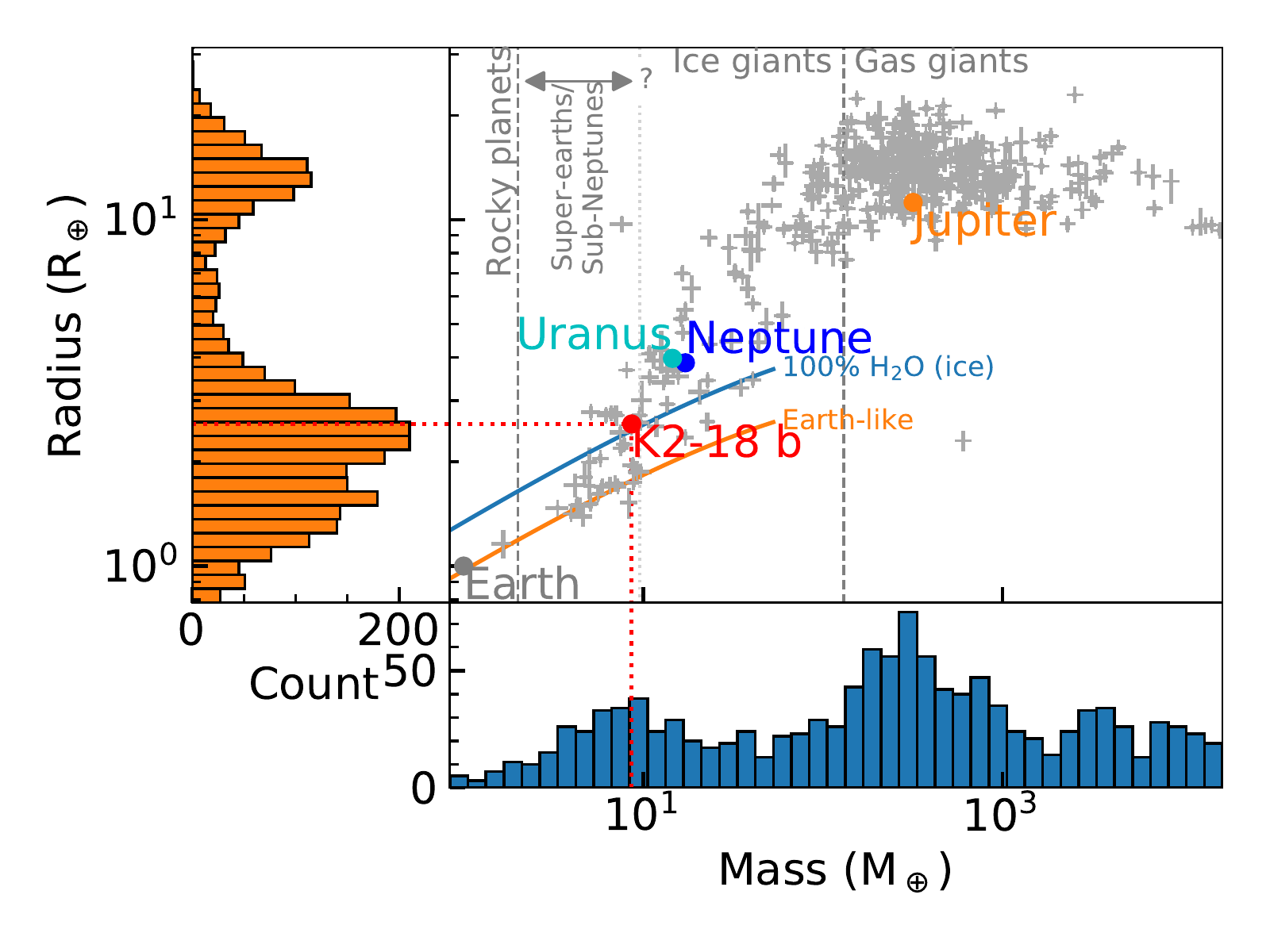}
\caption[]{Mass-radius distribution of the 4266\protect\footnotemark confirmed exoplanets to date (17 June 2020). The scatter plot shows only the 454 planets for which the mass and the radius are known within $\pm15\%$. The mass-radius curves for an Earth-like interior (orange) and a pure water ice planet (blue) are taken from \citet{Seager2007}.}
\label{fig:exoplanet_distribution}
\end{figure}

\footnotetext{Not all the confirmed planets have a measured radius and/or mass, thus the total count in the histograms is less than $4266$.}
}

\section{Model}
Exo-REM is a 1D radiative-equilibrium model first developed for the simulation of young gas giants far from their star and brown dwarfs \citep{Baudino2015, Baudino2017, Charnay2018}. Fluxes are calculated using the two-stream approximation assuming hemispheric closure. The radiative-convective equilibrium is solved assuming that the net flux (radiative + convective) is conservative. The conservation of flux over the pressure grid is solved iteratively using a
constrained linear inversion method. We take into account Rayleigh scattering from H$_2$, He, and H$_2$O, as well as absorption and scattering by clouds -- calculated from the extinction coefficient, single scattering albedo, and asymmetry factor interpolated from pre-computed tables for a set of wavelengths and particle radii \citep[see][]{Charnay2018}.

We made several upgrades to this model in order to extend its simulation capabilities to high-metallicity (up to $\approx$ 1000 times the solar metallicity) and irradiated planets, namely sub-Neptunes and moderately hot Jupiters (equilibrium temperature lower than 2000 K). These upgrades are detailed in the following sections. The Exo-REM source code and the $k$-coefficients we used (see Section~\ref{sec:k_coeff}) are available online\footnote{\href{https://gitlab.obspm.fr/dblain/pytheas}{https://gitlab.obspm.fr/dblain/exorem}.}.

\subsection{Spectroscopic data}
\subsubsection{Collision-induced absorptions}
The collision-induced absorptions (CIA) included in our simulations and their references are given in \autoref{tab:cia}. We added the contributions of the H$_2$O--H$_2$O CIA and, since spectroscopic data for the H$_2$O--H$_2$ CIA were not available, of the H$_2$O--air CIA. The reasons for this addition are detailed in Section~\ref{sec:effect_of_the_H2O_CIA}.

\subsubsection{Absorption cross-sections}
\label{sec:absorption_cross_sections}
\begin{table*}[pt]
\centering
\caption{\label{tab:cia}Collision-induced absorption references}
\begin{tabular}{@{}lc@{}}
\hline
\hline
CIA                                     & References \\
\hline
H$_2$ -- H$_2$          & 1, 2, 3 \\
H$_2$ -- He             & 3 \\
H$_2$O -- H$_2$O        & 4 \\
\hline
\end{tabular}
\tablebib{
(1)~\citet{Borysow2001}; (2) \citet{Borysow2002}; (3) \citet{Richard2012}; (4) \citet{Mlawer2012}.
}
\end{table*}

\begin{table*}[pt]
\centering
\caption{\label{tab:species_xsec}Species cross-section parameters}
\begin{tabular}{@{}lcccl@{}} 
\hline
\hline
\multirow{2}{*}{Species}        & Wavenumber            & $\Delta\nu$           & Intensity cutoff                                & \multirow{2}{*}{Line list}         \\
                                                        & range (cm$^{-1}$) & (cm$^{-1}$)   & (cm$\cdot$molecule$^{-1}$)    &       \\ 
\hline
CH$_4$                                          & 30 -- 13330           & 250                     & $10^{-36}$ at 2000 K  & TheoReTS (1)  \\
CO                                                      & 30 -- 8330            & 120                     & $10^{-27}$ at 3000 K  & HITEMP (2)    \\
CO$_2$                                          & 30 -- 8130            & 120                     & $10^{-25}$ at 3000 K  & HITEMP (2)    \\
FeH                                                     & 30 -- 14830           & 120                     & $10^{-30}$ at 4000 K  & ExoMol (3)    \\
H$_2$O                                          & 30 -- 26430           & 120                     & $10^{-27}$ at 2000 K  & HITEMP (2)    \\
H$_2$S                                          & 30 -- 10830           & 120                     & $10^{-27}$ at 2000 K  & ExoMol (4)            \\
HCN                                             & 30 -- 12530           & 120                     & $10^{-25}$ at 3000 K  & ExoMol (5)    \\
K                                                       & 1030 -- 50030         & 9000                    & $10^{-27}$ at 2500 K  & NIST (6)              \\
Na                                                      & 1030 -- 50030         & 9000                    & $10^{-27}$ at 2500 K  & NIST (6)      \\
NH$_3$                                          & 30 -- 11830           & 120                     & $10^{-30}$ at 1500 K  & ExoMol (7, 8)         \\
PH$_3$                                          & 30 -- 9830            & 120                     & $10^{-30}$ at 2500 K  & ExoMol (9) \\ 
TiO                                                     & 230 -- 29230          & 120                     & $10^{-30}$ at 4000 K  & ExoMol (10)   \\
VO                                                      & 30 -- 19830           & 120                     & $10^{-30}$ at 4000 K  & ExoMol (11)   \\  
\hline
\end{tabular}
\tablebib{
(1)~\citet{Rey2017}; (2) \citet{Rothman2010}; (3) \citet{Bernath2020}; (4) \citet{Azzam2016}; (5) \citet{Harris2006}; (6) \citet{NIST_ASD}; (7) \citet{Coles2019}; (8) \citet{Yurchenko2015}; (9) \citet{Sousa-Silva2014}; (10) \citet{Schwenke1998}; (11) \citet{McKemmish2016}.
}
\end{table*}

\begin{table*}[pt]
\centering
\caption{\label{tab:species_broadening_ref}Line broadening references}
\begin{tabular}{@{}lll@{}} 
\hline
\hline
\multirow{2}{*}{Species}        & $\gamma$                                                                              & $n$             \\
                                                        & references                                                                    & references      \\ 
\hline                      
CH$_4$                                          & H$_2$ -- He (1, 2)    & Idem \\
CH$_3$D                                         & H$_2$ (3)                                     & Idem \\
CO                                                      & H$_2$ -- He  (4)                         & Idem \\
CO$_2$                                          & H$_2$ -- He (5)                                        & Air (6) \\
FeH                                                     & Same as CO                                                                    & Idem \\
H$_2$O                                          & H$_2$ (7)                                     & Air (6)  \\
H$_2$S                                          & H$_2$ (8)                                             & Air (6)  \\
HCN                                             & N$_2$ (9)                                     & Idem \\
K                                                       & H$_2$ (10)                                             & Idem \\
Na                                                      & H$_2$ (11)                                            & Idem \\
NH$_3$                                          & H$_2$ -- He (12, 13) & Idem \\
PH$_3$                                          &  H$_2$ -- He (14)             & Idem\\
TiO                                                     & Same as CO                                                                    & Idem \\
VO                                                      & Same as CO                                                                    & Idem\\
\hline
\end{tabular}
\tablebib{
(1)~\citet{Pine1992}; (2) \citet{Margolis1996}; (3)  \citet{Lerot2003}; (4) \citet{Wilzewski2016}; (5) \citet{Burch1969}; (6) \citet{Rothman2010}; (7) \citet{Langlois1994}; (8) \citet{Kissel2002}; (9) \citet{Rinsland2003}; (10) \citet{Allard2016}; (11) \citet{Allard2012}; (12) \citet{Nemtchinov2004}; (13) \citet{Brown1994}; (14) see \citet{Baudino2015}.
}
\end{table*}

We calculated the absorption cross-sections of CH$_4$, CO, CO$_2$, FeH, H$_2$O, H$_2$S, HCN, K, Na, NH$_3$, PH$_3$, TiO, and VO at 25 pressure levels equally spaced in the log-space between 0.1 and 10$^7$ Pa. At each of these pressure levels, we calculated the cross-sections at temperatures 100, 150, 200, 250, 300, 400, 500, 600, 800, 1000, 1200, 1500, 2000, 2500, and 3000 K -- except for NH$_3$, the reasons for which are detailed below. We calculated line absorption up to a given distance ($\Delta\nu$) from the line center, using the same procedure as described in \citet{Baudino2015}. We used a sub-Lorentzian line profile with a $\chi$ factor, based on \cite{Burch1969} and \citet{Hartmann2002}, for, respectively, CO$_2$ and then all the other species. The cross-sections were calculated over the wavenumber range displayed in \autoref{tab:species_xsec}, at a resolution that is similar to the line half width.

When line lists of individual isotopes were available, we merged them by multiplying line intensities in order to reproduce the isotopic ratio found for Jupiter. Otherwise, we used the default isotopic ratio given by the database. We used H$_2$ and He pressure-broadened halfwidths ($\gamma$) and temperature exponents ($n$) whenever they were available. When both were available, we used a mixture of 90$\%$ H$_2$ and 10$\%$ He, which roughly corresponds to the standard Solar System He/H ratio \citep{Lodders2019}. More details are given in \autoref{tab:species_xsec} and \autoref{tab:species_broadening_ref}. The specificities of some species are listed below.
\paragraph{CH$_4$:} We included the contribution from $^{12}$CH$_4$, $^{13}$CH$_4,$ and CH$_3$D, all taken from the TheoReTS database \citep{Rey2017}, which is more accurate than the ExoMol database. We used a $^{12}$C/$^{13}$C ratio of 89 \citep{Niemann1998} and a D/H ratio of 2$\times$10$^{-5}$ \citep{Lellouch2001}. We recall that the TheoReTS line list of CH$_3$D stops at 6500 cm$^{-1}$. The line list provided by the TheoReTS database separates the 'strong lines', which have an intensity of $>$ 10$^{-26}$ cm$\cdot$molecule$^{-1}$, from the very-high-spectral-density 'weak lines', which are regrouped in 'super lines'. This separation allows for much faster cross-section calculations, but at the price of losing information on the  quantum numbers of individual lines. Hence, we had to use the same $\gamma$ and $n$ for all lines.
\paragraph{FeH, TiO and VO:} H$_2$ or He $\gamma$ and $n$ were not available for these molecules and we used the parameters from CO.
\paragraph{K and Na :} Given that these species have very high intensity lines (up to $\approx 10 ^{-13}$ cm$\cdot$molecule$^{-1}$), we needed to extend our $\Delta\nu$ up to 9000 cm$^{-1}$ in order to correctly account for far wing absorption. Following \citet{Burrows2000}, we used a Voigt profile up to a detuning frequency of $20\times(T/500)^{0.6}$ cm$^{-1}$ for K and $30\times(T/500)^{0.6}$ cm$^{-1}$ for Na. Beyond that detuning frequency, we used the profile described by \citet{Baudino2015}.
\paragraph{NH$_3$: } Cross-sections were calculated up to 1500 K because the line list we used lacks completeness above this temperature. We took a $^{14}$N/$^{15}$N ratio of 500 \citep{Furi2015}.
\paragraph{PH$_3$: } The line list we used is not complete above 1000 K. However, \citet{Sousa-Silva2014} provide the PH$_3$ partition function up to 3000 K. This allows us to use the percentual loss of completeness to estimate the proportion of missing opacity, as suggested in the cited work.

\subsubsection{\textit{k}-coefficients}
\label{sec:k_coeff}
We used these high-resolution absorption coefficients to calculate $k$-coefficients according to the method described in \citet{Baudino2015}, using 16 Gauss-Legendre quadrature points (eight for values of the cumulative distribution of the absorption coefficients between 0 and 0.95, and eight between 0.95 and 1). Four sets of $k$-coefficients were calculated, at resolutions of 0.5, 20, 200, and 2000 cm$^{-1}$, although only the 20 cm$^{-1}$ step one was used in the main part of this study.

\subsection{Radiative-convective equilibrium model: stellar irradiance}
We added the planetary averaged stellar irradiance $E_{\downarrow,\,\text{e},\nu}$ reaching the planet to Exo-REM. We used stellar spectra from the BT-Settl model \citep{Allard2012}. We chose a spectrum modelled at an effective temperature $T_\text{S} = 3500$ K, $\log_{10}(g[\text{cm}\cdot\text{s}^{-2}]) = 5$, a null metallicity and no alpha enhancement. We neglected the spectral dependency on surface gravity and metallicity. Then we interpolated the BT-Settl spectrum on a spectral grid with a 0.1 cm$^{-1}$ wavenumber step. This interpolated spectrum was then convolved at the resolution of Exo-REM flux calculations (see Section~\ref{sec:methodology}) to obtain the modelled radiosity $J_{\text{S},\,\text{e},\nu}$. Then we obtained $E_{\downarrow,\,\text{e},\nu}$ from:

\begin{equation}
\label{eq:irradiance}
\begin{aligned}
E_{\downarrow,\,\text{e},\nu} & = \frac{1}{4}
J_{\text{S},\,\text{e},\nu} \frac{J_{\text{B},\,\text{e},\nu}(T_{\ast,\,\text{eff}})}{J_{\text{B},\,\text{e},\nu}(T_{\text{S},\,\text{eff}})} \left(\frac{R_\ast}{a_p}\right)^2,
\end{aligned}
\end{equation}
where $J_{\text{B},\,\text{e},\nu}(T)$ is the radiosity of a black body at temperature, $T$, and $T_{\ast,\,\text{eff}}$ is the effective temperature of the star, $R_\ast$ is the radius of the star, $a_p$ is the distance between the star and the planet, and the geometric factor $1/4$ is used to represent planet-averaged conditions. The ratio of the $J_{\text{B},\,\text{e},\nu}$ terms is used to obtain a stellar spectrum at the effective temperature of K2-18. This was done instead of interpolating on the BT-Settl grid as $T_{\text{S},\,\text{eff}}$ is close to $T_{\ast,\,\text{eff}}$.

\subsection{Atmospheric model: thermochemistry}
\label{sec:thermochemical_equilibrium}
We used the recommended present atomic solar system abundances given by \citet{Lodders2019} to define our standard abundances and our reference metallicity ((Z/H)$_\odot$). The atomic abundances at a metallicity Z/H are obtained by keeping the H, He, Ne, Ar, Kr, and Xe standard abundances constant while multiplying the standard abundances of other elements by Z/H.
The list of all the species included in our thermochemical calculations is displayed in \autoref{tab:species_chemistry}. It is possible to adjust the abundances of all individual elements listed in this table, although it is always assumed that the atmosphere is dominated by H$_2$, limiting Exo-REM capabilities to metallicities $\lessapprox$ 1000 (Z/H)$_\odot$. We included species that either: significantly affect the volume mixing ratio (VMR) profiles of our absorbing species (see Section~\ref{sec:absorption_cross_sections}), are susceptible to the formation of a major cloud layer, or substantially impact the molar mass of the atmosphere \citep[according to][]{Lodders2010}.

Most of the equilibrium abundances are derived from the equation of conservation of each species, using the standard Gibbs free energy of formation $\Delta G_f^\circ$ listed in \citet{Chase1998}. For PH$_3$(g), we used the $\Delta G_f^\circ$ from \citet{Lodders1999}, while for MnS(cr,l) and ZnS(cr,l), we used the values from \citet{Robie1995}, and for CaTiO$_3$(cr) we used values from \citet{Woodfield1999}. We used a grid of temperatures between 200 and 4000 K, with a 100 K step, to map our $\Delta G_f^\circ$. When the temperature range of the references used were smaller than our target temperature range, we used a linear extrapolation to fill up our grid. The $\Delta G_f^\circ$ at a given temperature are linearly interpolated from this grid. We calculated   the saturation pressure of the following species directly: H$_2$O \citep{Wagner1993, Wagner1994, Lin2004, Fray2009}, NH$_3$ \citep{Fray2009, Lide2009}, and NH$_4$SH \citep{Stull1947}. We used a very simplified Ca-Al-Ti chemistry compared  to, for instance, \citet{Lodders2002}, but the impact on our simulations was expected to be limited. We simulated the formation of Fe--Ni alloys by treating the Fe independently while keeping  the Ni(g)/Fe(g) ratio constant. We considered the formation of H$_3$PO$_4$ instead of P$_4$O$_6$, based on the results of \citet{Wang2016}. 

We included non-equilibrium processes in the CH$_4$--CO, CO--CO$_2$, N$_2$--NH$_3$--HCN chemical systems following the approach of \citet{Zahnle2014}. The chemical quenching level is determined by equating a reaction timescale to the mixing time $H^2/K_{zz}$, where $H$ is the atmospheric scale height and $K_{zz}$ the eddy mixing coefficient. Below the quenching level, the abundance profiles of the relevant species are governed by thermochemical equilibrium while, above this level, the mixing ratios of the quenched species are held constant. We used the chemical reaction timescales given by Eqs.~12--14 for CH$_4$--CO, Eq.~44 for CO--CO$_2$, Eq.~32 for NH$_3$--N$_2,$ and Eq.~40 for HCN--NH$_3$--N$_2$ in \citet{Zahnle2014}. In the case of PH$_3$, we assume that its conversion to H$_3$PO$_4$ is inhibited, as observed in the giant planets of our solar system. We compared our chemical model in the case of K2-18b (taking identical temperature profile, elemental abundances, and eddy diffusion coefficient) against the model described in \citet{Venot2020}\footnote{O. Venot, private communication. The comparison was made using the "nominal model" described in Section~\ref{sec:metallicity_irradiance}}. For CO, HCN, and NH$_3$, we found VMR differences lower than 20$\%$ with our model in the HST sensitivity region, and lower than 2$\%$ in the case of CH$_4$ and H$_2$O. The difference is larger for CO$_2$, with our VMR being 25$\%$ higher than the value found in O. Venot's model. We consider these differences to be satisfactory.

\begin{table*}[pt]
\centering
\caption{\label{tab:species_chemistry}Species included in thermochemical calculations.}
\begin{tabular}{@{}llll@{}} 
\hline
\hline
\multirow{3}{*}{Element}        & Standard                                              & \multirow{3}{*}{Gases}  & \multirow{3}{*}{Condensates}  \\
                                                        & abundance                                     &                                                                       &                                               \\
                                                        & ($\times10^{-6}$)                             &                                                                       &                                               \\                                              
\hline                      
H               & 10$^{6}$                              & H$_2$\tablefootmark{$\ast$}, H, H$_2$O\tablefootmark{$\ast$}, CH$_4$\tablefootmark{$\ast$}, ...                      &                                 \\
He              & 83950                                 & He\tablefootmark{$\ast$}                                      &                                \\
C\tablefootmark{a}& 295                                 & CH$_4$\tablefootmark{$\ast$}, CO\tablefootmark{$\ast$}, CO$_2$\tablefootmark{$\ast$}, HCN\tablefootmark{$\ast$}       &                                 \\
N\tablefootmark{a}& 70.8                                        & NH$_3$\tablefootmark{$\ast$}, N$_2$, HCN\tablefootmark{$\ast$}                                                        & NH$_3$, NH$_4$SH, NH$_4$Cl                              \\
O\tablefootmark{a}& 537                                 & H$_2$O\tablefootmark{$\ast$}, CO\tablefootmark{$\ast$}, CO$_2$\tablefootmark{$\ast$}, SiO                     & H$_2$O, Mg$_2$SiO$_4$, MgSiO$_3$, SiO$_2$, Al$_2$O$_3$, Cr$_2$O$_3$ \\
Ne              & 141                                   & Ne                                                    &                               \\
Na              & 1.26                                  & Na\tablefootmark{$\ast$}, NaCl                                            & Na$_2$S               \\
Mg              & 33.1                                  & Mg                                                    & Mg$_2$SiO$_4$, MgSiO$_3$                \\
Al              & 2.63                                  & Al                                                    & Al$_2$O$_3$     \\
Si              & 32.4                                  & SiO, SiH$_4$                                  & Mg$_2$SiO$_4$, MgSiO$_3$, SiO$_2$       \\
P               & 0.269                                 & PH$_3$\tablefootmark{$\ast$}, PH$_2$, PO, P$_2$       & H$_3$PO$_4$\tablefootmark{b}          \\
S               & 14.1                                  & H$_2$S\tablefootmark{$\ast$}                          & NH$_4$SH, Na$_2$S, MnS, ZnS             \\
Cl              & 0.17                                  & HCl, NaCl, KCl                        & NH$_4$Cl, KCl           \\
Ar              & 3.16                                  & Ar                                                    &                               \\      
K               & 0.117                                 & K\tablefootmark{$\ast$}, KCl                                             & KCl                   \\
Ca              & 1.86                                  & Ca                                                    & CaTiO$_3$\\
Ti              & 0.0794                                & Ti, TiO\tablefootmark{$\ast$}, TiO$_2$                         & TiN, CaTiO$_3$                \\
V               & 0.00891                               & V, VO\tablefootmark{$\ast$}, VO$_2$                                  & VO, CaTiO$_3$\tablefootmark{c}        \\
Cr              & 0.427                                 & Cr                                                    & Cr, Cr$_2$O$_3$                 \\
Mn              & 0.295                                 & Mn                                                    & MnS                     \\
Fe              & 28.2                                  & Fe, FeH\tablefootmark{$\ast$}                                         & Fe                      \\
Ni              & 1.58                                  & Ni                                                    & Fe\tablefootmark{d} \\
Zn              & 0.0407                                & Zn                                                    & ZnS             \\
Kr              & 0.00166                               & Kr                                                    &                         \\
Xe              & 0.00018                               & Xe                                                    &                         \\
\hline
\end{tabular}
\tablefoot{
Elements are classed by increasing atomic number. Elements not displayed are not taken into account. Aside from H, are mentioned only the species affecting the element equilibrium (i.e. TiO, VO and PO do not affect the abundance of O).
\tablefoottext{a}{Non-equilibrium chemistry based on a comparison of chemical time constants with vertical mixing time from \citet{Zahnle2014}.}
\tablefoottext{b}{Only at equilibrium.}
\tablefoottext{c}{Dissolution of VO into CaTiO$_3$, assuming an ideal solid solution and Henri's law.}
\tablefoottext{d}{Formation of Fe-Ni alloys.}
\tablefoottext{$\ast$}{Species for which lines and/or CIA are taken into account.}
}
\end{table*}

\section{Methodology}
\label{sec:methodology}

\begin{table*}[pt]
\centering
\caption{\label{tab:atmospheric_grid} Model grid}
\begin{tabular}{@{}lcc@{}} 
\hline
\hline
Parameter                                               & Range                         & Nominal value \\ 
\hline
Metallicity ((Z/H)$_\odot$)             & 1--1000                       & 175 \\
C/H ((C/H)$_\odot$)                             & 0.3--1000                     & 175 \\
Irradiance ($E_{p,\,\text{e}}$) & 0.5--1.5                      & 1 \\
$K_{zz}$ (cm$^2\cdot$s$^{-1}$)  & $10^5$--$10^{10}$     & $10^6$ \\
$r$ ($\mu$m)                                    & 20--600                       & 50 \\
$T_{p,\,\text{int}}$ (K)                & 45--200                       & 80 \\
 \hline
\end{tabular}
\end{table*}

The star and planetary parameters used in our simulations are displayed in \autoref{tab:general_parameters}. A summary of our model grid is displayed in \autoref{tab:atmospheric_grid}. We took a planetary radius at $10^5$ Pa of $R_p = 16400$ km, which is slightly different from the value used by \citet{Benneke2019} (who define $R_p$ as the radius at $10^3$ Pa), because it is closer to the radius we find when fitting the observed data. To compare our model spectrum with the data, we applied an offset on $R_p$ in the calculation of the transit depth, such that the $\chi^2$ is minimised.

The internal temperature resulting from the residual heat of formation of the planet was calculated following \citep{Rogers2010}:
\begin{equation}
\label{eq:effetive_temperature}
\begin{aligned}
T_{p,\,\text{int}} &= \left( \frac{L_{\text{int}}}{4\pi\sigma R_p^2} \right)^{1/4},
\end{aligned}
\end{equation}
where $L_{\text{int}}$ is obtained from \citep{Rogers2010}:
\begin{equation}
\begin{aligned}
\log_{10}\left( \frac{L_{\text{int}}}{L_\odot} \right) &= a_1 + a_{M_p}\log_{10}\left( \frac{M_p}{M_\oplus} \right) \\
&\quad+ a_{R_p}\log_{10}\left( \frac{R_p}{R_{\jupiter}} \right) \\
&\quad + a_{t_p}\log_{10}\left( \frac{t_p}{\text{1 Gyr}} \right),
\end{aligned}
\end{equation}
where $a_1 = -12.46 \pm 0.05$, $a_{M_p} = 1.74 \pm 0.03$, $a_{R_p} = -0.94 \pm 0.09$, $a_{t_p} = -1.04 \pm 0.04$, $t_p$ is the age of the planet, and the astronomical constants are defined by the International Astronomical Union \citep[IAU,][]{Mamajek2015}. According to a gyrochronological model from \citet{Guinan2019}, the age of K2-18 can be estimated at 2.4 $\pm$ 0.6 Gyr. Assuming that K2-18b ended its formation a few Myr after the formation of its star (so that $t_\ast \approx t_p$), like what happened in our solar system, we obtained from \autoref{eq:effetive_temperature} $T_{p,\,\text{int}} \approx 83^{+8}_{-6}$ K. Rounding down this value, we chose 80 K as our nominal internal temperature. 

The net fluxes are calculated from 40 to 30000 cm$^{-1}$, with a step of 20 cm$^{-1}$. The atmospheric grid consists of 81 levels equally spaced in the log-space between 0.1 and $10^7$ Pa. We imposed a correlation length of 0.5 pressure scale height to the solution temperature profile in order to avoid non physical oscillations \citep[see][] {Baudino2015}. We considered cloud radiative effects and scattering only from H$_2$O clouds. The NH$_3$ and NH$_4$SH clouds would form in atmospheres that are colder than what is expected for K2-18b, while NH$_4$Cl clouds are too thin to have a significant impact, and other clouds are condensing too deeply into the atmosphere (see Section~\ref{sec:clouds}). Cloud vertical mass mixing ratios are calculated assuming equilibrium between the downward flux of falling particles and the upward flux of gas and cloud particles due to advection and turbulent mixing \citep[see][]{Charnay2018}.

When we change the irradiance of the planet by a factor of $k$, we do so by multiplying $a_p$ by $1/\sqrt{k}$ (see \autoref{eq:irradiance}). We proceed this way to make model comparison easier. Indeed, $a_p$ impacts only the irradiance, while the two other parameters, $T_{\ast,\,\text{eff}}$ and $R_\ast$, affects respectively the stellar spectrum and both the stellar spectrum and the transmission spectrum.

We estimated $K_{zz}$ using the values derived by Exo-REM \citep[see][]{Charnay2018}. We found that typical $K_{zz}$ values for K2-18b range from $10^6$ to $10^9$ cm$^2\cdot$s$^{-1}$, with the highest values found in the convective layers. Given the uncertainties on the estimation of $K_{zz}$, we enlarged this range to $10^5$--$10^{10}$ cm$^2\cdot$s$^{-1}$. We also found that a quenching of our species often occurs just below the uppermost convective layer (see Section~\ref{sec:kzz_tint_effect}), hence, we set our nominal $K_{zz}$ value at 10$^6$ cm$^2\cdot$s$^{-1}$. 

We used a fractional area covered by clouds $f_c = 0.15$ for the calculation of the temperature profile, and $f_c = 1.0$ for the calculation of the transmission spectra. To ensure numerical stability, the cloud mean particle radius, $r,$ was fixed at a constant value (50 $\mu$m) that roughly corresponds to the maximum one predicted by the Exo-REM self-consistent cloud model at 300 (Z/H)$_\odot$ and nominal irradiation \citep[see][and Section~\ref{sec:clouds}]{Charnay2018}, unless stated otherwise. We call this our nominal model. We explore the effects of changing some of these parameters in the following sections.

\section{Results}
\label{sec:results}
\subsection{Metallicity and irradiance}
\label{sec:metallicity_irradiance}
Here, we assume that K2-18b has retained a relatively thick H$_2$--He atmosphere. This atmosphere can be enriched in heavy elements -- compared to the initial proto-stellar nebula -- during the formation of the planet via collisions with planetesimals \citep{Fortney2013}. There could also be a contribution from the erosion of the primordial core \citep{Iaroslavitz2007}.

\begin{figure*}[pt]
\centering
\includegraphics[width=1.0\linewidth]{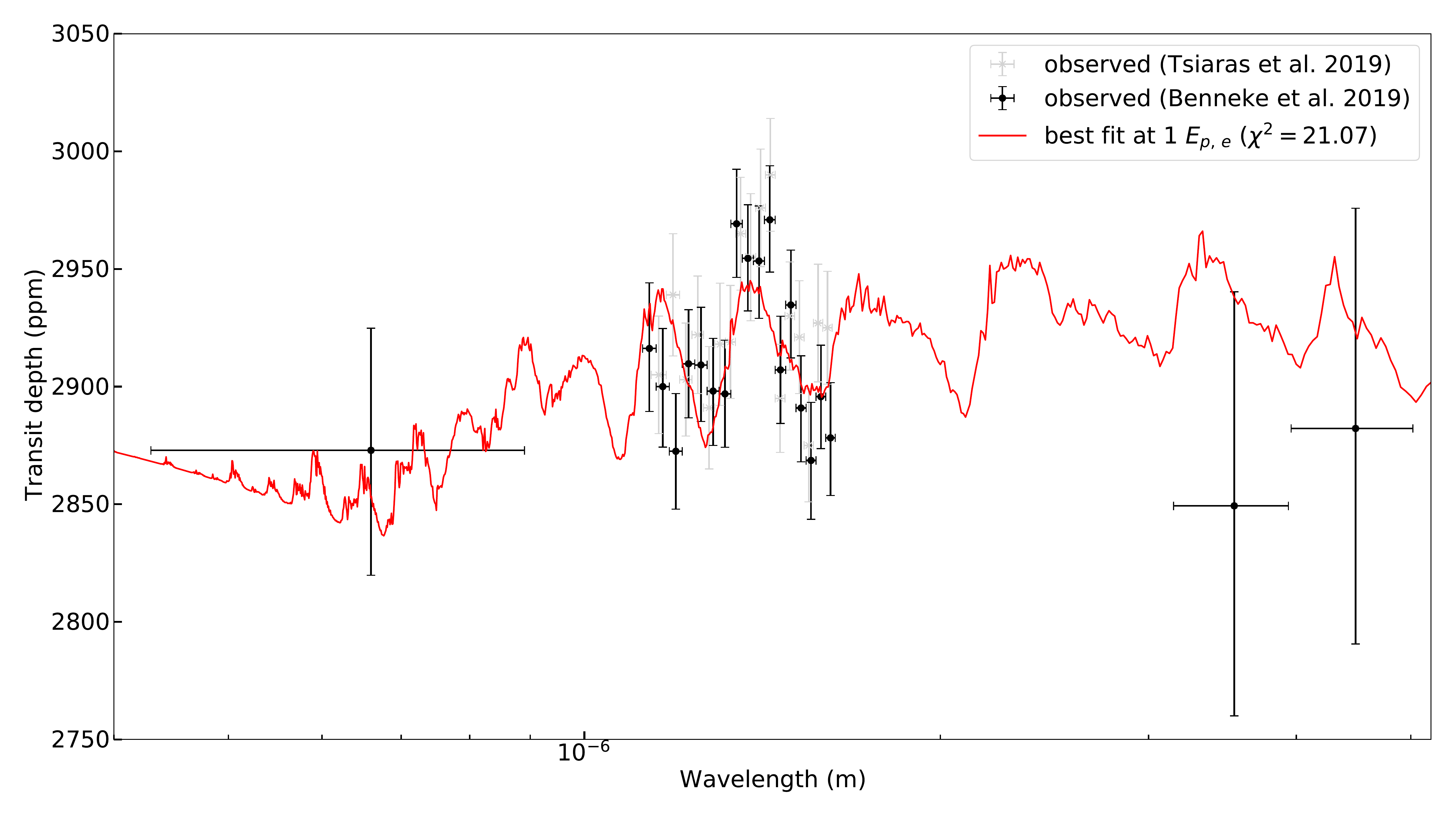}
\caption[Best fit]{Best fit of our model to the dataset of \citet{Benneke2019} at nominal irradiation (1368 W$\cdot$m$^{-2}$). Black: K2, HST and Spitzer data from \citet{Benneke2019}. Grey: HST data from \citet{Tsiaras2019}. Red: Exo-REM transmission spectrum at 175 (Z/H)$_\odot$, with an offset of the $10^5$-Pa level of $-5$ km. The $\chi^2$ of this spectrum against \citet{Benneke2019} data is indicated in parentheses.}
\label{fig:best_fit}
\end{figure*}

\begin{figure*}[pt]
\centering
\includegraphics[width=1.0\linewidth]{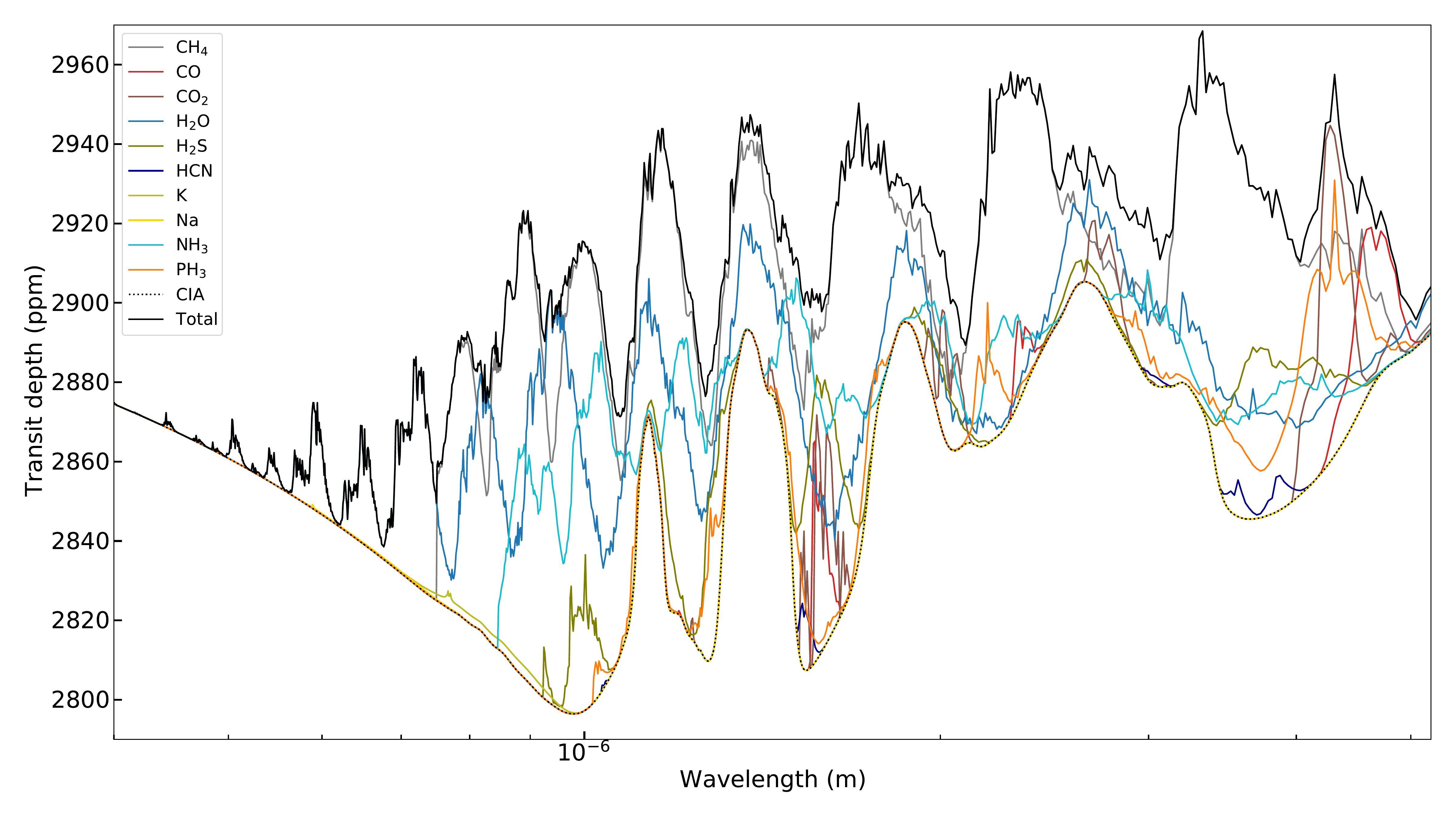}
\caption[Contributions]{Contributions of the absorbing species to our best-fit transmission spectrum at nominal irradiation within the K2 to Spitzer spectral range. The H$_2$O cloud is not forming in this case, so there is no cloud contribution. The spectral contribution of individual species takes the CIA and Rayleigh scattering (represented as a dotted curve) into account. The spectral contribution of FeH, TiO, and VO are not represented here as they are insignificant.}
\label{fig:spectrum_contribution}
\end{figure*}

\begin{figure}[pt]
\centering
\includegraphics[width=1.0\linewidth]{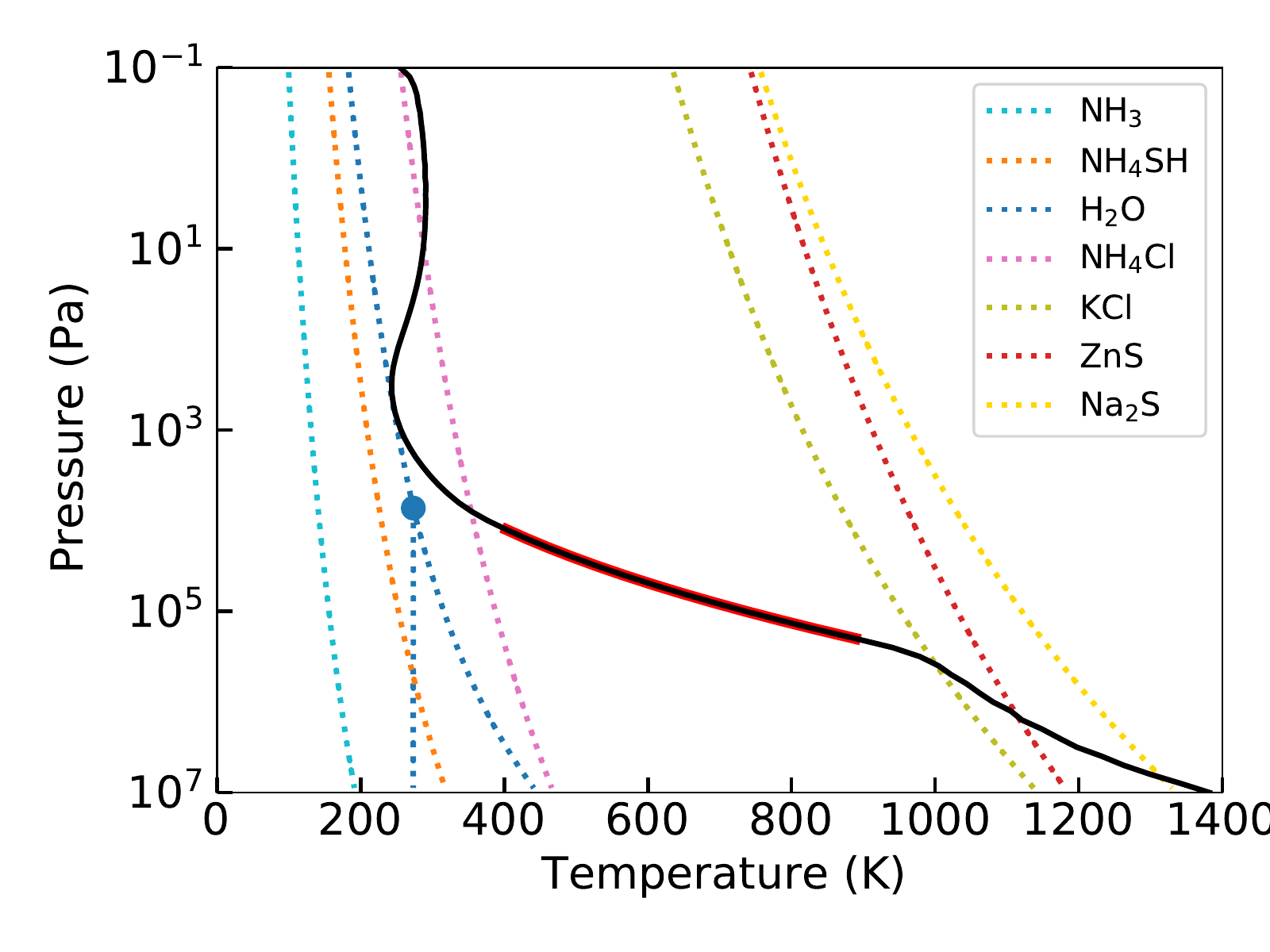}
\caption[Best fit temperature profile]{Temperature profile of our best fit atmospheric model against the dataset from \citet{Benneke2019} at nominal irradiation. Solid line: Temperature profile. Dotted lines: Condensation profiles of selected species. Red line: Convective layers. Blue dot: H$_2$O ice-Ih--liquid--gas triple point.}
\label{fig:best_fit_temperature}
\end{figure}

\begin{figure}[pt]
\centering
\includegraphics[width=1.0\linewidth]{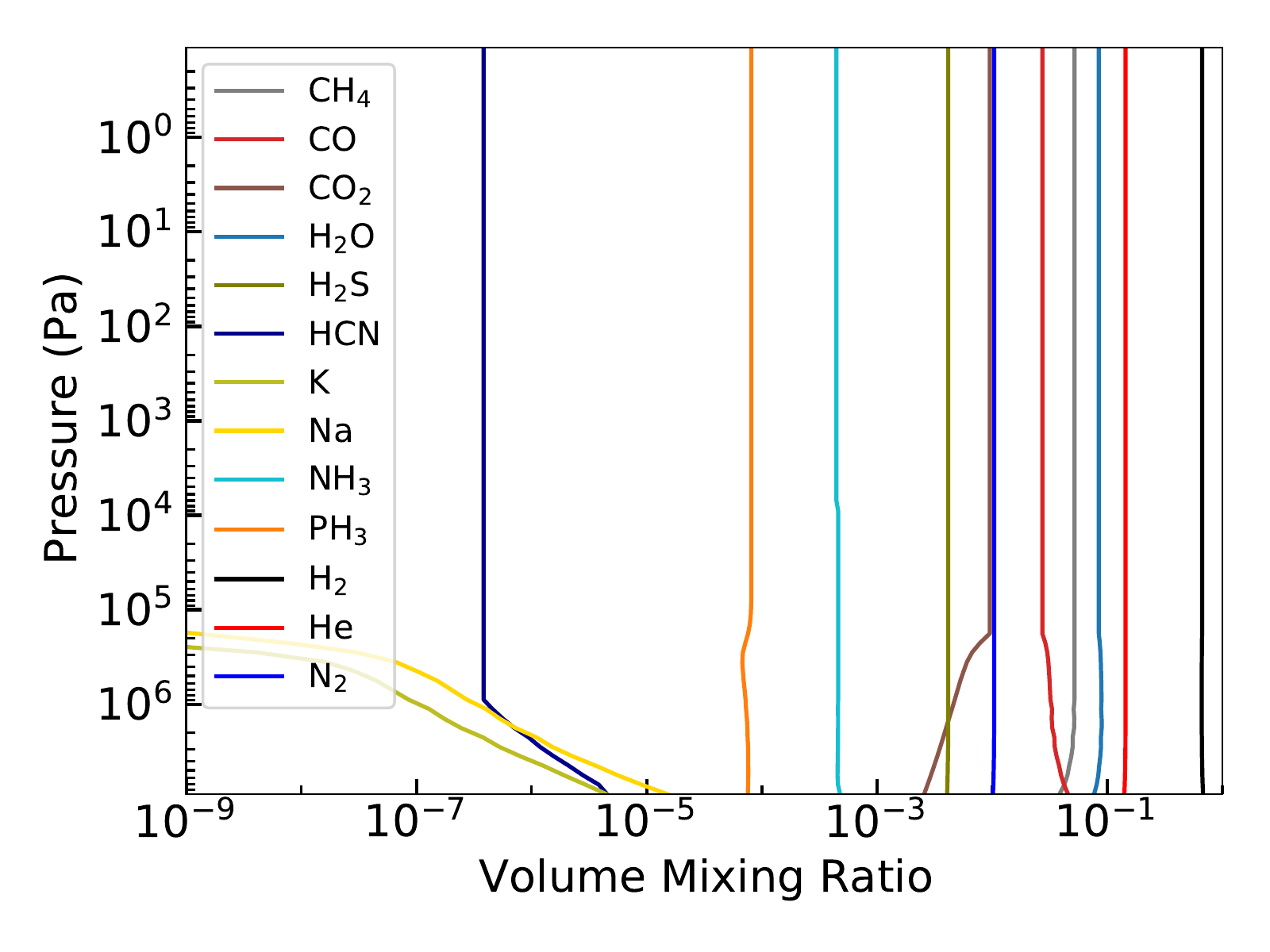}
\caption[Best fit Volume Mixing Ratio]{VMR of our best fit atmospheric model against the dataset from \citet{Benneke2019} at nominal irradiation (175 (Z/H)$_\odot$, $K_{zz} = 10^6$ cm$^2\cdot$s$^{-1}$). For clarity, only absorbing species and N$_2$ are represented. FeH, TiO, and VO are not represented as their respective abundances are, respectively, $< 10^{-12}$, $< 10^{-35}$ , and $< 10^{-36}$ within our pressure grid. The bump of the PH$_3$ VMR at $200$ kPa is due to the chemical equilibrium between PH$_3$ and P$_2$ peaking in favour of P$_2$ at this pressure.}
\label{fig:best_fit_vmr}
\end{figure}

In \autoref{fig:best_fit}, we show our best fit for the dataset of \citet{Benneke2019} at nominal irradiation, along with the HST data reduced by \citet{Tsiaras2019}. The contributions of the different opacity sources to this spectrum are represented in \autoref{fig:spectrum_contribution}. The temperature profile and VMR of this spectrum are represented respectively in \autoref{fig:best_fit_temperature} and \autoref{fig:best_fit_vmr}. Because the stellar spectrum we used is slightly more intense than a black body between $\approx$ 1 and 2.5 $\mu$m, the temperatures we obtain in the upper atmosphere are slightly higher ($\approx$ 5 K at 1 kPa) than what would be obtained with a black body at the effective temperature of the star. The prevalence of CH$_4$ absorptions over H$_2$O absorptions is discussed in Section~\ref{sec:h2o_vs_ch4}.

\begin{figure*}[pt]
\begin{subfigure}{0.5\textwidth}
\centering
\includegraphics[width=1.0\linewidth]{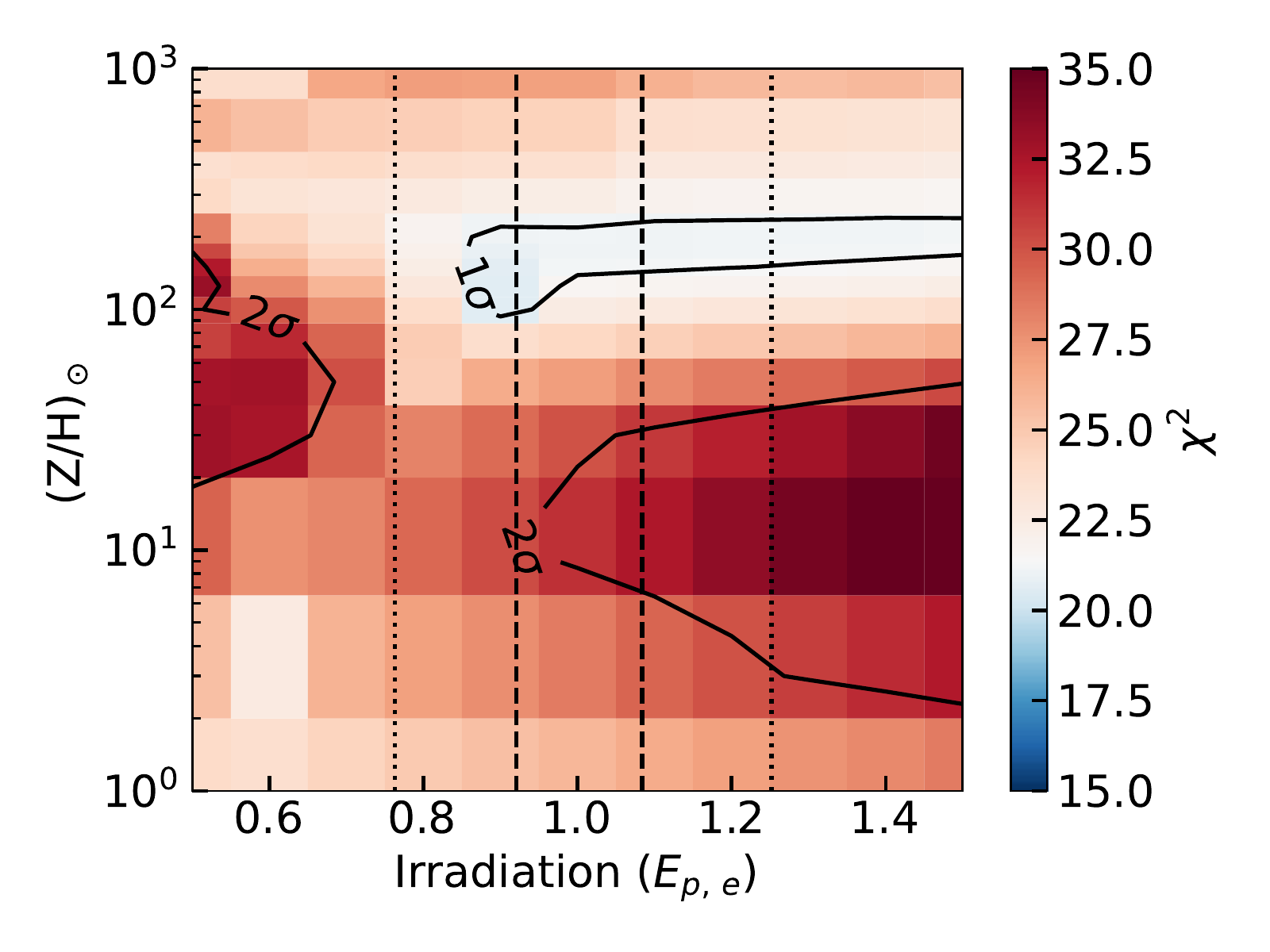}
\end{subfigure}
\begin{subfigure}{0.5\textwidth}
\centering
\includegraphics[width=1.0\linewidth]{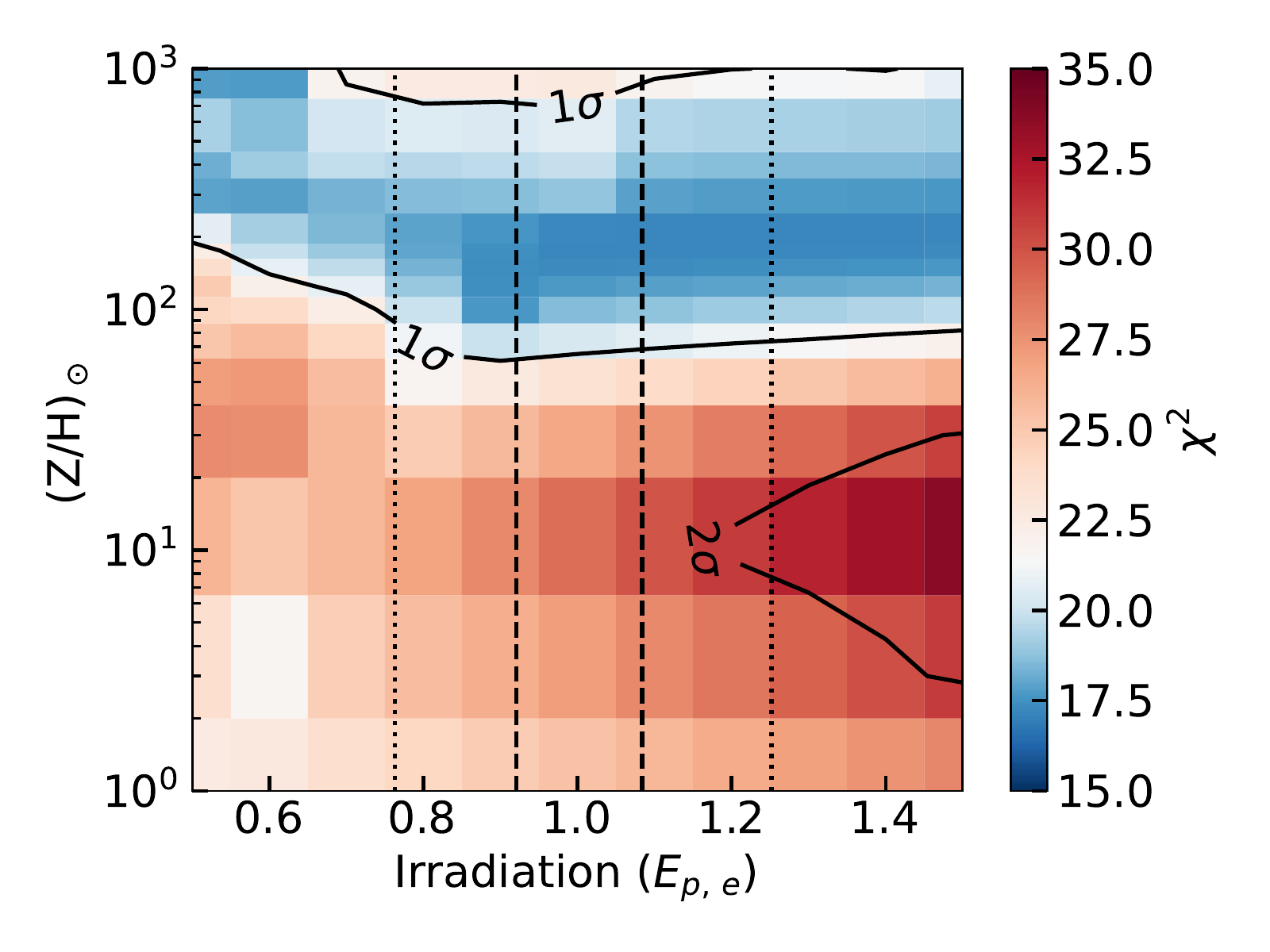}
\end{subfigure}
\caption[Goodness of fit]{Goodness of fit of our nominal models, varying only irradiation and metallicity. Left: Data from \citet{Benneke2019}, including HST, K2, and Spitzer data. Right: Data from \citet{Tsiaras2019}, including HST, K2, and Spitzer data. The white colour corresponds to the value of $\chi^2$ of 21.36, indicating a fit of the datasets at the 1$\sigma$ confidence level. Accordingly, the blue colour indicates overfit and the red colour underfit of the datasets. The solid lines represents the 1 and 2$\sigma$ confidence levels of our fits. The dashed and dotted lines represents, respectively, the 1 and 3$\sigma$ uncertainties on K2-18b irradiation.}
\label{fig:chi2}
\end{figure*}

In \autoref{fig:chi2}, we display the $\chi^2$ of our nominal models for the datasets from \citet{Benneke2019} and \citet{Tsiaras2019}, including their respective reduction of the same HST raw data, as well as in both cases the K2 and Spitzer data from \citet{Benneke2019}. We performed simulations of K2-18b for 1, 3, 10, 30, 50, 75, 100, 125, 150, 175, 200, 300, 400, 500, and 1000 times (Z/H)$_\odot$ and for irradiations between 0.5 and 1.5 time the nominal irradiation, with a step of 0.1. We consider a model as statistically accurate if it can reproduce the data within the 1$\sigma$ confidence level (68$\%$). Since there are 20 data points, we have 20 to 1 degrees of freedom, so the 1$\sigma$ confidence level corresponds to a $\chi^2$ of 21.36. While the dataset from \citet{Tsiaras2019} allows, according to our interpolation, for metallicities $\geq 65$ (Z/H)$_\odot$ within K2-18b irradiation $3\sigma$ uncertainties, the dataset from \citet{Benneke2019} is much more restrictive, allowing only a metallicity between 100 and 200 (Z/H)$_\odot$ at 0.9 times the nominal irradiation, or a metallicity between 150 and 200 (Z/H)$_\odot$ at or above the nominal irradiation. These values are within the most common range of metallicity predicted by \citet{Fortney2013} for planets with radius in the 2--4 R$_\oplus$ range (between 100 and 400$+$ (Z/H)$_\odot$). At nominal irradiation, our best fit against \citet{Benneke2019} and \citet{Tsiaras2019} data is located respectively at 175 (Z/H)$_\odot$ ($\chi^2 = 21.07$) and 150 (Z/H)$_\odot$ ($\chi^2 = 17.24$). In both cases, there is no H$_2$O cloud formation, and we derive a Bond albedo of respectively 0.017 and 0.018.

\begin{figure}[pt]
\centering
\includegraphics[width=1.0\linewidth]{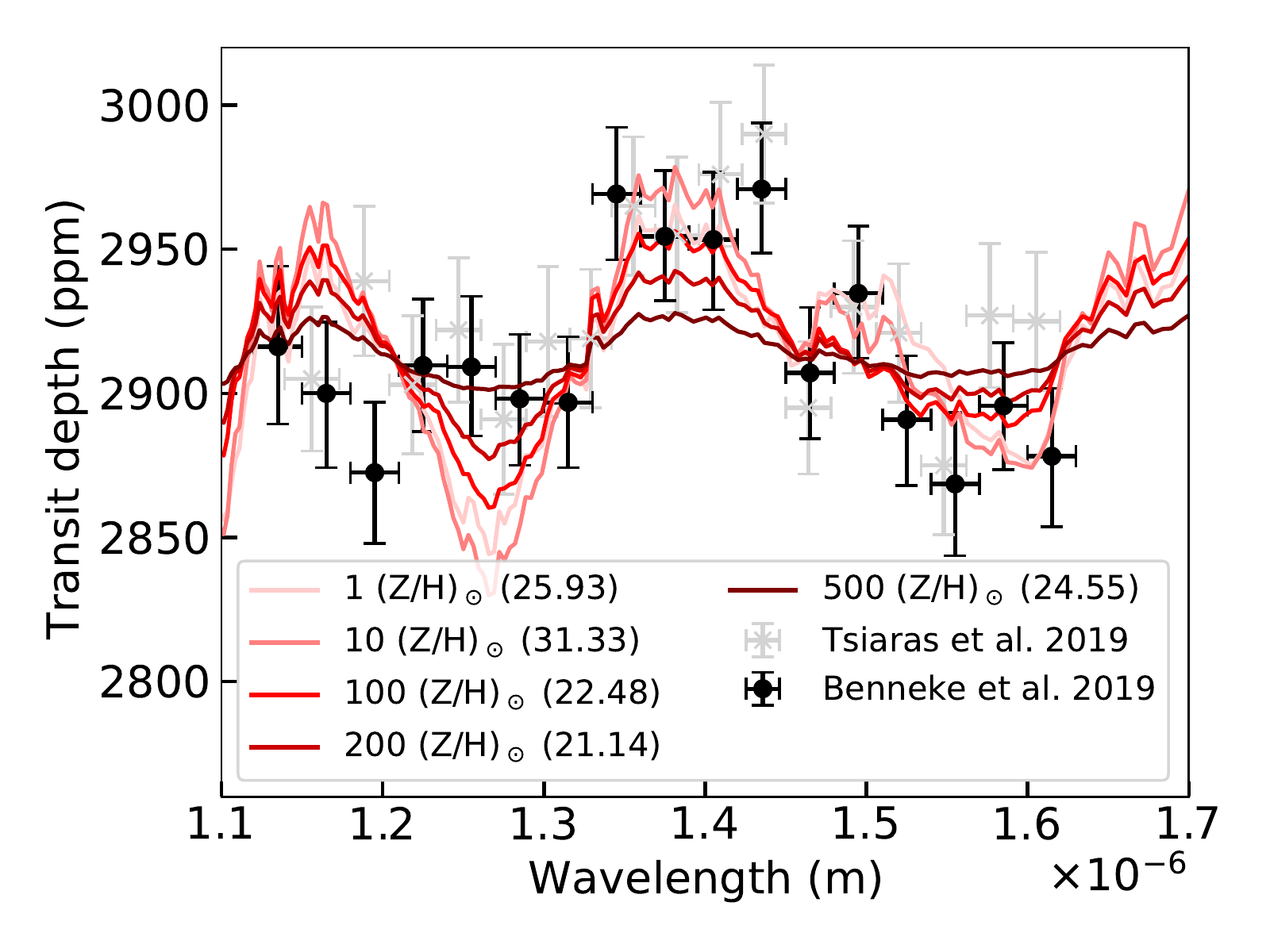}
\caption[Transmission spectra and metallicity]{Effect of metallicity on the transmission spectrum at nominal irradiation. The corresponding value of $\chi^2$ against \citet{Benneke2019} data is indicated in parentheses.}
\label{fig:transmission_metallicity}
\end{figure}

\begin{figure}[pt]
\centering
\includegraphics[width=1.0\linewidth]{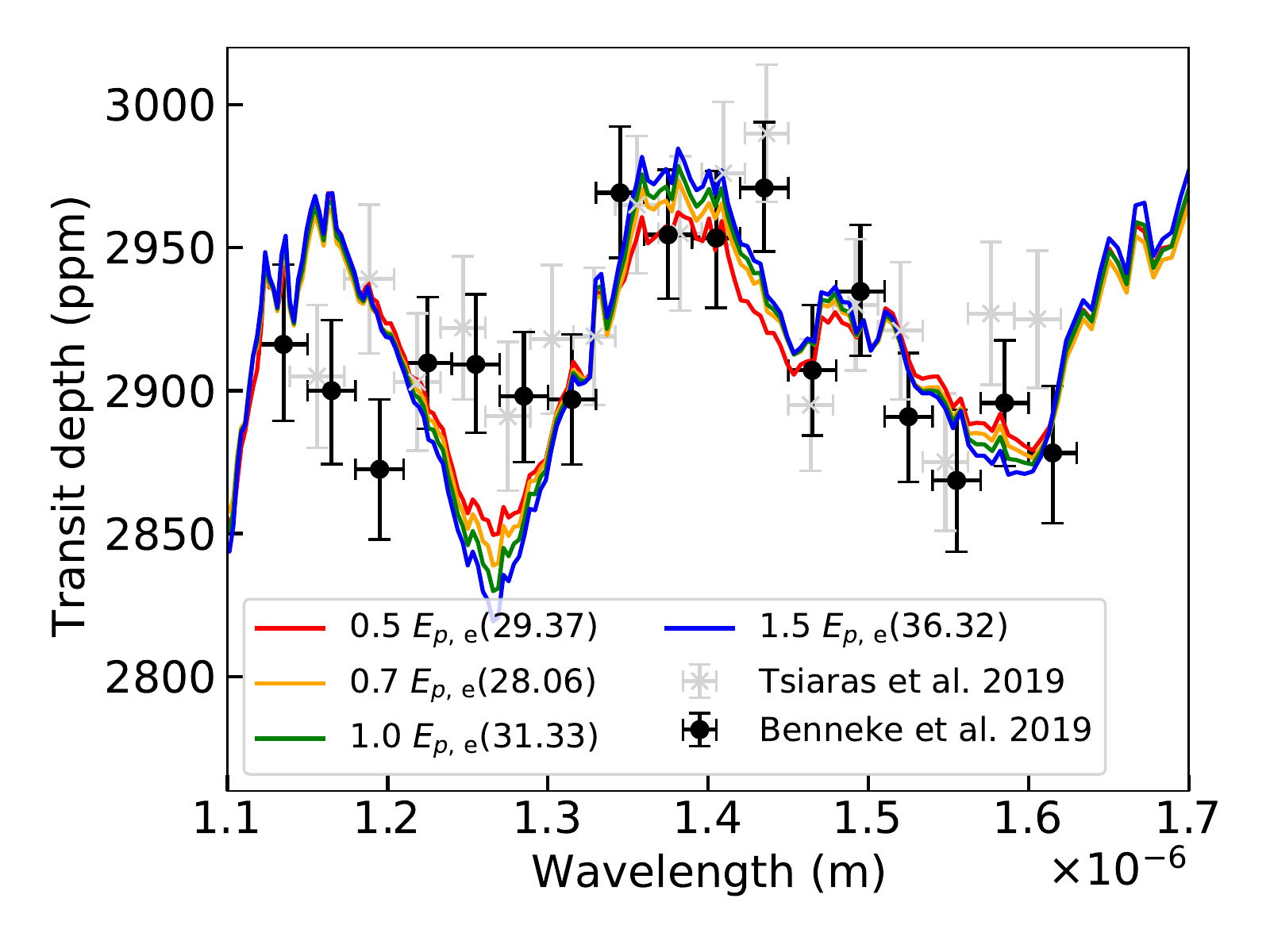}
\caption[Transmission spectra and irradiation]{Effect of irradiation on the transmission spectrum at 10 (Z/H)$_\odot$. The corresponding value of $\chi^2$ against \citet{Benneke2019} data is indicated in parentheses.}
\label{fig:transmission_irradiance}
\end{figure}

The relationship between temperature, metallicity, and goodness of fit presented in \autoref{fig:chi2} can be explained as follows. The amplitudes of the features in the transmission spectra are correlated with the abundance of absorbers in the atmosphere as well as to its scale height, which can be written as
\begin{equation}
\label{eq:scale_height}
\begin{aligned}
H & = \frac{R T}{\mu g},
\end{aligned}
\end{equation}
where $R$ is the gas constant, and $T$, $\mu$ and, $g$ are respectively the temperature, the molar mass and the gravity in an atmospheric layer.
As the metallicity decreases, two concurrent effects are occurring. The diminution of the VMR of absorbers in the atmosphere will of course decrease the spectrum amplitude. However, decreasing the VMR of heavy species will also leads to a decrease of $\mu$, and thus to an increase of the amplitudes. This latter effect is dominant in our simulations for metallicities $\geq$ 10 (Z/H)$_\odot$, as illustrated in \autoref{fig:transmission_metallicity}. At lower metallicities, heavy elements contribute to less than 10$\%$ of $\mu$, so the effect of less intense gas absorptions becomes dominant, flattening the transmission spectrum. If we decrease irradiation, the atmosphere gets colder and H$_2$O clouds start to form and to thicken, flattening the spectrum at  transparent wavelengths where part of the transmitted stellar light reaches the condensation level. As stellar irradiation decreases, the cloud forms lower and lower in the atmosphere, eventually ending below the region probed by transit spectroscopy where it can no longer directly affect the spectrum. The condensation also removes more gaseous H$_2$O, decreasing $\mu$ and H$_2$O absorption. On the other hand, increasing the irradiation leads to an increase of the temperatures, and, hence, of the scale height and of the absorption amplitudes, as shown in \autoref{fig:transmission_irradiance}.

\subsection{Clouds}
\label{sec:clouds}
\begin{figure}[pt]
\centering
\includegraphics[width=1.0\linewidth]{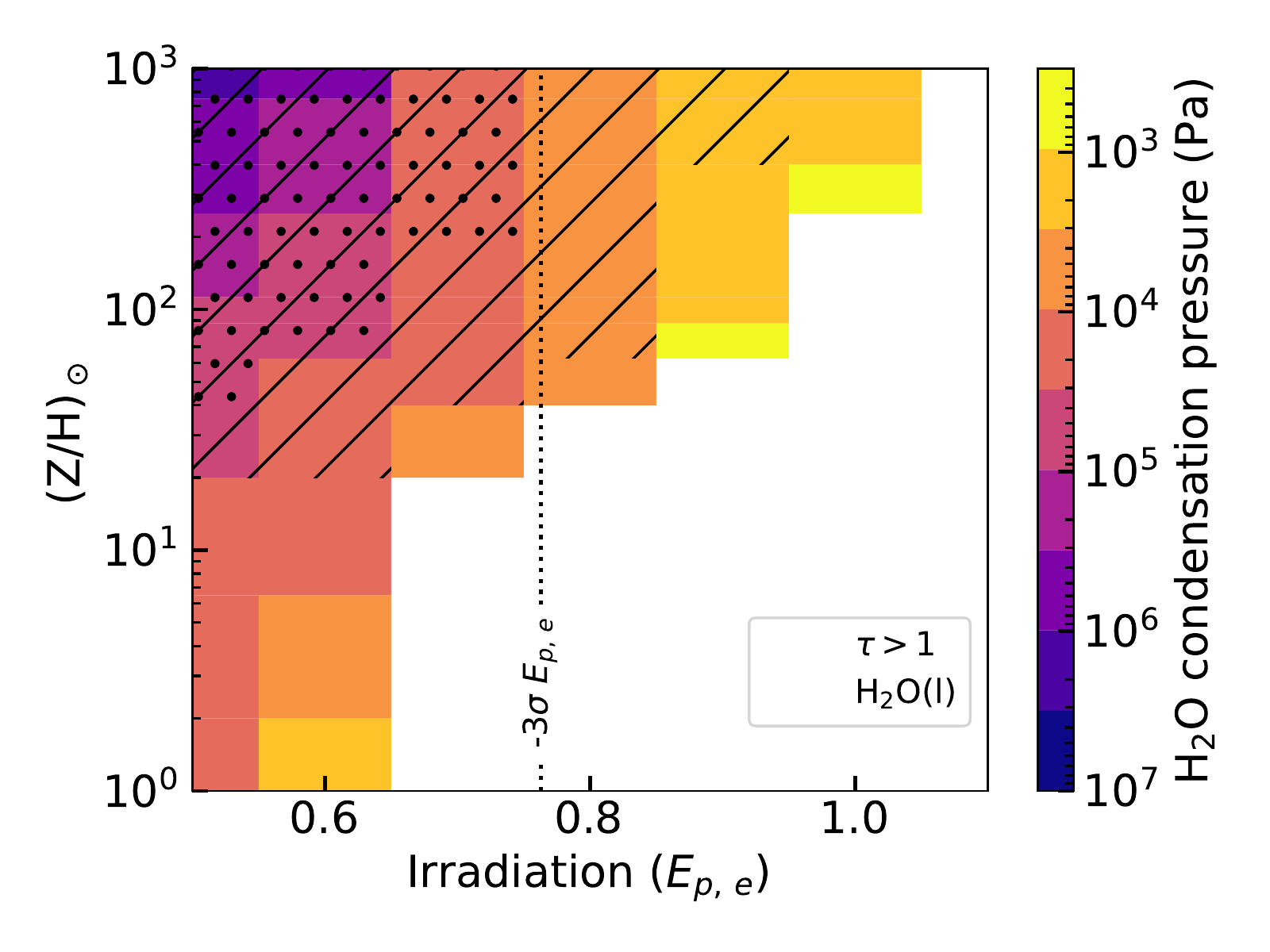}
\caption[H$_2$O condensation pressure]{H$_2$O condensation pressure level of our nominal models, varying only irradiation and metallicity. White squares: No cloud formation. Strips: Simulation where the H$_2$O clouds are optically thick (normal optical depth $\tau > 1$ at 1 $\mu$m ). Dots: Simulations where H$_2$O condenses into its liquid phase. Dotted line: 3$\sigma$ lower limit of K2-18b irradiation.}
\label{fig:h2o_condensation_pressure}
\end{figure}

\begin{figure*}[pt]
\centering
\includegraphics[width=1.0\linewidth]{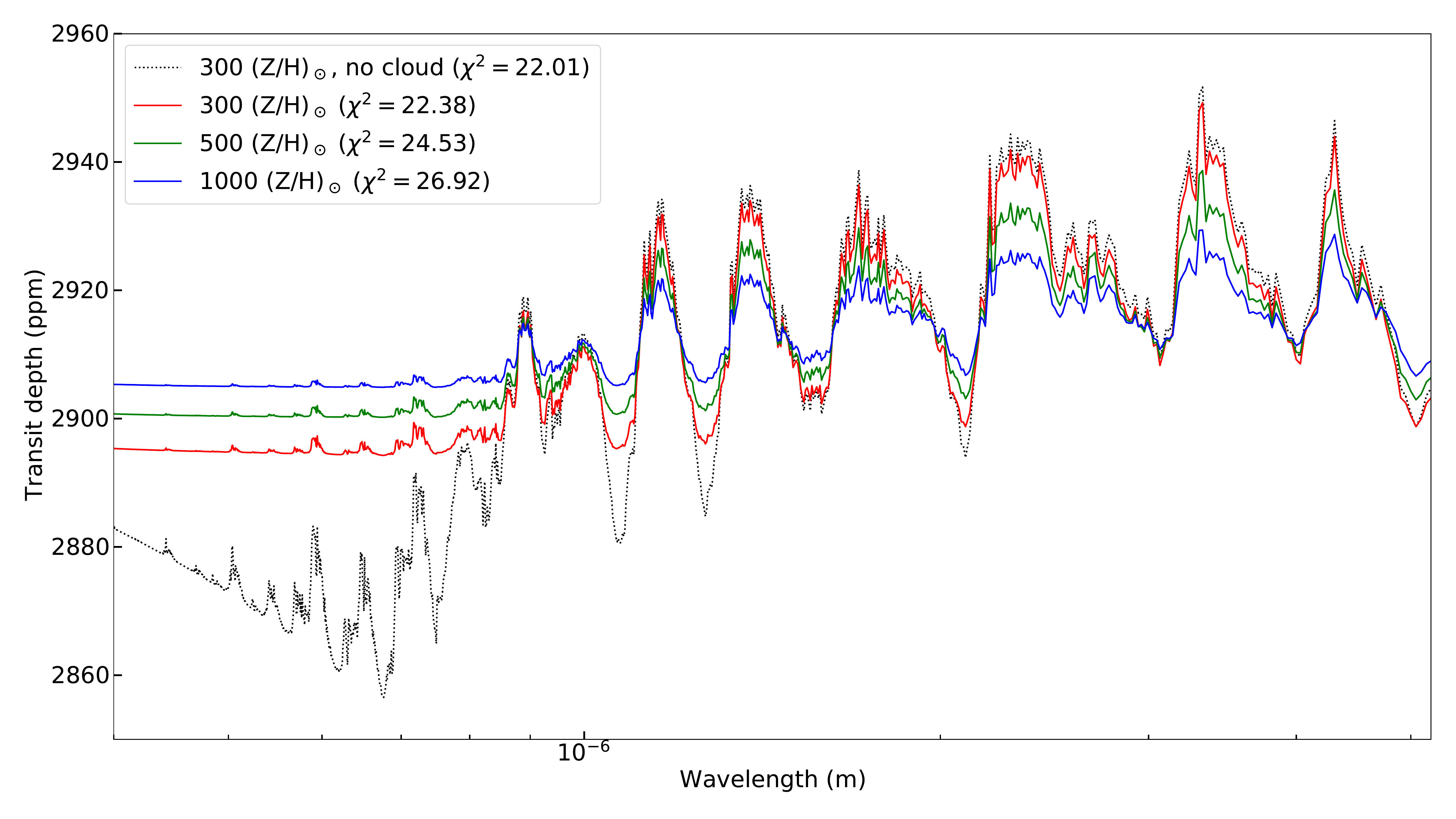}
\caption[Cloud effect on transmission spectrum]{H$_2$O cloud and metallicity effect on the transmission spectrum, at nominal irradiation. The $\chi^2$ against \citet{Benneke2019} data is indicated in parentheses.}
\label{fig:transmission_spectrum_cloud_effect}
\end{figure*}

\begin{figure}[pt]
\centering
\includegraphics[width=1.0\linewidth]{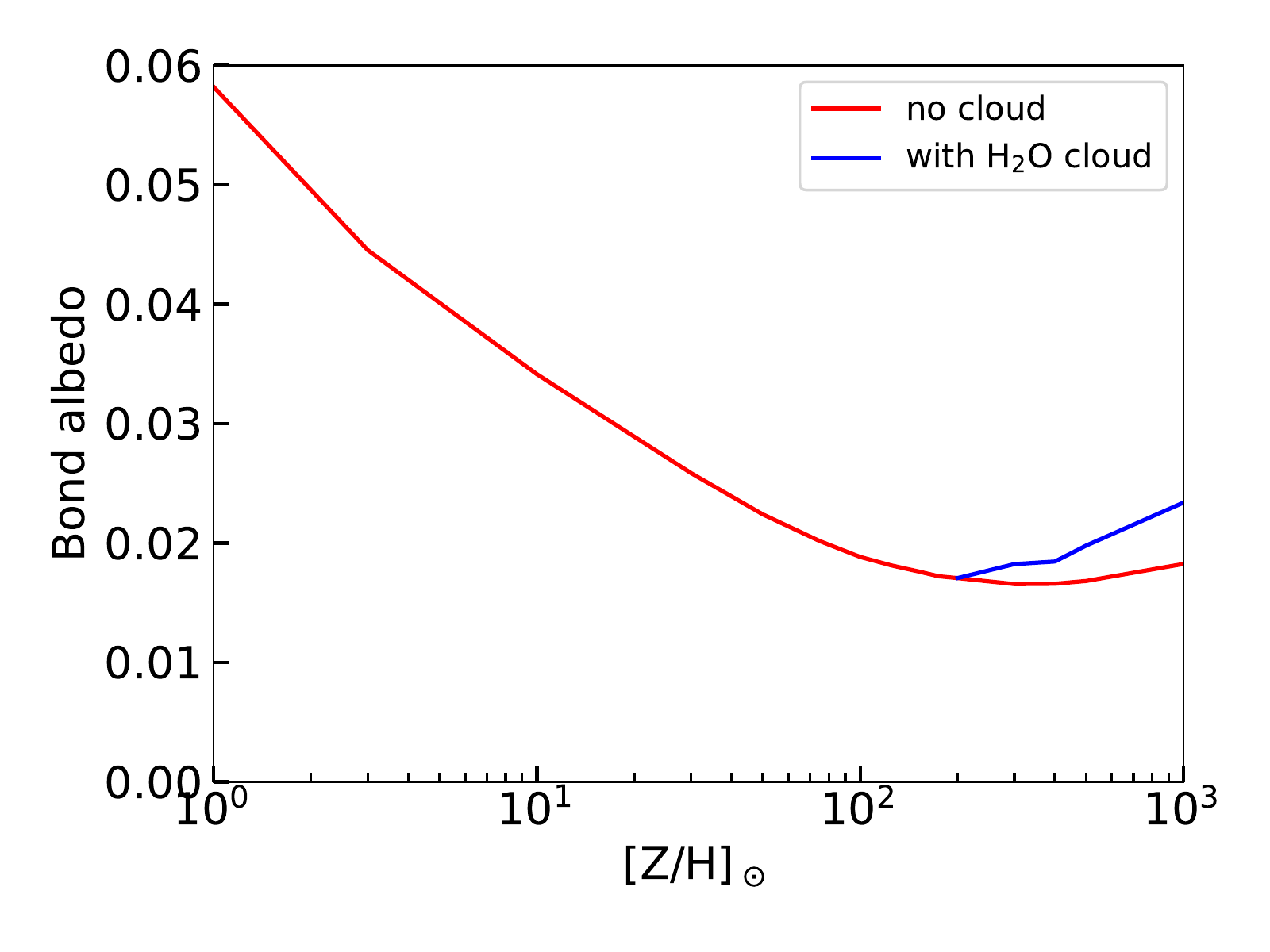}
\caption[Bond albedo and metallicity]{H$_2$O cloud and metallicity effect on the Bond albedo of K2-18b, at nominal irradiation. Cloud condensation occurs only above 200 (Z/H)$_\odot$.}
\label{fig:bond_albedo_metallicity}
\end{figure}

\begin{figure*}[pt]
\centering
\includegraphics[width=1.0\linewidth]{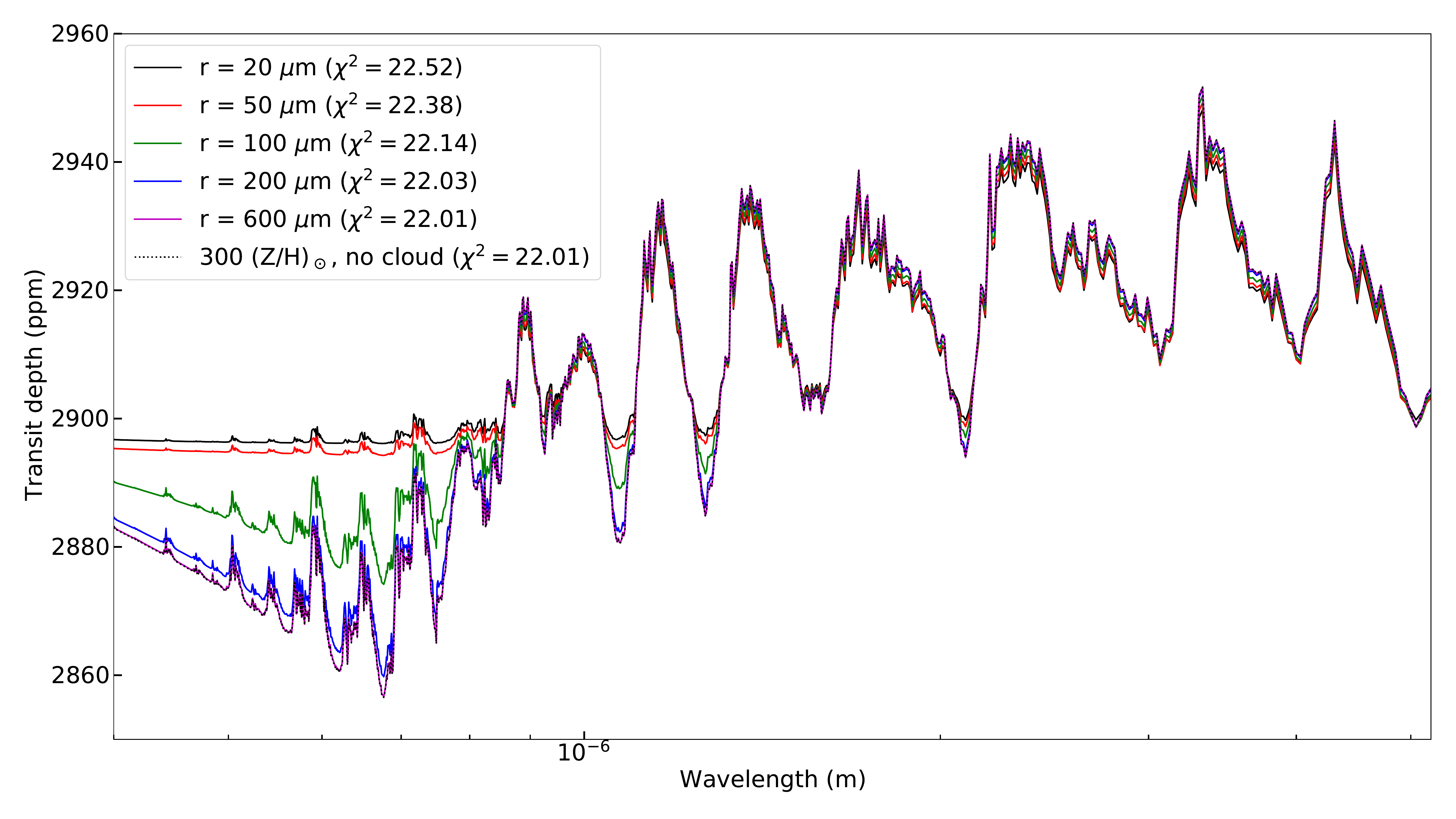}
\caption[Cloud particle radius effect on transmission spectrum]{H$_2$O cloud particle radius effect on the transmission spectrum, at 300 (Z/H)$_\odot$ and at nominal irradiation. The $\chi^2$ against \citet{Benneke2019} data is indicated in parentheses.}
\label{fig:transmission_spectrum_cloud_particle_size_effect}
\end{figure*}

In \autoref{fig:h2o_condensation_pressure}, we show that H$_2$O cloud can form at nominal irradiation and metallicities above 200 (Z/H)$_\odot$. At 300 (Z/H)$_\odot$, the cloud forms at 1 kPa. However, this cloud cannot form even at 1000 (Z/H)$_\odot$ if the irradiation is more than 10$\%$ higher than the nominal irradiation. As a general tendency, the H$_2$O cloud forms lower in the atmosphere as the irradiation decreases and the metallicity increases. This is because as the irradiation decreases, the temperature decreases, so the temperature profile crosses the H$_2$O condensation profile at higher pressures. Also, as the metallicity increases, the partial pressure of H$_2$O increases, shifting its condensation profile towards higher temperatures. 

The effect of H$_2$O clouds at nominal irradiation on the temperature profile is insignificant, but its effect on the transmission spectrum is quite important, as shown in \autoref{fig:transmission_spectrum_cloud_effect}. The cloud shields the transit spectrum from the star light passing below its altitude, with a particularly strong effect in the visible range. The cloud also has a slight effect on the Bond albedo, as shown in \autoref{fig:bond_albedo_metallicity}. Under $200$ (Z/H)$_\odot$, increasing the metallicity decreases the Bond albedo, because there is more and more gas absorption. At 300 (Z/H)$_\odot$, adding clouds increases the Bond albedo from 0.017 to 0.018. This effect then increases with metallicity, because the cloud thickens and, thus, it is able to reflect more light. We note that above 300 (Z/H)$_\odot$, the Bond albedo increases with metallicity  even without clouds. This is due to the removal of gaseous H$_2$O from the atmosphere.

We were unable to make Exo-REM simulations converge at nominal irradiation when using its self-consistent cloud mode \citep{Charnay2018}. In this mode, the mean particle radius is determined from a comparison of the timescales of the microphysical processes governing the formation and growth of cloud particles. This is likely due to the temperature profile barely crossing the H$_2$O saturation profile, so that a slight variation of temperature in the solution profile can have a major effect on cloud formation. Nevertheless, we were able to determine that the H$_2$O cloud particles, according to Exo-REM, should be between 20 and 50 $\mu$m at 300 (Z/H)$_\odot$, for a calculated $K_{zz}$ of $\approx$ $9\times10^7$ cm$^2\cdot$s$^{-1}$ at the layer of condensation. The result is essentially the same if we fix the sedimentation parameter ($f_\text{sed}$) between 1 and 5, which is typical of the clouds in the solar system \citep[see][]{Charnay2018}. If we fix $K_{zz}$ at $10^{10}$cm$^2\cdot$s$^{-1}$ and $f_\text{sed}$ at 5, we obtain a maximum mean cloud particle radius of 600 $\mu$m. In \autoref{fig:transmission_spectrum_cloud_particle_size_effect}, we show the effect of the mean cloud particle radius on the transmission spectrum. As the mean radius increases, the effect of the cloud is less and less visible. At 600 $\mu$m, it is almost indistinguishable from the spectrum without clouds.

We found that in our simulations, liquid H$_2$O clouds can form at irradiations lower than 80$\%$ of the nominal irradiation, which is slightly lower than the 3$\sigma$ lower uncertainty of K2-18b irradiation. However, according to \autoref{fig:chi2}, this case is not favoured by the data, and is above the $1\sigma$ confidence level against \citet{Benneke2019} data. It is not surprising that our results for the likelihood of liquid H$_2$O differ from \citet{Benneke2019}, even though they also used a self-consistent model in their demonstration. Indeed, they assumed an albedo of 0.3, which is much higher than what we found. They probably also included a much lower amount of CH$_4$, which significantly reduces the stellar heating.

We note from \autoref{fig:best_fit_temperature} that aside from H$_2$O, clouds of NH$_4$Cl, KCl, ZnS and Na$_2$S are condensing within our pressure grid. Clouds below KCl (included) form too low in the atmosphere to significantly impact the temperature profile above the convective layers, so they can safely be ignored. This is not the case for NH$_4$Cl. According to our simulations, NH$_4$Cl removes $\approx$ 30$\%$ of the total amount of Cl in the atmosphere, with the remaining 70$\%$ being removed mainly via the condensation of KCl, RbCl, and CsCl. The latter two are not included in Exo-REM, the condensation of RbCl and CsCl removing less than 0.15$\%$ of the Cl, assuming a standard composition. While the NH$_4$Cl cloud is, to our knowledge, rarely mentioned in the exoplanet literature, it offers an explanation to the lack of Cl-bearing species in the upper atmosphere of the giant planets of our solar system \citep{Fouchet2004, Teanby2006}. Nevertheless, we found that with our nominal model, NH$_4$Cl clouds with $r = 5$ $\mu$m -- roughly the minimum value found by the Exo-REM self-consistent cloud model on the layer with the most particles -- have essentially no impact on the temperature profile and a marginal impact on the visual part of the transmission spectrum\footnote{NH$_4$Cl optical constants were not available, and we used the parameters from NH$_4$SH. Both molecules show similar strong absorptions due to NH$_4^{+}$ and minor contributions from Cl$^{-}$ or HS$^{-}$ \citep[NIST chemical WebBook and][]{Howett2007}.}, lowering the $\chi^2$ against \citet{Benneke2019} data by $\approx$ 0.2. Therefore, they can probably be neglected for K2-18b, but they might play a more important role in the transmission spectra of slightly hotter planets.

\subsection{Eddy diffusion coefficient and internal temperature}
\label{sec:kzz_tint_effect}
\begin{table}[pt]
\centering
\caption{\label{tab:kzz_t_int}Best fits against \citet{Benneke2019} dataset as a function of internal temperature.}
\begin{tabular}{@{}llll@{}}
\hline
\hline
$T_{p,\,\text{int}}$ (K)        & (Z/H)$_\odot$ & $K_{zz}$ (cm$^2\cdot$s$^{-1}$)        & $\chi^2$        \\
\hline
45                                                      & 175                   & $10^{8}$                                                & 20.61 \\
80                                                      & 175                   & $10^{10}$                                                       & 21.00 \\
115                                                     & 175                   & $10^{10}$                                                       & 21.00 \\
\hline
\end{tabular}
\end{table}

\begin{figure*}[pt]
\begin{subfigure}{0.5\textwidth}
\centering
\includegraphics[width=1.0\linewidth]{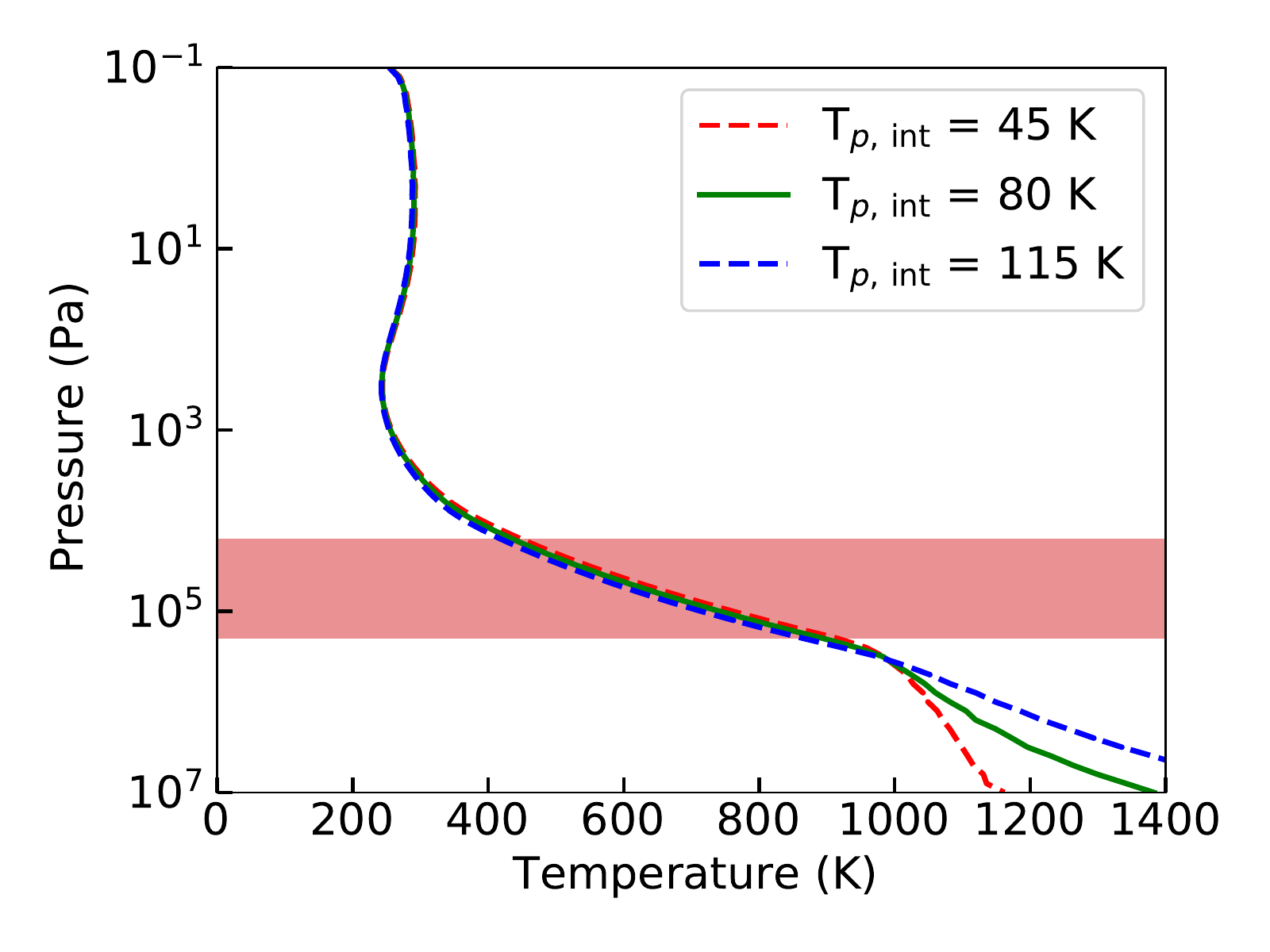}
\end{subfigure}
\begin{subfigure}{0.5\textwidth}
\centering
\includegraphics[width=1.0\linewidth]{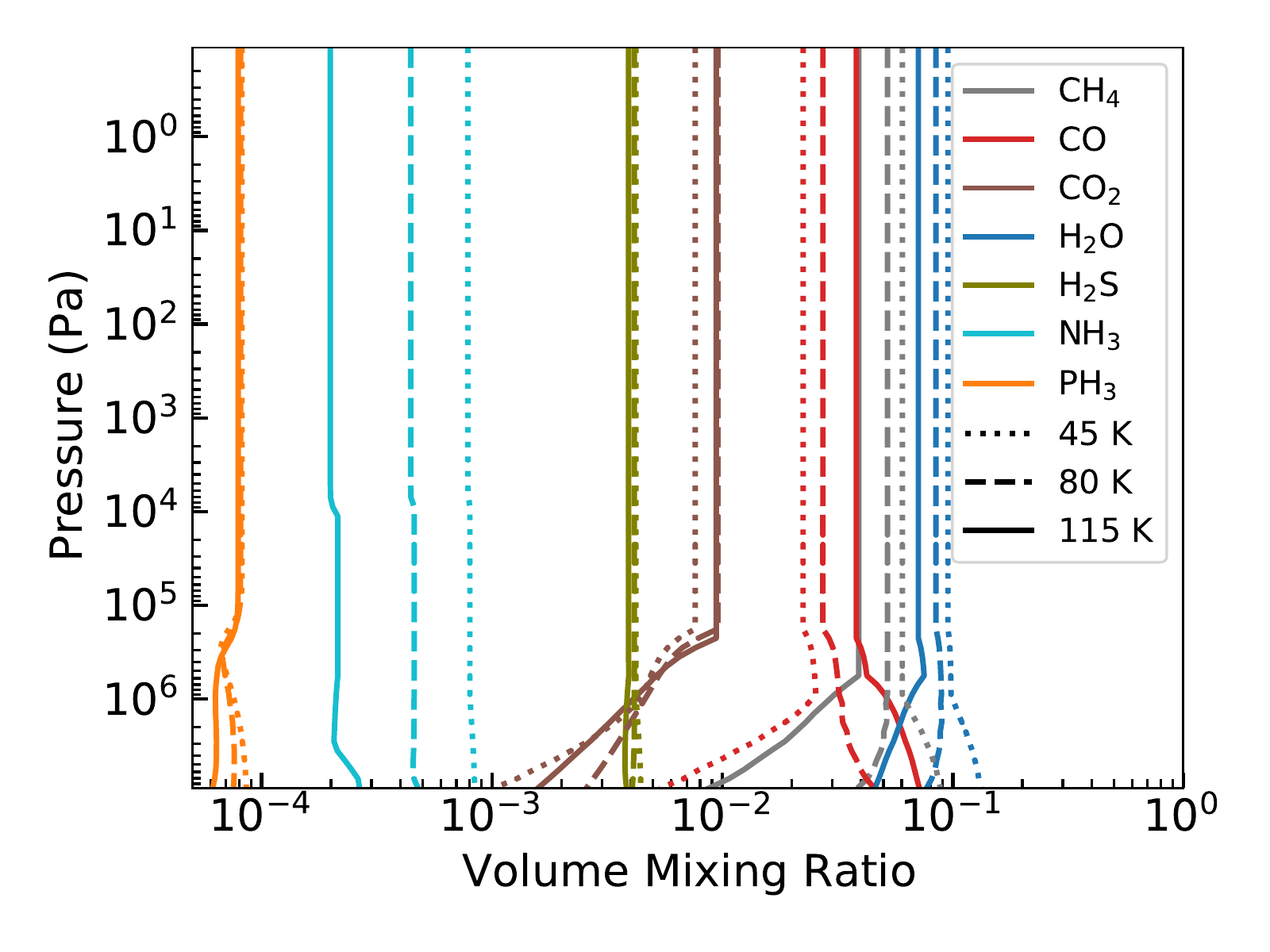}
\end{subfigure}
\caption[Effect of the internal temperature]{Effects of $T_{p,\,\text{int}}$ on K2-18b, at 175 (Z/H)$_\odot$, nominal irradiation, and $K_{zz}$ = $10^6$ cm$^2\cdot$s$^{-1}$. Left: Temperature profiles. The red area represents the convective layers. Right: VMR of the most abundant absorbers. Dotted: $T_{p,\,\text{int}} = 45$ K. Dashed: $T_{p,\,\text{int}} = 80$ K. Solid: $T_{p,\,\text{int}} = 115$ K.}
\label{fig:temperature_profile_teff_effect}
\end{figure*}

\begin{figure}[pt]
\centering
\includegraphics[width=1.0\linewidth]{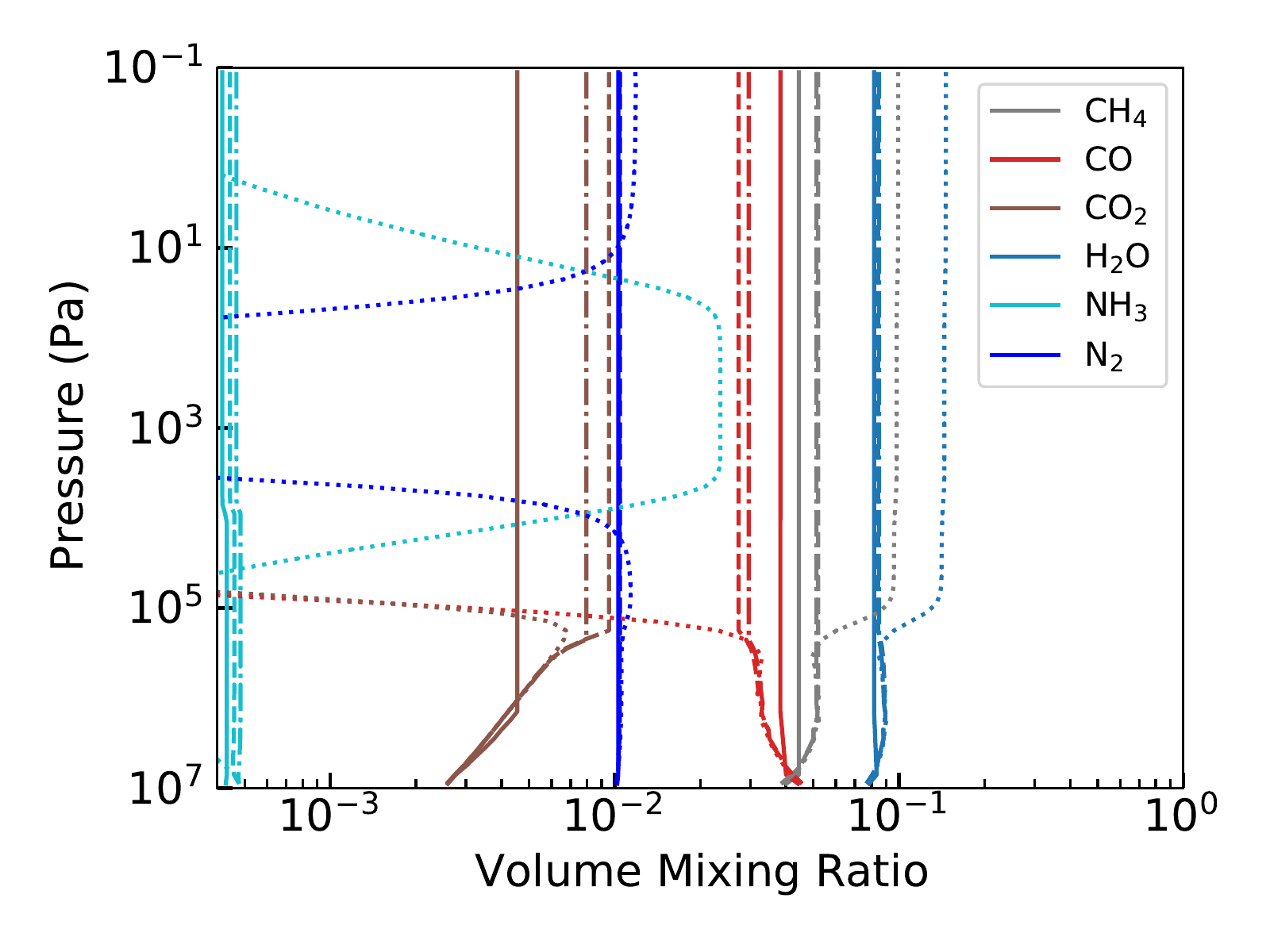}
\caption[Effect of the eddy diffusion coefficient]{Effects of $K_{zz}$ on the VMR of K2-18b main C, N, O bearing species, at 175 (Z/H)$_\odot$. Dashed: $K_{zz} = 10^{6}$ cm$^2\cdot$s$^{-1}$. Dotted-dashed: $K_{zz} = 10^{8}$ cm$^2\cdot$s$^{-1}$. Solid: $K_{zz} = 10^{10}$ cm$^2\cdot$s$^{-1}$. Dotted: Thermochemical equilibrium.}
\label{fig:vmr_kzz_effect}
\end{figure}

\begin{figure}[pt]
\centering
\includegraphics[width=1.0\linewidth]{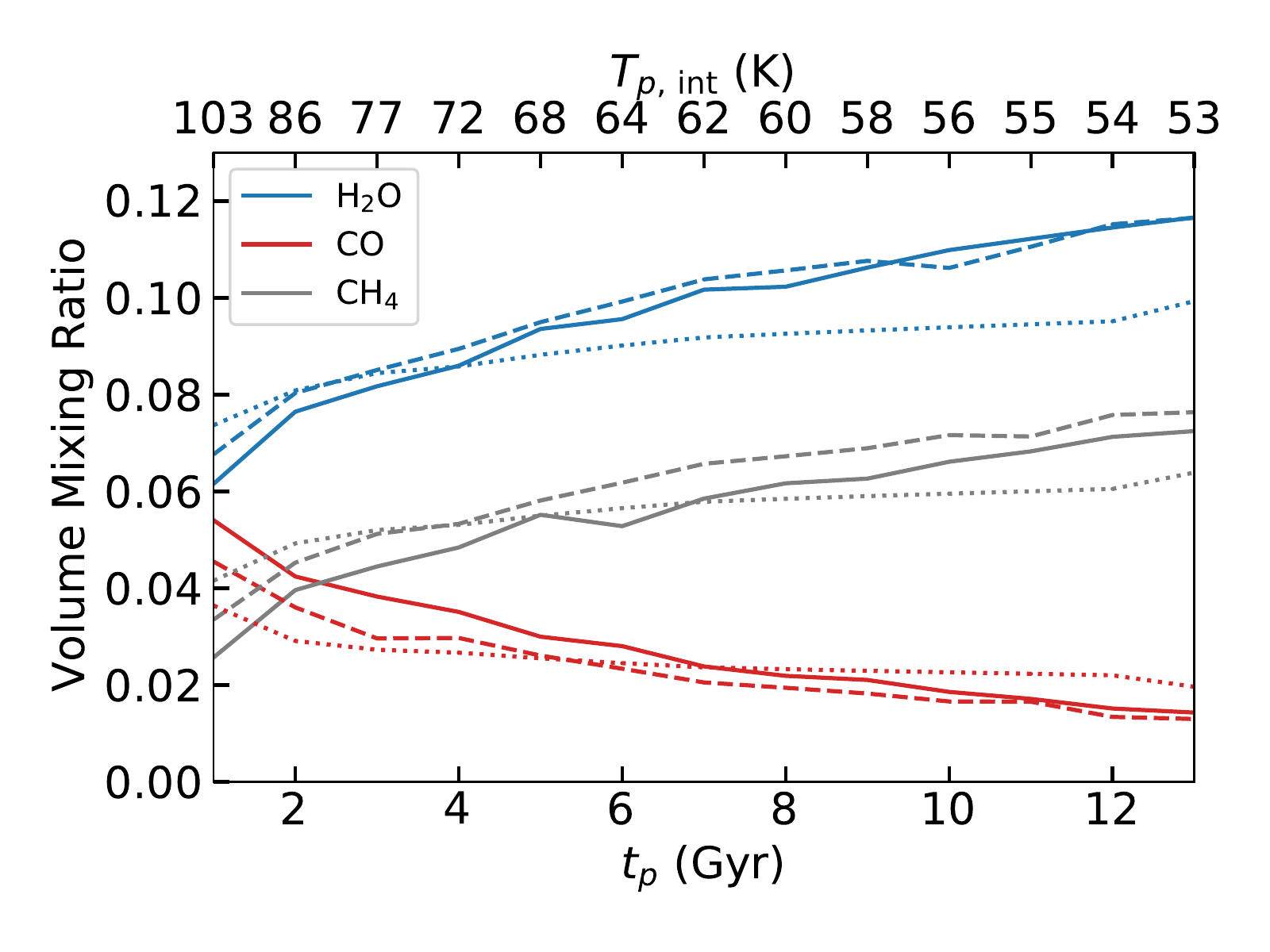}
\caption[Chemical evolution]{Chemical evolution of the upper atmosphere of K2-18b for the most abundant C- and O-bearing species over time, depending on the value of $K_{zz}$, at 175 (Z/H)$_\odot$. Dotted: $K_{zz} = 10^{6}$ cm$^2\cdot$s$^{-1}$. Dashed: $K_{zz} = 10^{8}$ cm$^2\cdot$s$^{-1}$. Solid: $K_{zz} = 10^{10}$ cm$^2\cdot$s$^{-1}$.}
\label{fig:vmr_kzz_teff_effect}
\end{figure}

In \autoref{fig:temperature_profile_teff_effect}, we represented the temperature profiles and VMRs obtained with our upper and lower uncertainty boundaries on the internal temperature of K2-18b (assuming $t_p$ between 1 and 13 Gyr), as well as our selected nominal value (80 K). It appears that the internal temperature has only a minor effect on the temperature profile, except in the deepest layers of the atmosphere. This effect was also noted by other  authors such as \citet{Morley2017}. This is because the received flux (the stellar irradiance) is two to three orders of magnitude higher than the internal flux. Hence, above the convective layers (here at $\approx 20$ kPa), the difference of temperature between our two extreme simulations is less than 40 K (8 K on average). However, below the convective layers ($\approx 200$ kPa), the atmosphere is opaque, thus, the stellar irradiance can no longer heat the atmosphere and the internal heating become prominent. Consequently, at the bottom of our pressure grid, the difference in temperature between our two extreme models reaches $\approx$ 500 K. This affects the species abundances, in the upper atmosphere as well if the quench level of a species is at a pressure level below the region of stellar heating.

In \autoref{fig:vmr_kzz_effect}, we show the effect of the eddy diffusion coefficient on the VMR of the main C-, N-, and O-bearing species at 175 (Z/H)$_\odot$. The VMR at thermochemical equilibrium are also displayed. The main effect of increasing $K_{zz}$ is to shift down the quench levels of the species.

In \autoref{fig:vmr_kzz_teff_effect}, we display the VMR of CH$_4$, CO, and H$_2$O for $K_{zz} = 10^{6}$, $10^{8}$ , and 10$^{10}$ cm$^2\cdot$s$^{-1}$ as a function of the age of the planet and the corresponding internal temperature (see \autoref{eq:effetive_temperature}). As the internal temperature decreases (or as the planet gets older), the impact of the $K_{zz}$ on these VMR gets smaller. If the $K_{zz}$ is low ($\lessapprox 10^6$ cm$^2\cdot$s$^{-1}$), the chemical composition of the atmosphere essentially does not change. On the other hand, under vigorous mixing, the chemical composition of the upper atmosphere can be significantly altered over time. With $K_{zz} \gtrapprox 10^{10}$ cm$^2\cdot$s$^{-1}$, if the planet is young ($\lessapprox$ 1 Gyr or $T_{p,\,\text{int}} \gtrapprox 100$ K), CO could become the dominant O-bearing species and remain more abundant than CH$_4$ for several Gyr. In parallel, $K_{zz}$ could also decrease over time, as the heat to be dissipated by convection decreases. Moreover, if the quench level switches from a convective layer to a radiative one, $K_{zz}$ could drop by several orders of magnitude, rapidly changing the chemistry of the upper atmosphere if this occurs when the planet is still young.

We compared our simulations within our range of $K_{zz}$, at $T_{p,\,\text{int}}$ of 45, 80, and 115 K and at metallicities of 50, 75, 100, 125, 150, 175, 200, and 300 (Z/H)$_\odot$. The best-fit parameters against \citet{Benneke2019} data for each tested internal temperature can be found in \autoref{tab:kzz_t_int}. We found no significant differences at constant metallicity in terms of goodness of fit for the tested $T_{p,\,\text{int}}$, the $\chi^2$ value varying by less than 0.7 across all $K_{zz}$ values. In each case, our best fit is located at 175 (Z/H)$_\odot$. The slight difference in goodness of fit at $T_{p,\,\text{int}} = 45$ K compared to the other tested internal temperatures is due to H$_2$O cloud condensation occurring in the former case and not in the latter. $\chi^2$ values above 200 and below 150 (Z/H)$_\odot$ are systematically above the 1$\sigma$ confidence level. We note that even at 175 (Z/H)$_\odot$, $T_{p,\,\text{int}} = 115$ K and $K_{zz} = 10^{10}$ cm$^2\cdot$s$^{-1}$, we retrieve a CH$_4$ VMR above 1.5$\%$ in the upper atmosphere. Hence, internal heating from a relatively young ($t_p \approx 1$ Gyr) K2-18b and vigorous vertical mixing alone cannot explain a CH$_4$-depleted atmosphere.

\subsection{C/O ratio}
\label{sec:co}
\begin{figure*}[pt]
\centering
\begin{subfigure}{0.49\textwidth}
\centering
\includegraphics[width=1.0\linewidth]{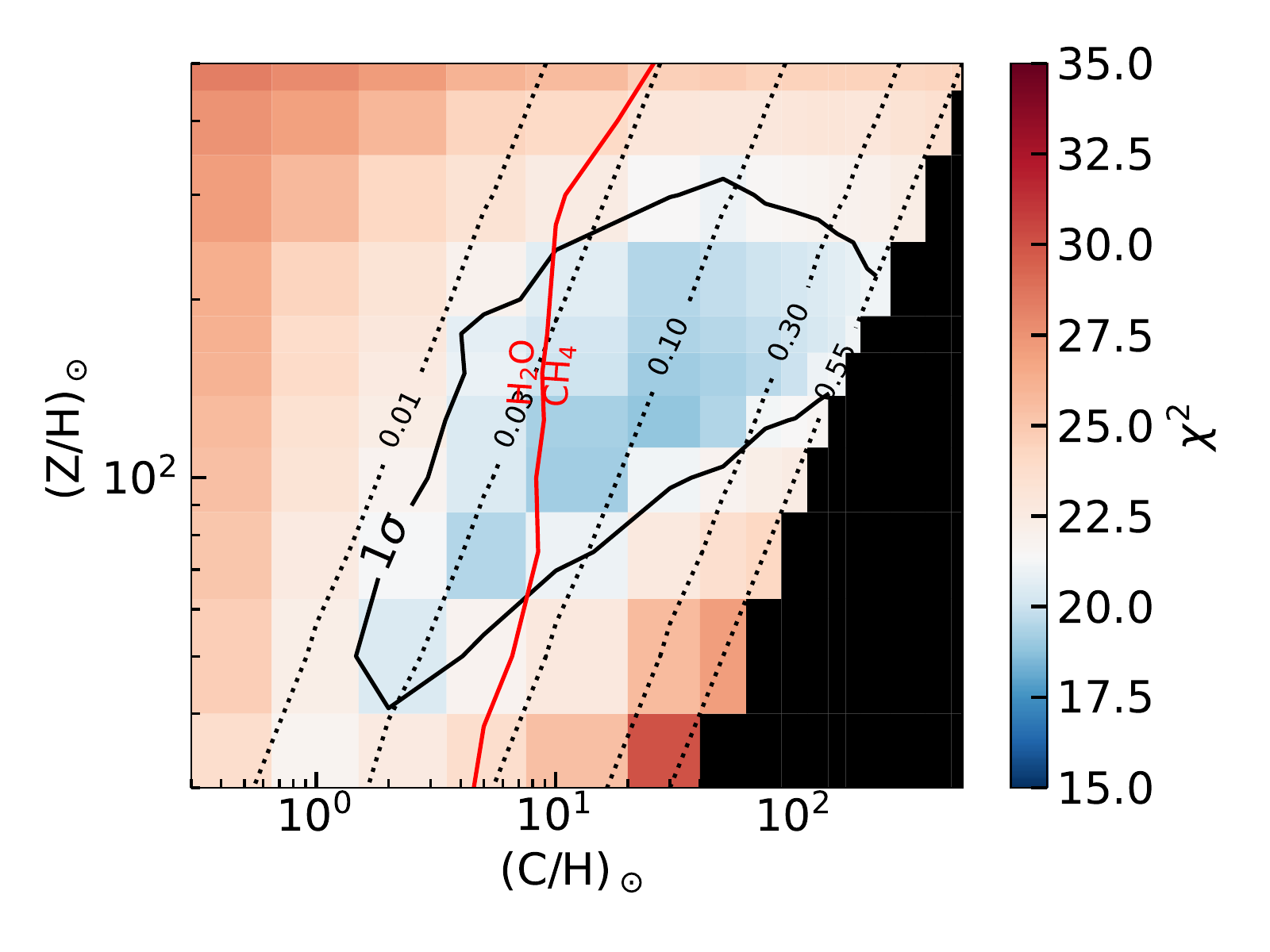}
\end{subfigure}
\begin{subfigure}{0.49\textwidth}
\centering
\includegraphics[width=1.0\linewidth]{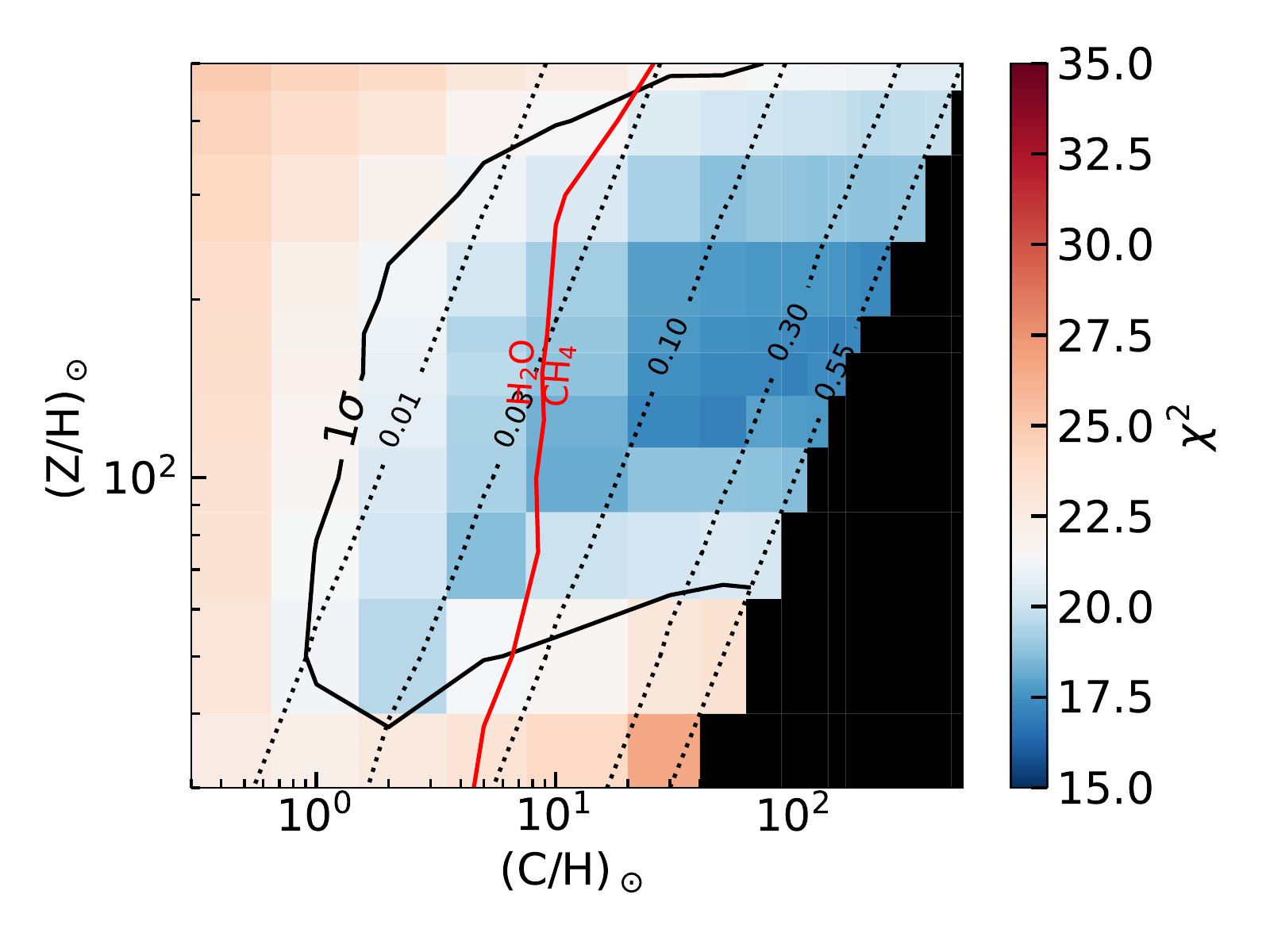}
\end{subfigure}
\caption[Goodness of fit in function of C/H]{Goodness of fit of our models for $T_{p,\,\text{int}} = 80$ K, $K_{zz} = 10^6$ cm$^2\cdot$s$^{-1}$ and nominal irradiation, as a function of the metallicity and C/H, compared to the data from \citet{Benneke2019} (left) and \citet{Tsiaras2019} (right). The white colour corresponds to the value of $\chi^2 = 21.36$, indicating a fit at the 1$\sigma$ confidence level. The solid black line represents the 1$\sigma$ confidence level. The dotted lines indicates the C/O ratio. The solid red line indicates which species dominates on average the transmission spectrum in the 1.355--1.415 $\mu$m range. The cases where C/O $>$ (C/O)$_\odot$ was not explored are represented as black rectangles.}
\label{fig:chi2_low_c_o}
\end{figure*}

We simulated atmospheres with 30, 50, 75, 100, 125, 150, 175, 200, 300, 400, and 500 times the solar system metallicity for all elements except C and the noble gases, and with a carbon-to-hydrogen abundance ratio (C/H) of 0.3, 1, 2, 5, 10, 30, 50, 75, 100, 150, 200, 300, 400, and 500 times the solar ratio. The $\chi^2$ of these simulations against \citet{Benneke2019} data are displayed in \autoref{fig:chi2_low_c_o}. We call that our 'free C/H' scenario. Our best fit against \citet{Tsiaras2019} data ($\chi^2 = 17.07$) is located at 125 (Z/H)$_\odot$ and 50 (C/H)$_\odot$ (C/O = 0.22\footnote{(C/O)$_\odot = 0.550^{+0.130}_{-0.108}$ \citep{Lodders2019}}), while our best fit against \citet{Benneke2019} data ($\chi^2 = 18.86$) is located at 125 (Z/H)$_\odot$ and 30 (C/H)$_\odot$ (C/O = 0.13). A discussion of these results is available in Section~\ref{sec:low_co}.

\section{Comparisons with previous studies and discussions}
\subsection{\texorpdfstring{Considering whether H$_2$O}{H2O} or \texorpdfstring{CH$_4$}{CH4} is the dominant absorber in the HST transit spectra}
\label{sec:h2o_vs_ch4}

\begin{figure*}[pt]
\centering
\includegraphics[width=1.0\linewidth]{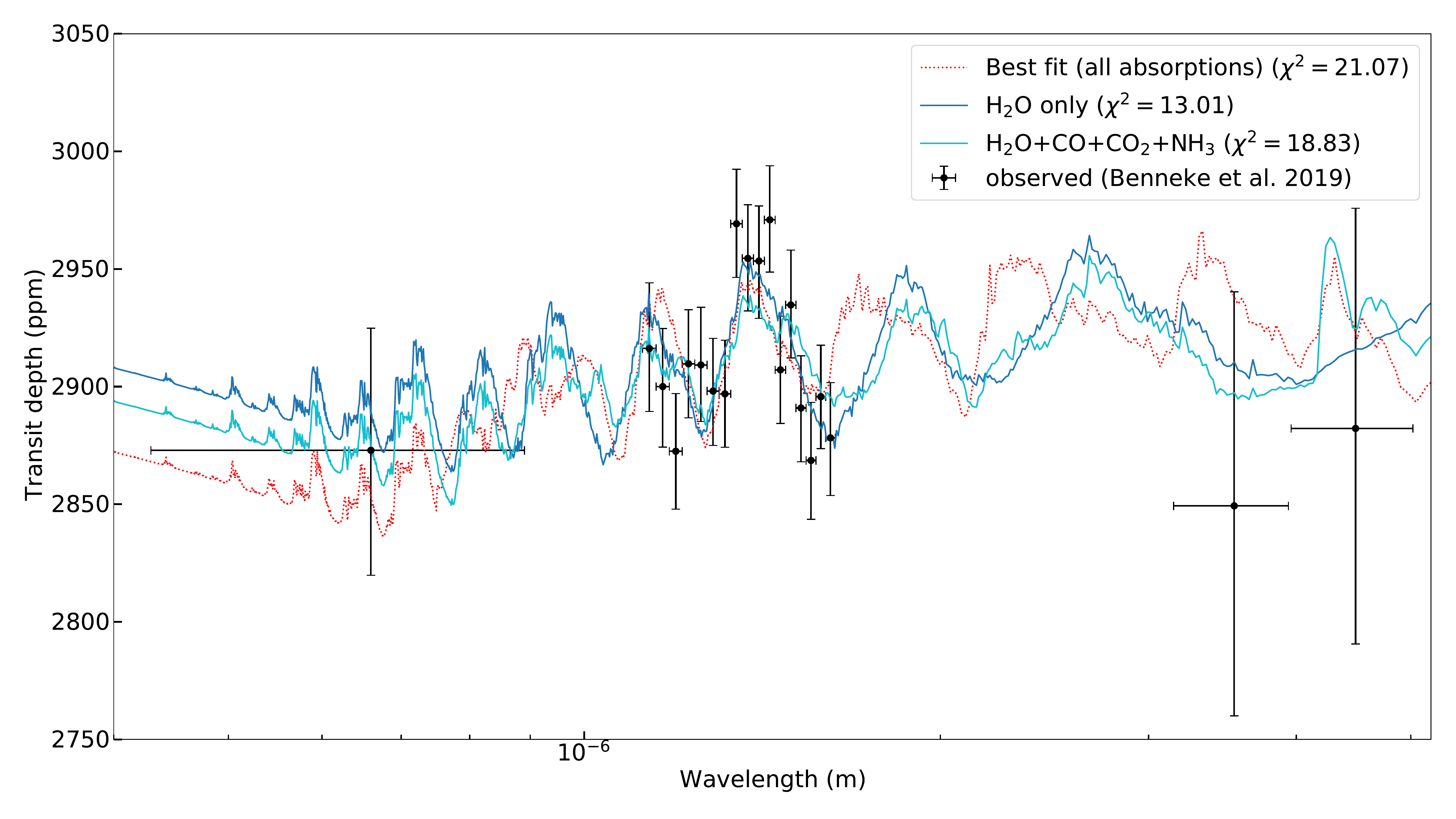}
\caption[H2O only]{Best fit against \citet{Benneke2019} dataset with transit spectra in which the spectral contributions of selected species has been removed. Blue: H$_2$O contribution of the transit spectrum displayed in \autoref{fig:spectrum_contribution}, with an offset of the $10^5$-Pa level of $+98$ km. Cyan: H$_2$O, CO, CO$_2$, and NH$_3$ total contribution of the transit spectrum displayed in \autoref{fig:spectrum_contribution}, with an offset of the $10^5$-Pa level of $+57$ km. Black: K2, HST and Spitzer from \citet{Benneke2019}. The dotted red line is our best-fit spectrum with all absorptions (see Section~\ref{sec:results}). The $\chi^2$ of these spectra against the data is indicated in parentheses.}
\label{fig:H2O_only}
\end{figure*}

\begin{figure*}[pt]
\centering
\includegraphics[width=1.0\linewidth]{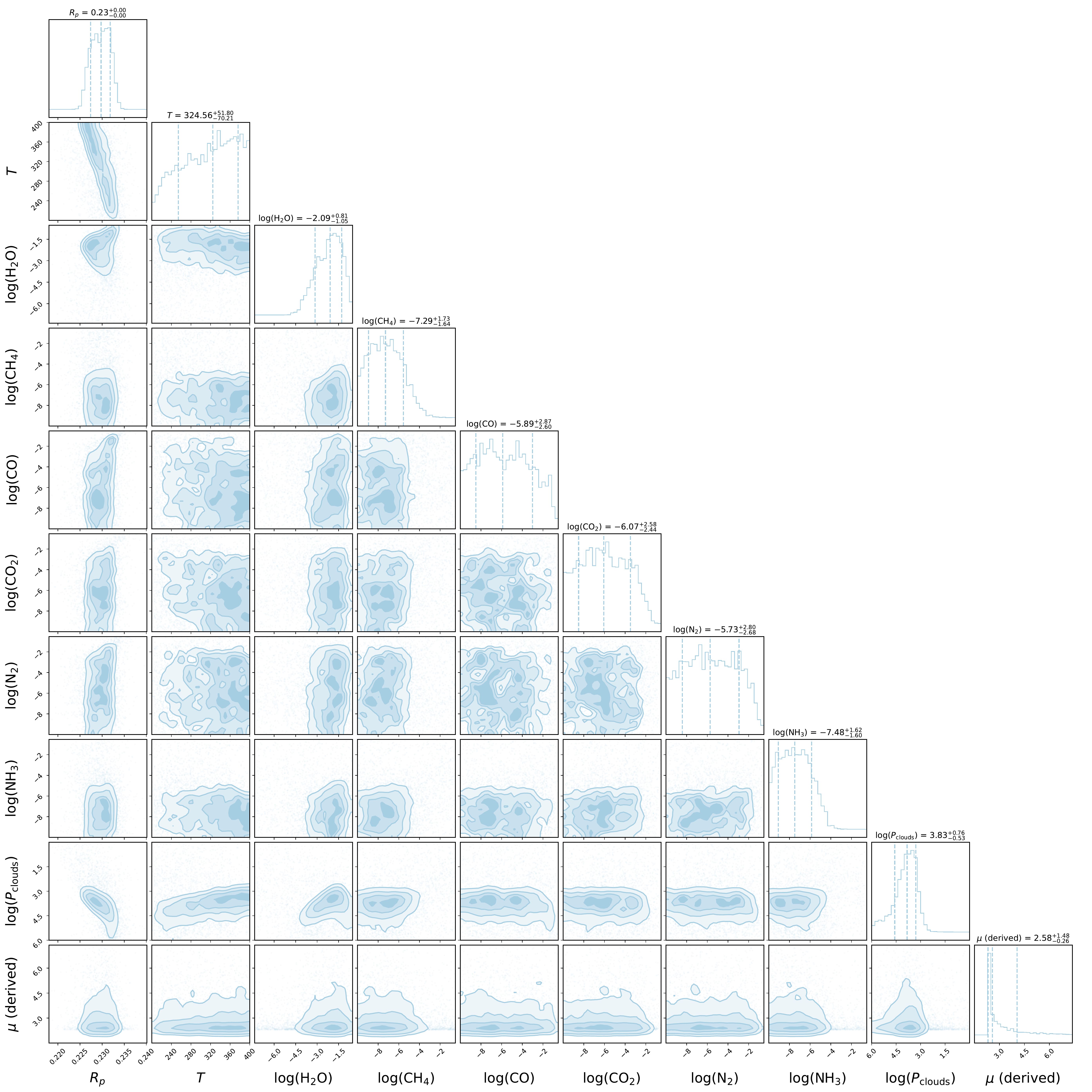}
\caption[TauREx]{TauREx 3 atmospheric retrieval posterior distributions against the dataset from \citet{Benneke2019}, with planetary radius ($R_{\jupiter}$), temperature (K), decimal logarithm of the VMR of H$_2$0, CH$_4$, CO, CO$_2$, N$_2$ , and NH$_3$, and decimal logarithm of the cloud top pressure (Pa) as free parameters.}
\label{fig:posteriors}
\end{figure*}

\begin{table*}[pt]
\centering
\caption{\label{tab:retrievals}Comparison of the Bayesian log-evidence of TauREx 3 against \citet{Benneke2019} dataset for different models.}
\begin{tabular}{@{}lcccccc@{}}
\hline
\hline
Setup                                                   & $\log(Z)$ & $\chi^2$  & $\log_{10}$(CH$_4$)             & $\log_{10}$(H$_2$O)           & $\log_{10}(P_\text{clouds})$(Pa)\\
\hline                                                              
No active gas\tablefootmark{a}  & 171.03        & 31.78         & -                                                     & -                                                       & 2.84$^{+1.97}_{-1.55}$        \\
\textbf{With clouds:} \\                                  
All absorbers\tablefootmark{b}  & 178.83        & 8.63          & -7.29$^{+1.74}_{-1.64}$       & -2.09$^{+0.81}_{-1.06}$         & 3.83$^{+0.75}_{-0.53}$ \\
No H$_2$O\tablefootmark{c}              & 173.22        & 21.34         & -2.60$^{+1.23}_{-1.15}$ & -                                                     & 3.04$^{+1.07}_{-1.06}$ \\
CH$_4$-only                                     & 173.24        & 20.95         & -2.53$^{+1.13}_{-1.31}$ & -                                                     & 2.94$^{+1.13}_{-0.96}$ \\
No CH$_4$\tablefootmark{c}              & 178.57        & 8.84          & -                                                       & -2.48$^{+0.94}_{-1.00}$       & 3.93$^{+0.73}_{-0.49}$ \\
H$_2$O-only                                             & 178.89        & 8.66            & -                                             & -2.51$^{+1.11}_{-1.14}$       & 3.95$^{+0.77}_{-0.61}$ \\
\textbf{No cloud:} \\                                     
All absorbers\tablefootmark{b}  & 177.40        & 10.50         & -7.55$^{+1.60}_{-1.48}$       & -3.28$^{+1.74}_{-0.75}$ & - \\
CH$_4$-only                                     & 172.20        & 21.12         & -1.83$^{+1.20}_{-2.07}$ & -                                                     & - \\
H$_2$O-only                                             & 178.60        & 9.50            & -                                                     & -3.34$^{+1.69}_{-0.79}$         & - \\
\hline
\end{tabular}
\tablefoot{
All the models, except the 'no active gas' model, include H$_2$--H$_2$ and H$_2$--He CIA, Rayleigh scattering and N$_2$ contribution to $\mu$.
\tablefoottext{a}{Grey cloud without molecular absorptions, CIA and Rayleigh scattering.}
\tablefoottext{b}{CH$_4$, CO, CO$_2$, H$_2$O and NH$_3$.}
\tablefoottext{c}{Same as \textit{(b)}, but without H$_2$O or CH$_4$.}
}
\end{table*}

Our nominal model, presented in Section~\ref{sec:results}, shows that CH$_4$ should be the main C-bearing species in the atmosphere, and, despite H$_2$O being more abundant than CH$_4$, the latter should be the main contributor to the absorption features of the transmission spectrum. In particular, CH$_4$ and H$_2$O have an overlapping band at 1.4 $\mu$m in the HST-WFC3 spectral range. This is outlined in \citet{Bezard2020}, who showed that at this wavelength and for sub-Neptunes with effective temperatures lower than $\approx$ 600 K and high atmospheric metallicities, CH$_4$ should be a stronger absorber than H$_2$O in transit spectroscopy due to the numerous weak lines of CH$_4$. However, this is in disagreement with \citet{Benneke2019} and \citet{Tsiaras2019} who claim that the HST data provide evidence for the presence of H$_2$O. 

To investigate this discrepancy, \citet{Bezard2020} compared their best fit Exo-REM spectrum (in which CH$_4$ absorption dominates over H$_2$O) and an H$_2$O-only spectrum (in which all other absorbers were removed) to the HST data reduced by \citet{Tsiaras2019}. The $\chi^2$ associated with the two models are essentially the same (0.91 and 0.93). In addition, applying the retrieval algorithm TauREx 3 \citep{AlRefaie2019} to these data, \citet{Bezard2020} found solutions favouring a CH$_4$-rich atmosphere whose absorption at 1.4 $\mu$m is dominated by CH$_4$. Thus, the apparent disagreement with the analysis of \citet{Tsiaras2019} simply arises from the fact that these authors did not consider an atmosphere with significant amounts of CH$_4$ in the three scenarios they investigated. 

Regarding the \citet{Benneke2019} dataset, we used the same technique as \citet{Bezard2020}: we took the H$_2$O contribution of our best-fit spectrum, displayed in \autoref{fig:spectrum_contribution}, and compared it with the data. In the same way, we made another spectrum with only the contributions of the absorbers included in the retrieval analysis of \citet{Benneke2019} except CH$_4$ (i.e. H$_2$O, CO, CO$_2$ , and NH$_3$, HCN contribution is negligible and was not included). We will refer to this latter spectrum as 'no CH$_4$'. We stress that none of these spectra are directly the results of self-consistent simulations. They are displayed in \autoref{fig:H2O_only}. We found that our H$_2$O-only spectrum does provide a superior fit compared with our best fit Exo-REM model. The fit is improved in the 1.15--1.20 $\mu$m spectral range, around 1.6 $\mu$m, and for the Spitzer data points. With our 'no CH$_4$' spectrum, the fit is also better than our best-fit model, but less so than our H$_2$O-only spectrum due to the absorptions of NH$_3$ in the HST spectral range and the CO and CO$_2$ absorptions in the Spitzer spectral range. We note that our VMRs of NH$_3$, N$_2$, CO, and CO$_2$ of respectively 0.04$\%$, 1.02$\%$, 2.73$\%,$ and 0.96$\%$ (see \autoref{fig:best_fit_vmr}), are all lower than the '2$\sigma$ (97.5$\%$)' upper limits given by \citet{Benneke2019} for these species (respectively $13.5\%$, $10.9\%$, $7.45\%$ and $2.4\%$). The spectrum with only H$_2$O absorption is strongly overfitting the \citet{Benneke2019} data, with a $\chi^2$ of 13.01. On the other hand, our best-fit model, with a $\chi^2$ of 21.07, is still within the 1-$\sigma$ confidence interval ($\chi^2 < 21.36$). The 'no CH$_4$' spectrum is characterised by a slight overfitting, providing a better $\chi^2$ (18.83) compared to our best fit, but the removal of CH$_4$ absorption is artificial.

We also applied TauREx 3 to the \citet{Benneke2019} data. This algorithm uses the nested sampling code Multinest \citep{Feroz2009} to explore the parameter space and find the best fit corresponding to a given spectrum. In this retrieval analysis, we used 500 live points, an evidence tolerance of 0.5, and the cross-sections provided by the TauREx website\footnote{\href{https://exoai.github.io/software/taurex/xsec}{https://exoai.github.io/software/taurex/xsec}} at a resolution power of 15000 and between 0.3 and 15 $\mu$m. We simulated the atmosphere of K2-18b using an isothermal temperature profile with 100 atmospheric layers between $10^{-3}$ and $10^6$ Pa. We took into account molecular absorptions from CH$_4$, CO, CO$_2$, H$_2$O, and NH$_3$, CIA from H$_2$--H$_2$ and H$_2$--He, the contribution to $\mu$ of N$_2$, Rayleigh scattering and spectrally gray clouds. The planet radius was allowed to vary by $\pm 10\%$ of its value determined by \citet{Benneke2019}, while the cloud top was allowed to be between $10^6$ and 1 Pa. We chose to set bounds on the temperature and the species VMR, respectively, between 200 and 400 K, and between $10^{-10}$ and 0.3. The corresponding posterior distributions are shown in \autoref{fig:posteriors}. We also tested models removing all molecular absorptions except H$_2$O or CH$_4$, removing clouds, or including all absorptions except that of H$_2$O. For comparison, we also computed a 'no active gas' model including only the cloud contribution, that is, a flat spectrum. Finally, we calculated the $\chi^2$ and corresponding $\sigma$ confidence of each model using the same technique as in Section~\ref{sec:results} and the 'binned' spectrum output from TauREx 3. Our results are summarised in \autoref{tab:retrievals}. A small discussion on the results is available in Appendix~\ref{sec:app_retrievals}.

We found that solutions with low amounts of CH$_4$ are unambiguously favoured: the 'all absorbers with clouds' model we used is favoured over the 'no H$_2$O with clouds' model at 273:1 \citep[3.79$\sigma$, see][]{Benneke2013}. The difference in log-evidence between our CH$_4$-only models and our 'H$_2$O-included' models (i.e. all models including at least the molecular absorption of H$_2$O) is, at worst of 4.16, indicating that the latter are favoured at $\geq$ 64:1 (3.35$\sigma$). Moreover, the 2$\sigma$ (95.4$\%$) upper limit of CH$_4$ is 0.009$\%$ with our 'all absorbers with clouds' model, even lower than what was inferred by \citet{Benneke2019} (0.248$\%$). This confirms that an H$_2$O-dominated spectrum is a far better fit to the \citet{Benneke2019} dataset than a CH$_4$-dominated spectrum, but not that the latter must be rejected. Indeed, looking at the $\chi^2$, we find that all of the spectra obtained from our TauREx 3 models that do not include H$_2$O are slightly below the 1$\sigma$ confidence level, with $\chi^2$ values and CH$_4$ VMR close to our Exo-REM best fit. In contrast, all the H$_2$O-included TauREx 3 spectra strongly overfit the spectrum, with $\chi^2 \leq 10.50$, which is much less than the number of data points, and likely much less than the number of effective degrees of freedom. We interpret this as an H$_2$O-only spectrum leading to a significant overfit of the dataset.

The overfit that we identified could be explained by an over-estimation of the HST error bars. However, the \citet{Benneke2019} and \citet{Tsiaras2019} HST spectra that we analysed come from independent reduction methods \citep{Tsiaras2018} and both show similar transit depth uncertainties. Moreover, the custom HST pipeline used by \citet{Tsiaras2019} is described in  \citet{Tsiaras2018} to exhibit 'nearly photon-noise limited' performance, meaning that the spectral uncertainties should be close to the theoretical minimum. It can also be seen, for example, in \autoref{fig:transmission_metallicity} that the data points near 1.2 $\muup$m from the two datasets are not compatible at the 1-$\sigma$ confidence level, which points to additional systematic errors. These are probably linked to uncertainties in some parameters used in the data reduction\citep[orbital parameters, limb-darkening coefficients, or calibration, as suggested in][]{Tsiaras2018}. For all of these reasons, we regard this possibility as unlikely.

To summarise, of the two datasets available to us, only the one from \citet{Benneke2019} strongly favours a CH$_4$-depleted atmosphere. This favouritism is likely due to free retrieval algorithms searching for the solution that best fits the data. But if such a solution is strongly overfitting, as is the case for an H$_2$O-only spectrum against the dataset from \citet{Benneke2019}, any other solution that is considered to be more physically acceptable may also appear as statistically unlikely. However, if a strong depletion of CH$_4$ in the atmosphere of K2-18b is not a straightforward scenario from a chemical standpoint and ends up overfitting the data, it is indisputably a statistically valid one and, thus, it must be considered. In the next section, we try to find a model that could explain such a depletion.

\subsection{\texorpdfstring{CH$_4$}{CH4}-depleted scenarios for K2-18b}
\label{sec:low_ch4}
While our standard models, presented in Section~\ref{sec:results}, are statistically able to reproduce the observed spectrum of K2-18b, here we test some scenarios allowing for CH$_4$ VMR to be as small as retrieved by all the other teams who analysed the data so far. Indeed, our nominal model gives a CH$_4$ VMR of $\approx 5^{+1}_{-2}\%$ (for metallicities between 65 and 500 (Z/H)$_\odot$), while \citet{Benneke2019}, \citet{Madhusudhan2020} and \citet{Scheucher2020} give, respectively, a $2\sigma$ upper limit of $0.248\%$, a $99\%$ upper limit of $3.47\%$, and an upper limit of 460 ppm. \citet{Benneke2019b} propose three explanations to this depletion: (i) a high internal temperature, either from residual heat of formation or tidal heating, (ii) a low C/O ratio resulting from planetary formation process, and (iii) a catalytic destruction of CH$_4$ by photolysis. The latter cannot be simulated using Exo-REM, so we will focus on the first two possibilities.

\subsubsection{High internal temperature}
\begin{figure*}[pt]
\centering
\begin{subfigure}{0.5\textwidth}
\centering
\includegraphics[width=1.0\linewidth]{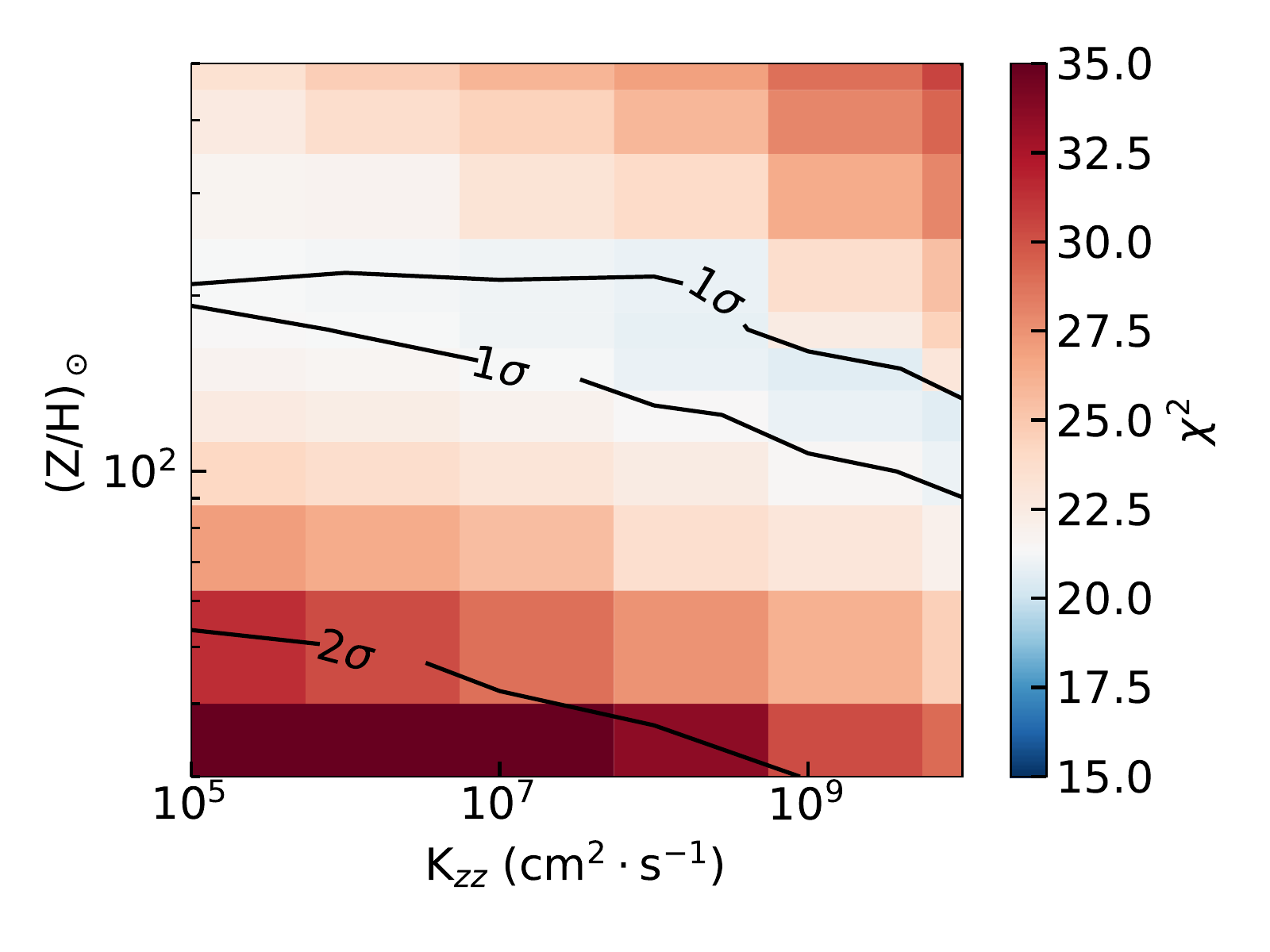}
\end{subfigure}\begin{subfigure}{0.5\textwidth}
\centering
\includegraphics[width=1.0\linewidth]{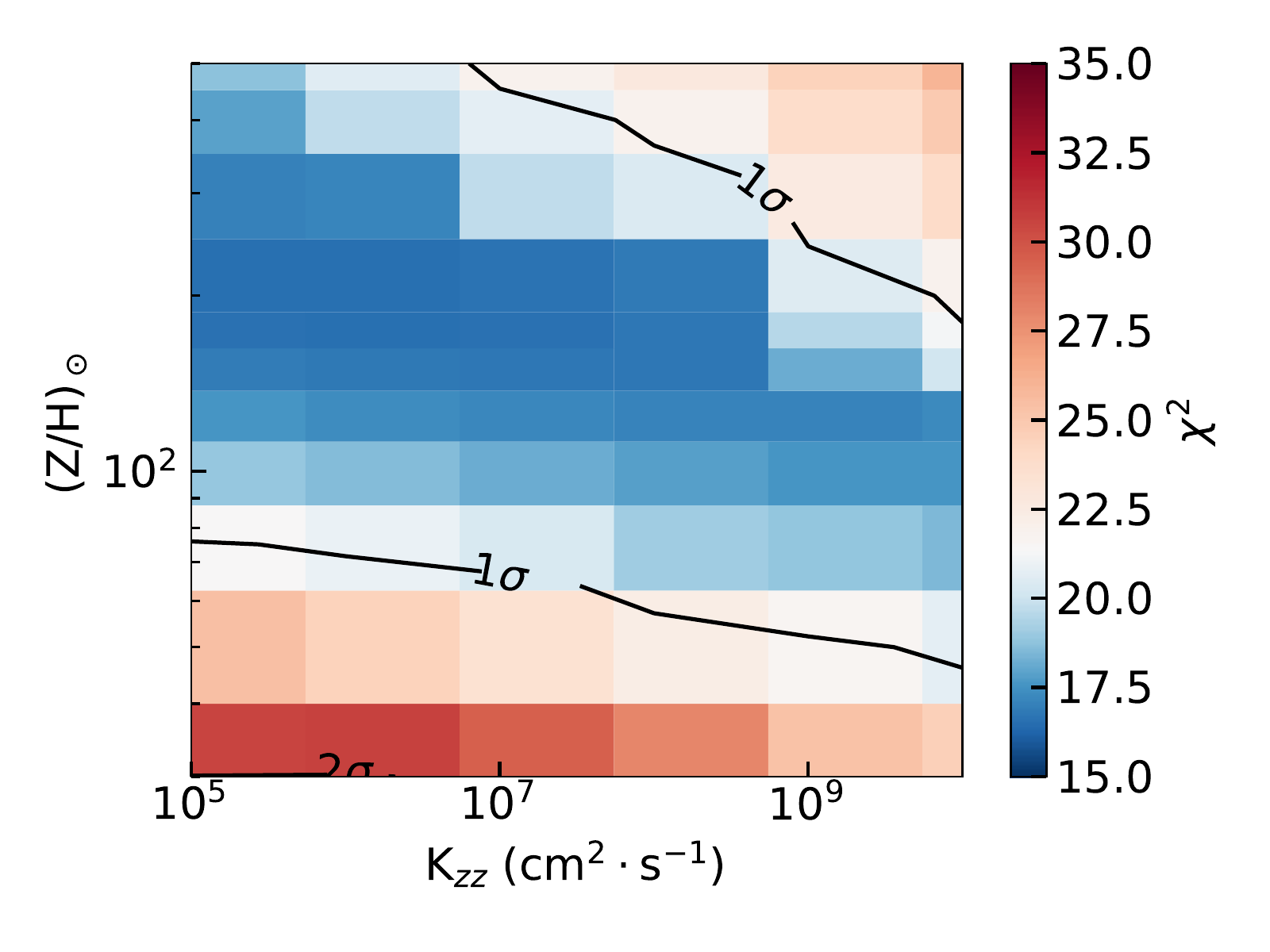}
\end{subfigure}
\caption[Goodness of fit at 200 K]{Goodness of fit of our models at $T_{p,\,\text{int}} = 200$ K and nominal irradiation as a function of metallicity and $K_{zz}$, compared to the data from \citet{Benneke2019} (left) and \citet{Tsiaras2019} (right). The white colour corresponds to the value of $\chi^2 = 21.36$, indicating a fit at the 1$\sigma$ confidence level. The best fit against \citet{Benneke2019} data ($\chi^2 = 20.64$) is located at (Z/H)$_\odot = 125$ and $K_{zz} = 10^{10}$ cm$^2\cdot$s$^{-1}$. The best fit against \citet{Tsiaras2019} data ($\chi^2 = 16.56$) is located at (Z/H)$_\odot = 200$ and $K_{zz} = 10^{5}$ cm$^2\cdot$s$^{-1}$.}
\label{fig:chi2_high_t_int}
\end{figure*}

\begin{figure}[pt]
\centering
\includegraphics[width=1.0\linewidth]{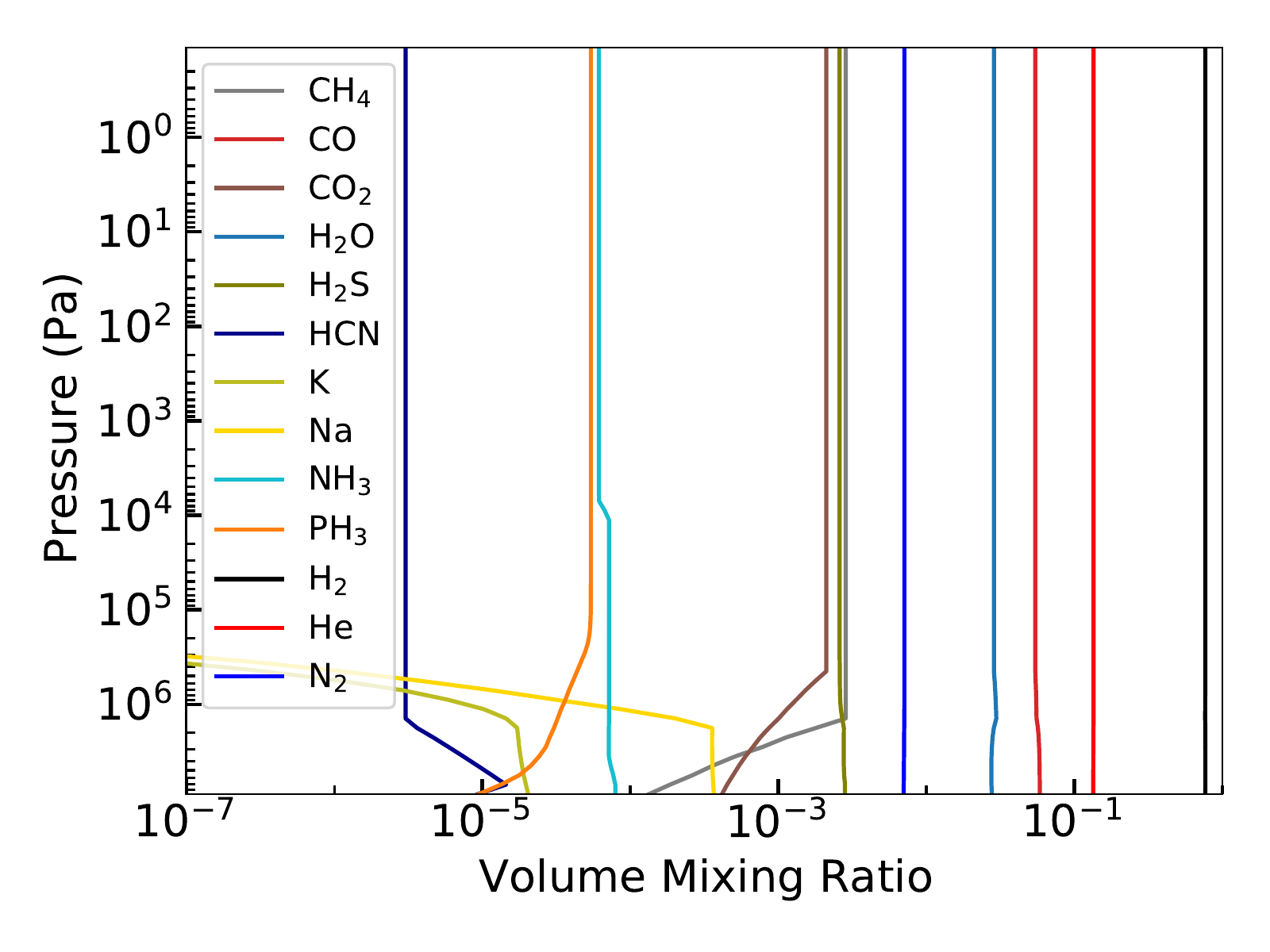}
\caption[VMR at 200 K]{VMR of selected species of our best-fit model against \citet{Benneke2019} data at $T_{p,\,\text{int}} = 200$ K and nominal irradiation.}
\label{fig:vmr_best_fit_200K_benneke2019}
\end{figure}

Following the trend presented in Section~\ref{sec:kzz_tint_effect}, it is possible to reach a CH$_4$ VMR that is compatible with what was found by \citet{Benneke2019} by considering $T_{p,\,\text{int}}$ as high as $200$ K. Using this internal temperature, we simulated atmospheres at 30, 50, 75, 100, 125, 150, 175, 200, 300, 400, and 500 (Z/H)$_\odot$ and within our selected range of $K_{zz}$. The goodness of fit for our models compared to the datasets from \citet{Benneke2019} and \citet{Tsiaras2019} is displayed in \autoref{fig:chi2_high_t_int}. We note that H$_2$O never dominates the transmission spectrum in the 1.355--1.415 $\mu$m window. Our best fit against \citet{Tsiaras2019} data ($\chi^2 = 16.78$) is located at 150 (Z/H)$_\odot$ and $10^{7}$ cm$^2\cdot$s$^{-1}$, while our best fit against \citet{Benneke2019} data ($\chi^2 = 20.70$) is located at 150 (Z/H)$_\odot$ and $10^{10}$ cm$^2\cdot$s$^{-1}$. There is no H$_2$O cloud condensation in the latter scenario. The goodness of fit is again slightly better than for our nominal model. The VMR are displayed in \autoref{fig:vmr_best_fit_200K_benneke2019}. In this configuration, we obtain VMR for CH$_4$, CO, CO$_2$, NH$_3$ , and H$_2$O in the upper atmosphere at levels of 0.283$\%$, 6.589$\%$, 0.280$\%$, 0.006$\%,$ and 1.657$\%$, respectively, which are all below or close to the 2$\sigma$ upper limit found by \citet{Benneke2019}. 

Reaching such a high internal temperature, however, would require K2-18b to be a very young planet ($\ll 1$ Gyr old according to \autoref{eq:effetive_temperature}) or a planet experiencing strong tidal heating, or both. According to the internal temperature model we used and the age of the system determined by \citet{Guinan2019}, it seems unlikely that the residual heat of formation could be sufficient to explain a high internal temperature. To assess the possibility of strong tidal heating on K2-18b, we can make a few considerations. The K2-18 system is composed of at least another planet: K2-18~c \citep{Cloutier2019}, orbiting closer to its star than K2-18b. Both planets may have a moderately high eccentricity that is rapidly evolving due to the secular interactions between the two objects \citep{Gomes2020}. In that configuration, a lot of orbital energy can indeed be dissipated, but only in the innermost planet\footnote{J. Leconte, private communication.}, that is, K2-18~c. Thus, there should be no tidal heating on K2-18b.

\subsubsection{Low C/O} 
\label{sec:low_co}

\begin{figure}[pt]
\centering
\includegraphics[width=1.0\linewidth]{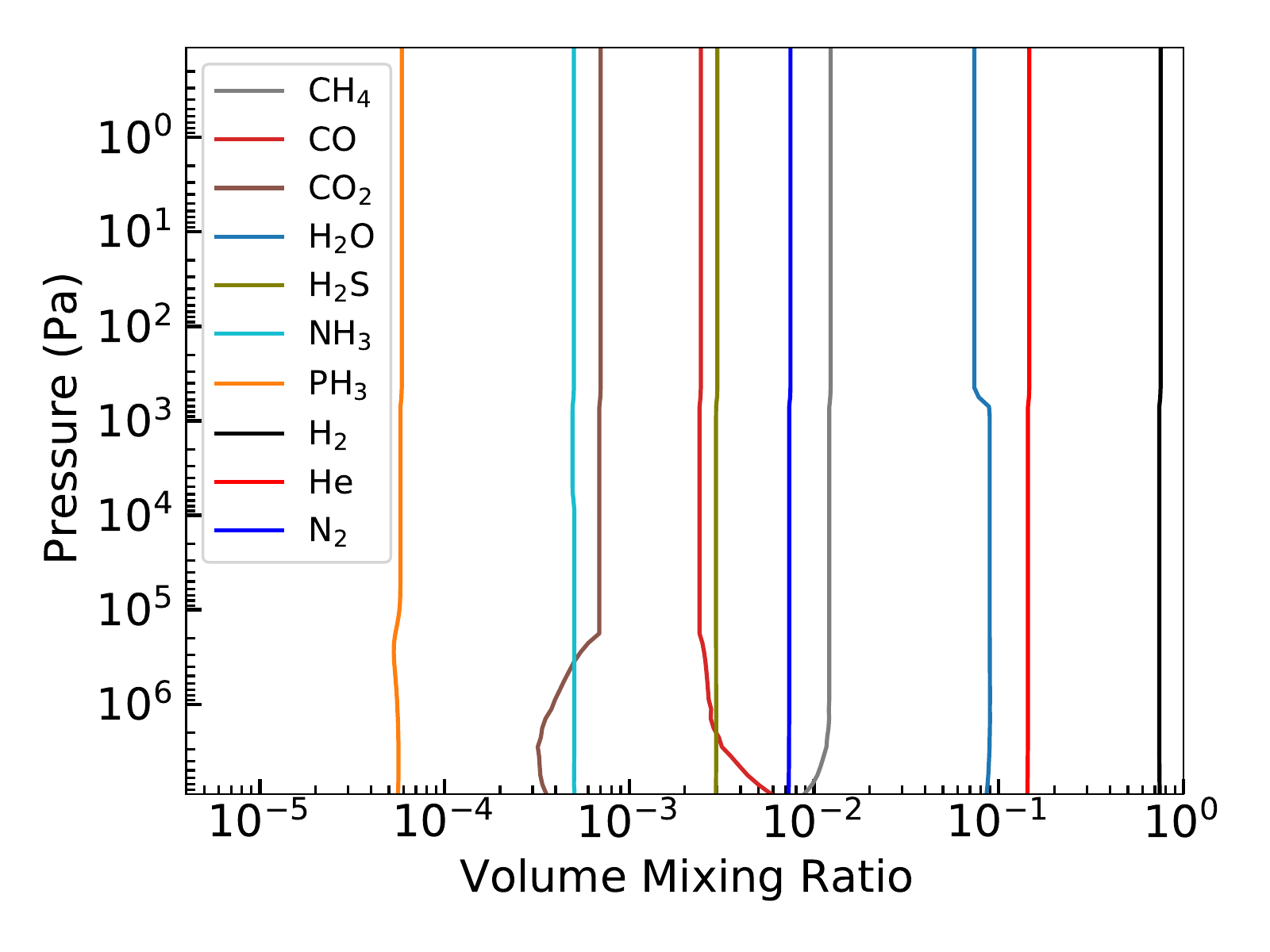}
\caption[VMR at C/O = 0.1]{VMR of selected species at $T_{p,\,\text{int}} = 80$ K, $K_{zz} = 10^6$ cm$^2\cdot$s$^{-1}$, nominal irradiation, 125 (Z/H)$_\odot,$ and 30 (C/H)$_\odot$.}
\label{fig:vmr_low_c_o}
\end{figure}

To test the possibility of a low atmospheric C/O ratio on K2-18b, we use the results from the simulations presented in Section~\ref{sec:co}. The VMR of the most abundant absorbers in our best fit against \citet{Benneke2019} are represented in \autoref{fig:vmr_low_c_o}. The amount of CH$_4$ in the upper atmosphere we obtain in this best-fit simulation is 1.23$\%$, incompatible with the 2$\sigma$ upper limit found by \citet{Benneke2019} (0.248$\%$), but compatible with the 99$\%$ upper limit found by \citet{Madhusudhan2020} (3.47$\%$). However, simulations at metallicities between 50 and 150 (Z/H)$_\odot$ at 5 (C/H)$_\odot$ and at 20 (Z/H)$_\odot$ and 1 (C/H)$_\odot$ are statistically able to reproduce the data while allowing the CH$_4$ VMR to be close or below 0.248$\%$, with a minimum at 0.101$\%$. There is no simulation within the 1$\sigma$ confidence level for C/H $\lessapprox$ 1 (C/H)$_\odot$ (CH$_4$ VMR $\lessapprox$ 0.05$\%$ at 50 (Z/H)$_\odot$). Accordingly, C/O must be $\gtrapprox$ 0.01 (0.02 (C/O)$_\odot$) in order to obtain a satisfactory fit of the data. We note that C/H must be $\lessapprox$ 5 (C/H)$_\odot$ (or C/O $\lessapprox$ 0.1 (C/O)$_\odot$) in order for H$_2$O to dominate the transmission spectrum in the 1.355--1.415 $\mu$m interval.

Heavily C-depleted atmospheres (C/O $\lessapprox$ 0.01) do not seem to agree with the data. Lowering the abundance of C in the atmosphere naturally decreases the contribution of CH$_4$ to the transmission spectrum, and improves the fit of \citet{Benneke2019} data, as discussed in Section~\ref{sec:low_ch4}. A side effect, however, is that because CH$_4$ has a strong greenhouse effect, the temperatures decrease significantly, favouring the condensation of H$_2$O, dampening the absorption features of the latter, and worsening the fit around 1.4 $\mu$m. This side effect explains why our results are quite different from the results of other teams who used free retrieval algorithms that do not include any physical constraint on the model atmosphere. Moderately-C-depleted atmospheres (C/O $\approx$ 0.10), on the contrary, are clearly favoured, but in that case CH$_4$ still dominates the absorption features of the transmission spectrum. Moreover, the precise phenomenon that could lead to such a depletion has, to our knowledge, yet to be discovered.

\section{Observational perspectives}
\begin{figure*}[pt]
\centering
\begin{subfigure}{0.49\textwidth}
\centering
\includegraphics[width=1.0\linewidth]{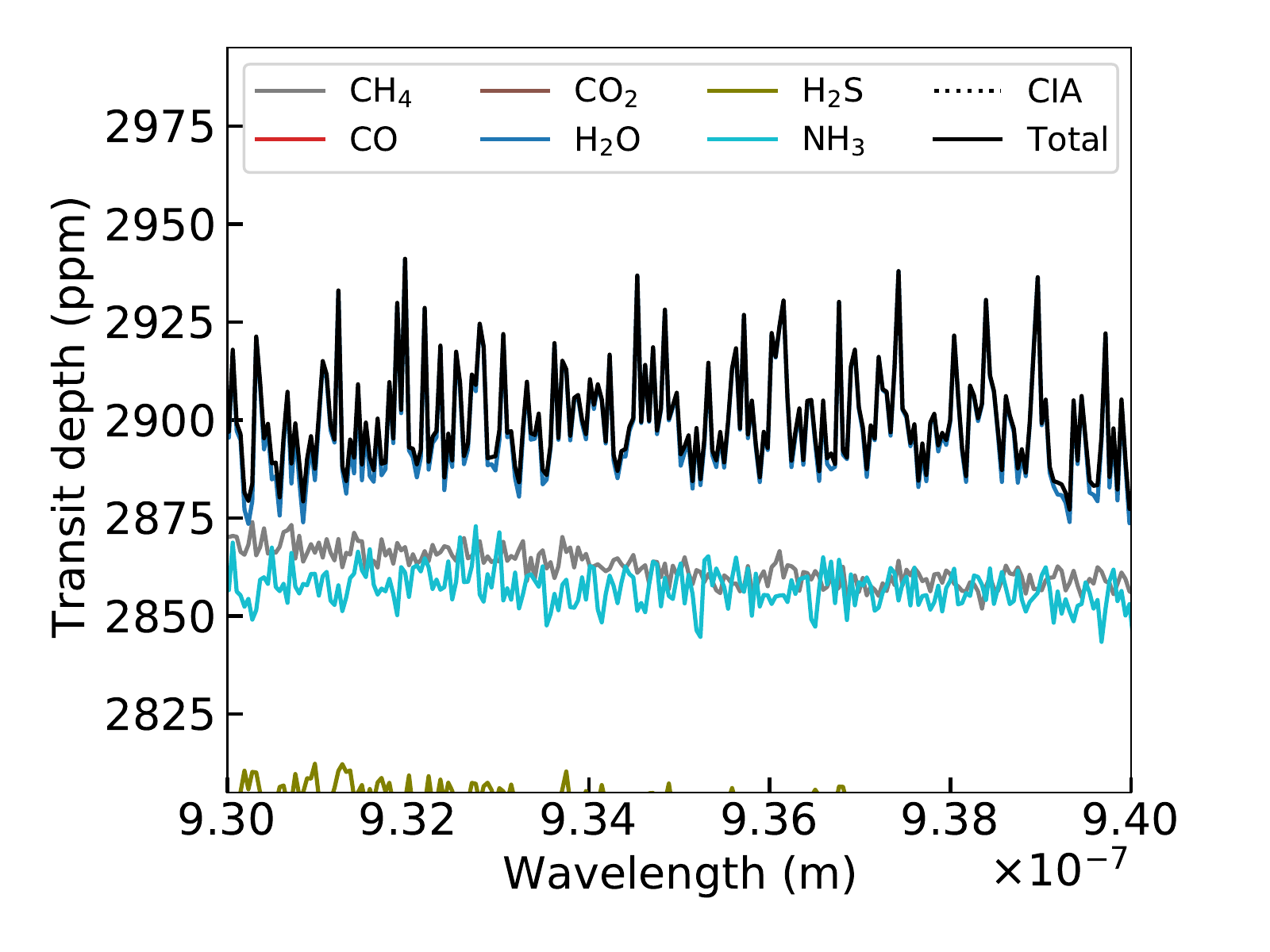}
\end{subfigure}
\begin{subfigure}{0.49\textwidth}
\centering
\includegraphics[width=1.0\linewidth]{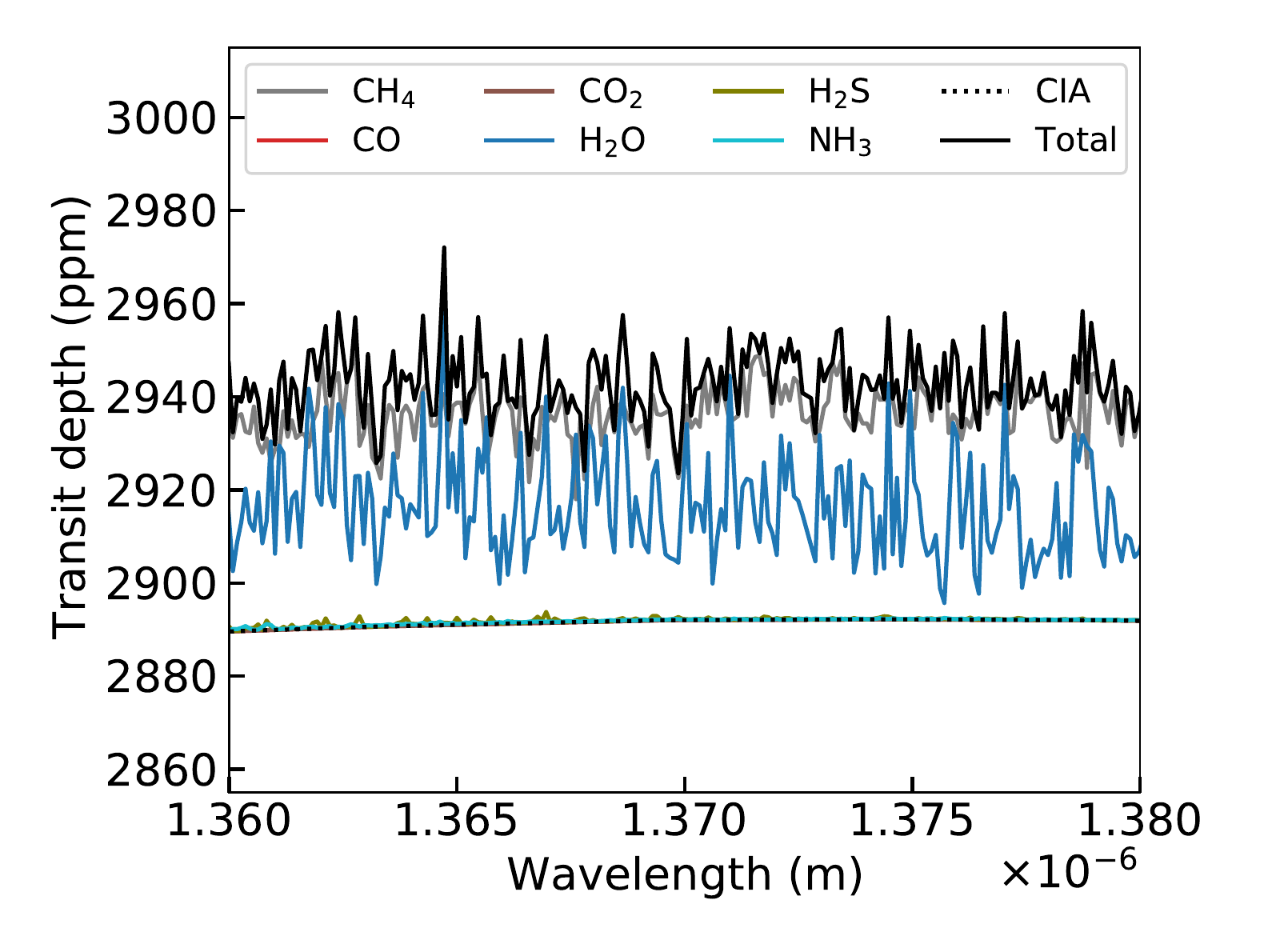}
\end{subfigure}
\begin{subfigure}{0.49\textwidth}
\centering
\includegraphics[width=1.0\linewidth]{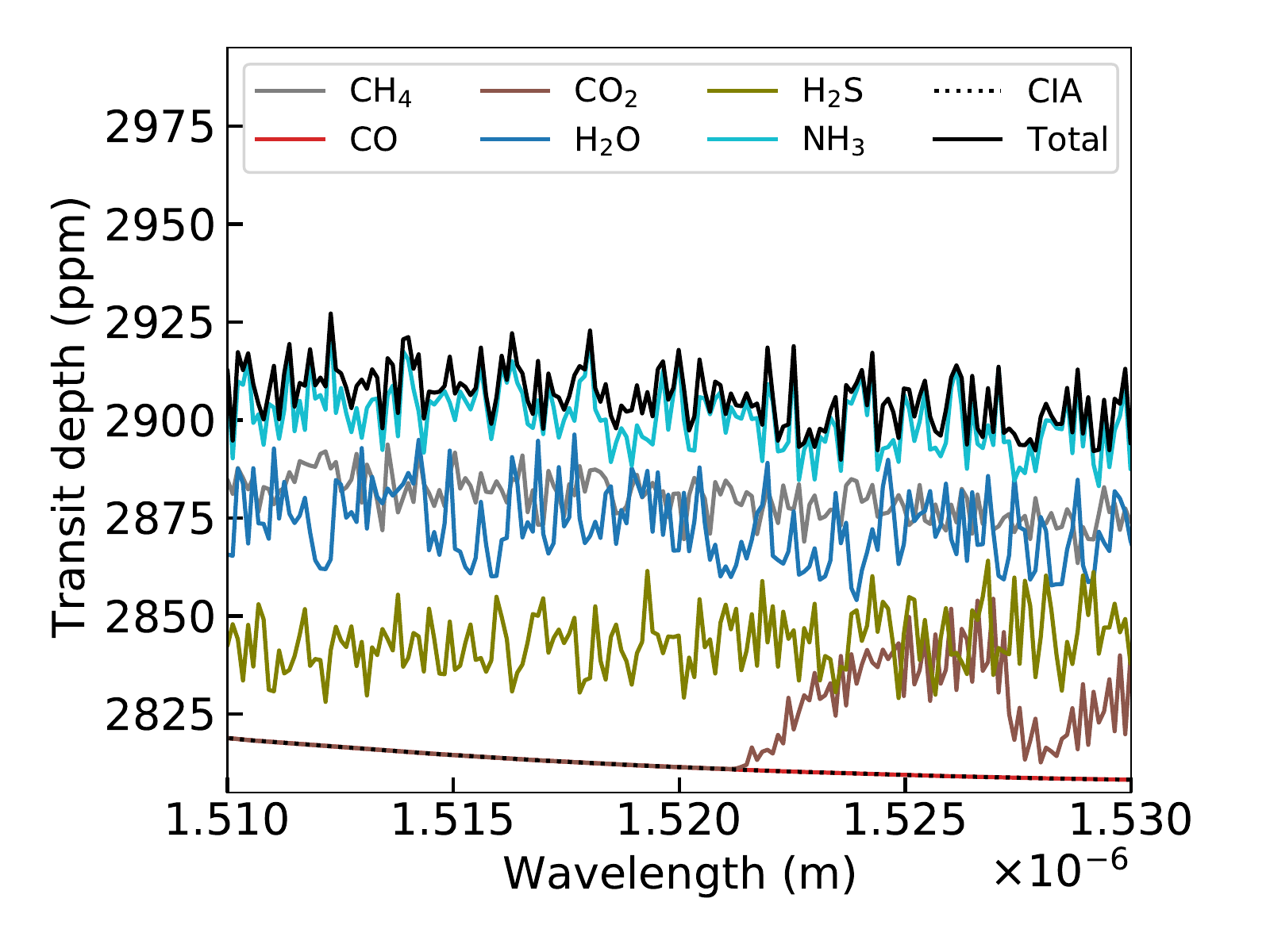}
\end{subfigure}
\begin{subfigure}{0.49\textwidth}
\centering
\includegraphics[width=1.0\linewidth]{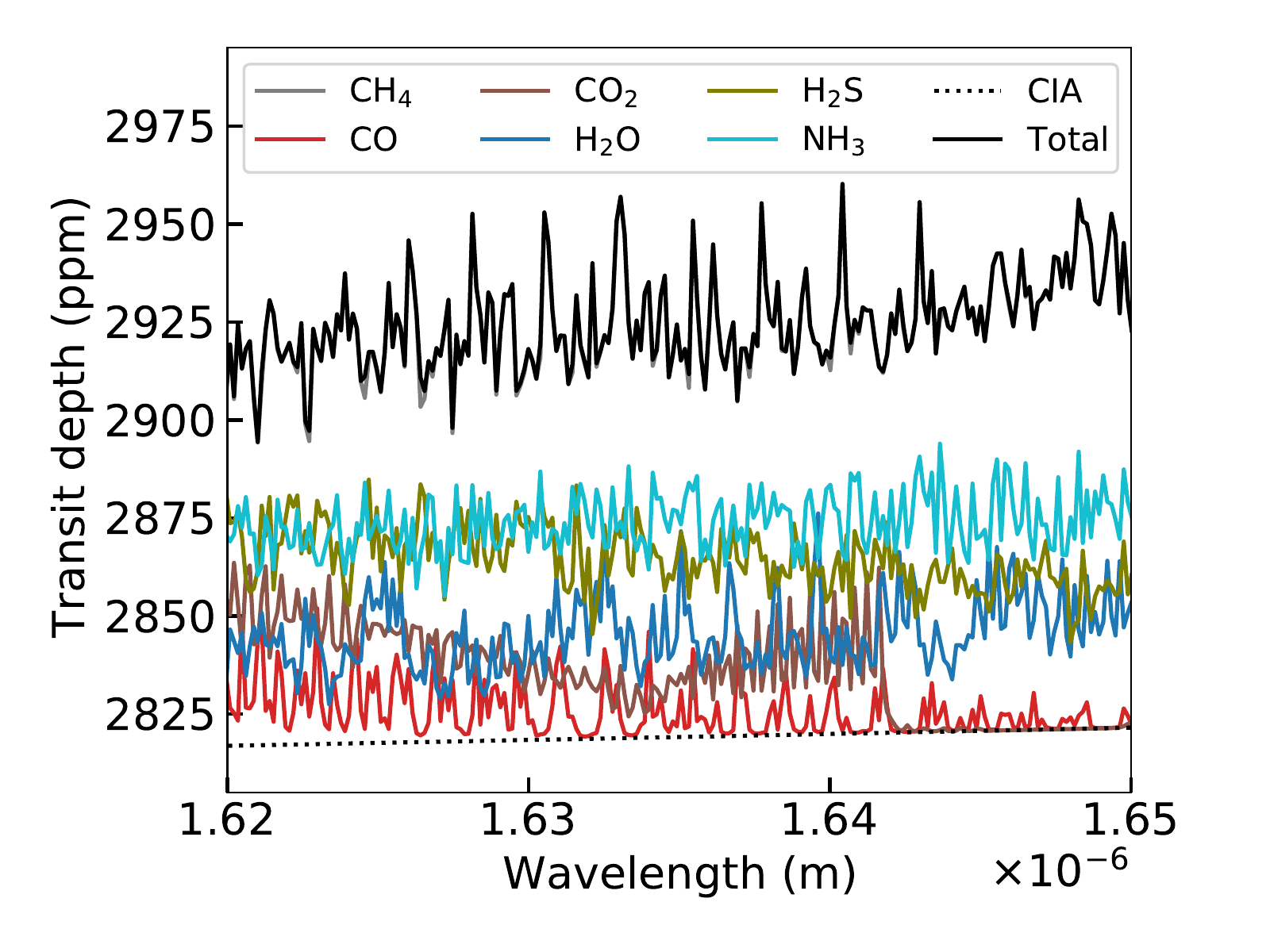}
\end{subfigure}
\begin{subfigure}{0.49\textwidth}
\centering
\includegraphics[width=1.0\linewidth]{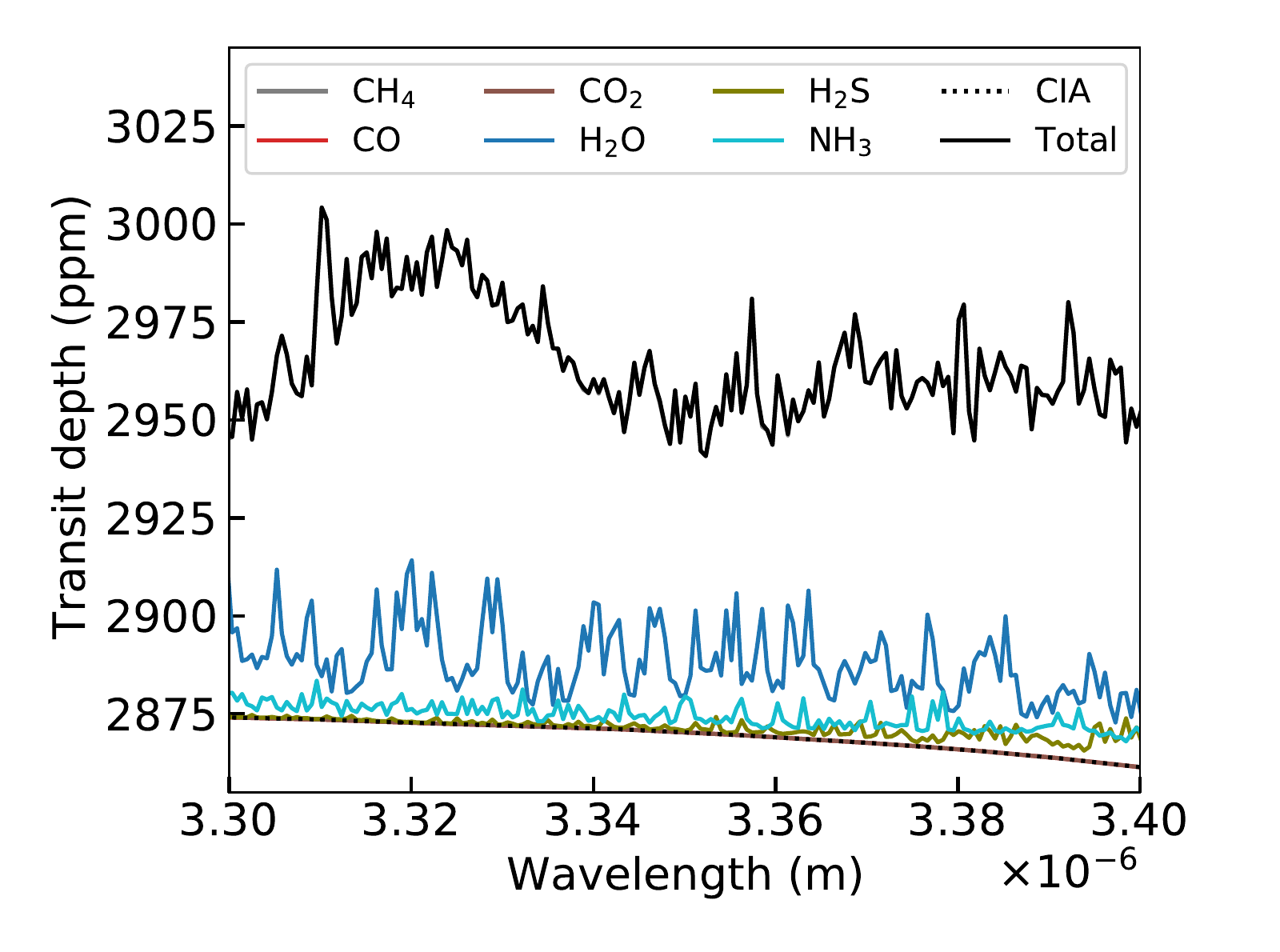}
\end{subfigure}
\begin{subfigure}{0.49\textwidth}
\centering
\includegraphics[width=1.0\linewidth]{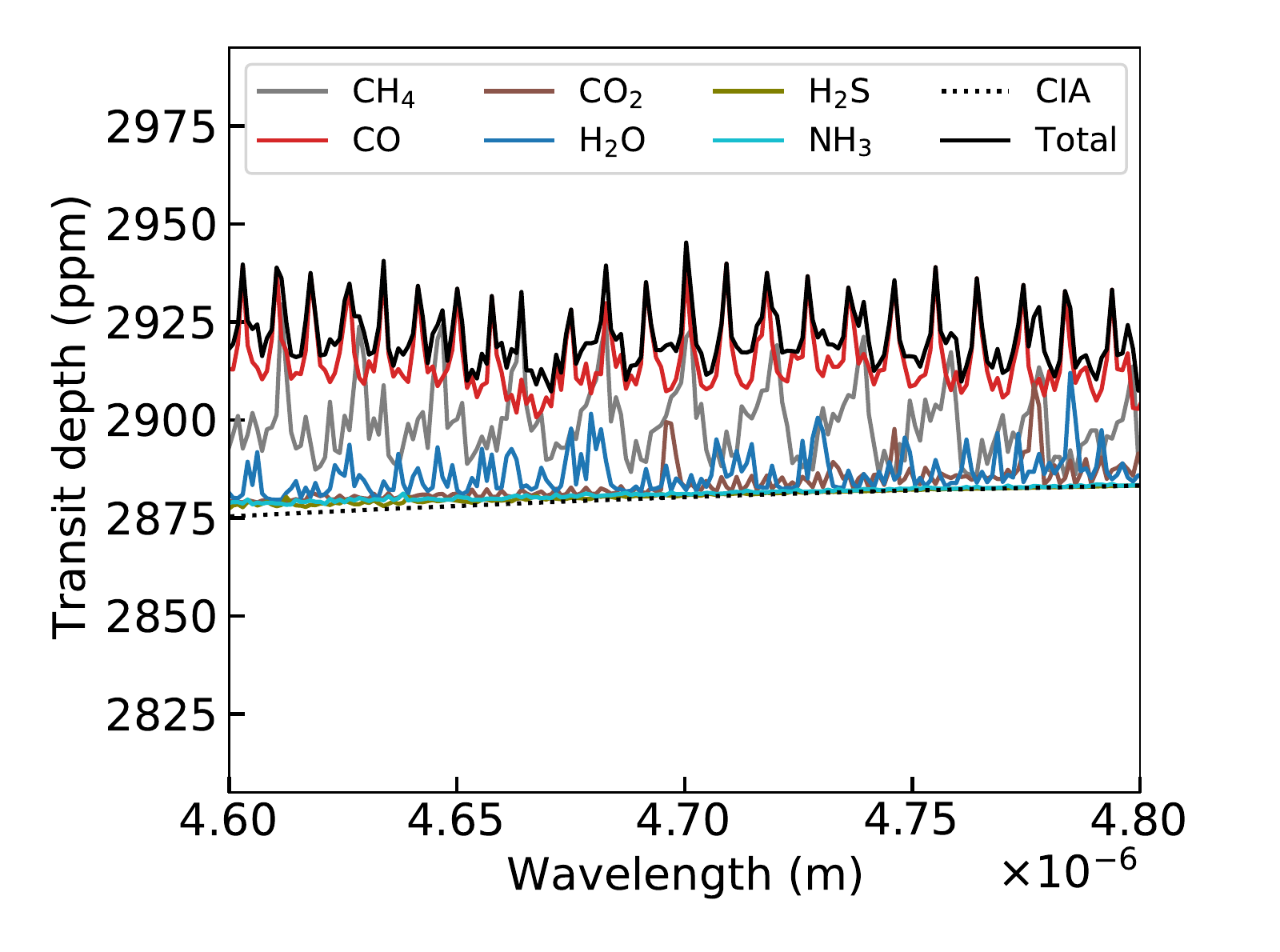}
\end{subfigure}
\caption[High resolution spectra]{Contributions within some spectral ranges of a selection of absorbing species to the transmission spectrum of our nominal model (175 (Z/H)$_\odot$) at nominal irradiation and a resolution of 0.5 cm$^{-1}$ (resolving power of $20\,000$ at 1 $\mu$m). The spectral contribution of individual species takes the CIA and Rayleigh scattering (dotted curve) into account.}
\label{fig:h_band}
\end{figure*}

For K2-18b, acquiring higher resolution spectra would permit to discriminate between the three scenarios we identified. We used our 0.5 cm$^{-1}$ resolution $k$-coefficient set to derive high-resolution spectra of K2-18b and show some spectral ranges in \autoref{fig:h_band} that could prove interesting in that regard. We note that NH$_3$ dominates the 1.45--1.56 $\mu$m spectral range for
any of the following three scenarios.

Firstly, in the nominal scenario, CH$_4$ should dominate over most of the infrared (IR) transmission spectrum, with particularly strong lines around 1.6 $\mu$m (end of the H band), or in the 3.0--4.0 $\mu$m spectral range (L band). CO would dominate the spectrum in the 4.6--5.0 $\mu$m spectral range (M band) but not in the 2.3--2.5 $\mu$m spectral range (end of the K band). Isolating H$_2$O lines would be challenging in the IR, but might be possible around 1.16 $\mu$m and 1.37 $\mu$m (beginning and end of the J band). However, it should dominate around 823 and 924 nm.

Secondly, in the low C/O scenario, contrary to the nominal scenario, H$_2$O should overall dominate the IR spectrum over CH$_4$. CH$_4$ may still dominate the spectrum at the end of the H band (1.6--1.8 $\mu$m), and in the K and L bands, depending on the intensity of the C-depletion. However, in contrast to the nominal scenario, it should have a minor contribution in the J band. CO might still dominate the M band, again depending on the extent of the of C-depletion.

Thirdly, in the scenario involving high internal temperature or incomplete chemistry, H$_2$O should again dominate most of the IR spectrum. CO should dominate in the M band and possibly even in the K band, in contrast to the low C/O scenario. The CH$_4$ spectral contribution should be similar to the low C/O scenario.

The  James Webb Telescope (JWST) and the  Atmospheric Remote-sensing Infrared Exoplanet Large-survey (ARIEL) mission, with their broader spectral range and precision higher than the data we studied here, will probably be very helpful to discriminate between these scenarios. In complement, we could acquire higher resolution spectra from ground-based instruments -- such as the CRyogenic InfraRed Echelle Spectrograph+ (CRIRES+) or the Echelle SPectrograph for Rocky Exoplanets and Stable Spectroscopic Observations (ESPRESSO) mounted on the Very Large Telescope (VLT) -- or by using the incoming generation of large telescopes (Extremely Large Telescope, Thirty Meter Telescope, etc.) in order to detect the absorption of individual lines and, thus, to unambiguously detect species. 

As an example, we calculated the signal to noise ratio (S/N)  we could obtain for a detection of CH$_4$ in our nominal case using CRIRES+ at a resolution power of $100\,000$ in the 1.5--1.8 $\mu$m spectral range. This spectral range has the advantage of presenting a lot of deep CH$_4$ lines (50--100 ppm) in a band for which CRIRES+ has a high sensitivity\footnote{\href{https://www.eso.org/sci/facilities/paranal/instruments/ocrires/inst.html}{eso.org.}}. We calculated a synthetic transmission spectrum at a resolving power of $100\,000$ using a line-by-line radiative transfer program, including absorption from H$_2$-H$_2$ and H$_2$O-H$_2$O CIA and lines from H$_2$O, CH$_4$, NH$_3$ and CO. We calculated the S/N, while neglecting the terrestrial absorptions and the photon noise of the star, using the following:
\begin{equation}
\label{eq:snr_ch4}
\text{S/N}_{\text{CH}_4} = \left( \frac{\sum_{\lambda=\lambda_\text{min}}^{\lambda_\text{max}} \left( \delta_{\text{CH}_4} - \bar{\delta}_{\text{CH}_4} \right) \left( \delta_\text{tot} -\bar{\delta}_{\text{tot}} \right)}{N} \right)^{1/2} \text{S/N}_\ast
\end{equation}
where $\lambda$ is the discretised wavelength, $\lambda_\text{min}$ and $\lambda_\text{max}$ are respectively the minimum and maximum wavelengths of the considered spectral range, $\delta_{\text{CH}_4}$ and $\delta_{\text{tot}}$ are, respectively, the transit depths that consider only the absorption from CH$_4$ and the absorption from all absorbers; the bar indicates a smoothing by a boxcar of width 16 resolution elements \citep[also used by][for their CRIRES observation of HD 189733]{Brogi2016}, $N$ is the number of calculated points per resolution element within the spectral range, and $\text{S/N}_\ast$ is the expected CRIRES+ S/N on K2-18 for one hour of observation at R = $100\,000$. From the H limiting magnitude given in \href{https://www.eso.org/sci/facilities/paranal/instruments/ocrires/inst.html}{eso.org}, we derive a S/N of 2500 for K2-18 at R = $50\,000$ per spectral dispersion element for a 1-hr integration. At R = $100\,000$, assuming no loss of the star signal, the S/N should thus be $2500 / \sqrt(2) = 1800$. Hence, we can obtain a total CH$_4$ signal of 1500 ppm and, thus, from \autoref{eq:snr_ch4}, we have an optimistic value for $\text{S/N}_{\text{CH}_4}$ of $\approx 2.7$ for one hour of observation with CRIRES+ at R = $100\,000$. We note that the transit duration of K2-18b is approximately three hours. A full transit would give S/N$_{\text{CH}_4} \approx$ 4.7, which is sufficient for detecting CH$_4$. However, this estimate is probably optimistic as we did not take into account the telluric absorption and the photon noise from the star.

Observing CO from the ground would, on the other hand, prove to be extremely difficult. Even at R = $100\,000$, the CO lines overlap with those of the star. The Doppler shift between the planet and the star during the transit is in the range $\pm$ 0.7 km$\cdot$s$^{-1}$, which is too low to use a cross-correlation method as described by, for example, \citet{Snellen2010}.

\section{Summary and conclusions}
\begin{figure}[pt]
\centering
\includegraphics[width=1.0\linewidth]{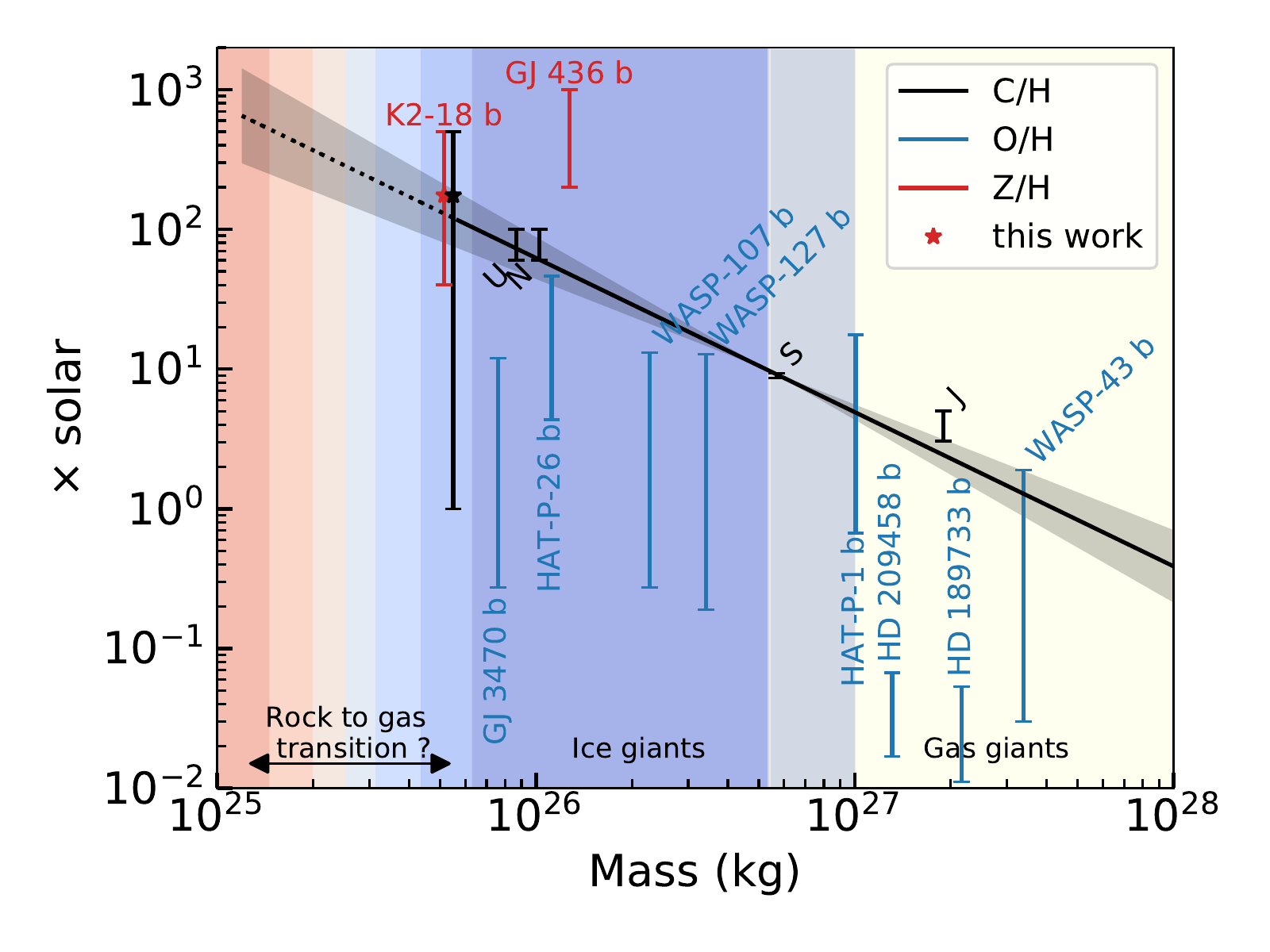}
\caption[Mass metallicity relationship]{Atmospheric metallicity as a function of mass for a selection of planets. The Jupiter (J), Saturn (S), Uranus (U), and Neptune (N) C/H are taken respectively from \citet{Wong2004}, \citet{Fletcher2009}, \citet{Sromovsky2011} and \citet{Karkoschka2011} \citep[values compiled in][]{Atreya2018}. The data for GJ436b and WASP-127b are taken respectively from \citet{Morley2017} and \citet{Skaf2020}, while the data for all the other planets are taken from \citet{Welbanks2019}. The black area represents the $1\sigma$ C/H against mass trend for Jupiter, Saturn, Uranus, and Neptune resulting from a linear fit. The fit is $\log_{10}(\textrm{(Z/H)}/\textrm{(Z/H)}_\odot) = -1.10 \pm 0.21 \log_{10}(M[\text{kg}]) + 30.5 \pm 5.5 $. The error bars of our results correspond to the simulations that were within the $1\sigma$ confidence interval against \citet{Tsiaras2019} data in the 'free C/H' scenario. Our C/H results were slightly shifted to the right for clarity. The O/H ratios in the other works were estimated assuming [H$_2$] + [He] + [H$_2$O] + [CH$_4$] = 1 and a solar He/H.}
\label{fig:mass_metallicity}
\end{figure}

\begin{figure}[pt]
\centering
\includegraphics[width=1.0\linewidth]{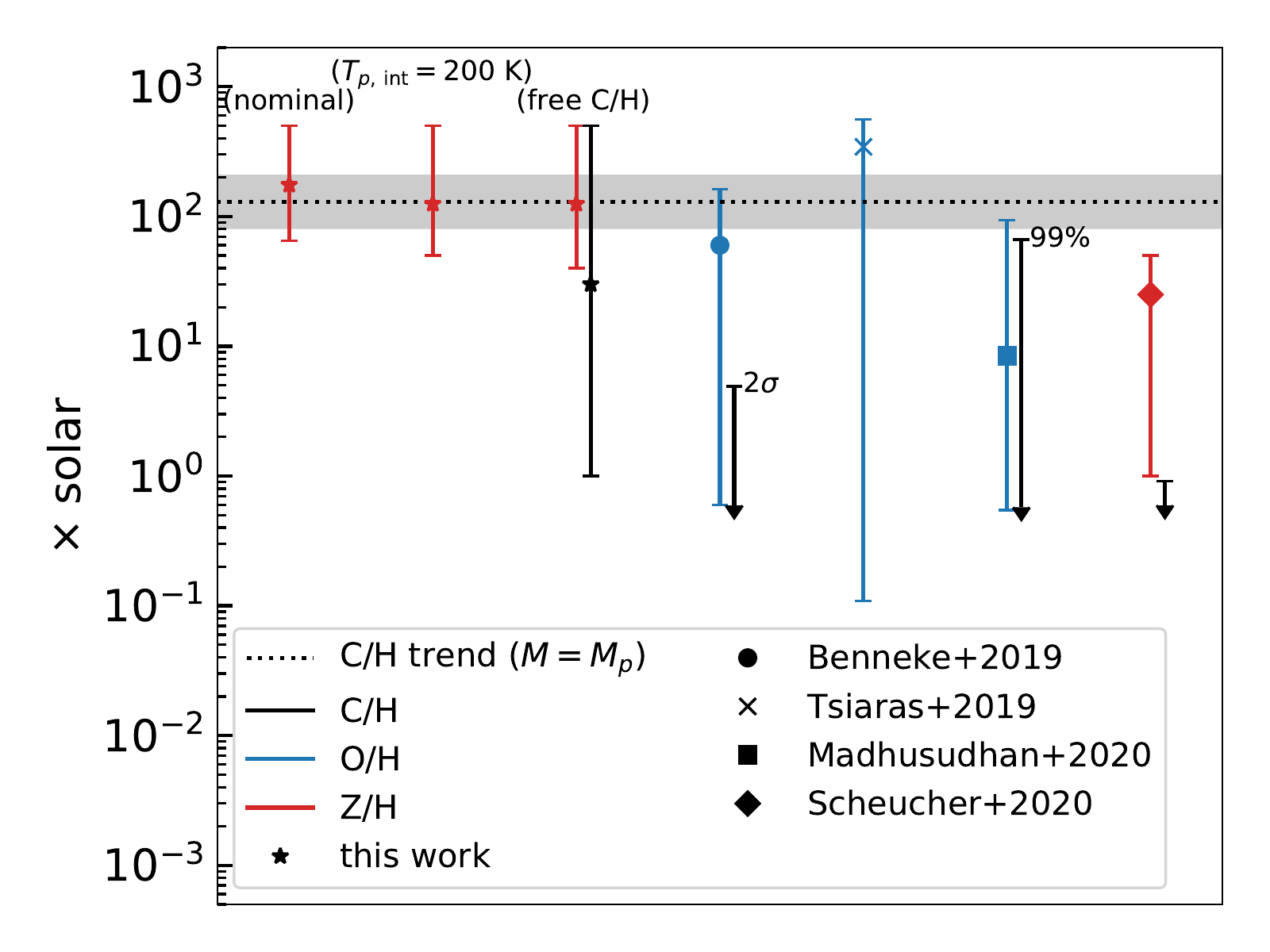}
\caption[Comparison]{Summary of our results and results of other teams on K2-18b metallicity. The error bars of our results corresponds to the simulations that were within the $1\sigma$ confidence level against \citet{Tsiaras2019} data. The O/H and C/H ratio of the other works were estimated assuming [H$_2$] + [He] + [H$_2$O] + [CH$_4$] = 1 and a solar He/H. The C/H upper limit of \citet{Benneke2019} and \citet{Madhusudhan2020} corresponds respectively to the $2\sigma$ and the 99$\%$ upper limit. The dotted line represents  the metallicity expected for the mass of K2-18 b using the linear fit in \autoref{fig:mass_metallicity}, based on solar-system C/H ratios.}
\label{fig:result_comparison}
\end{figure}

\begin{figure*}[pt]
\centering
\includegraphics[width=1.0\linewidth]{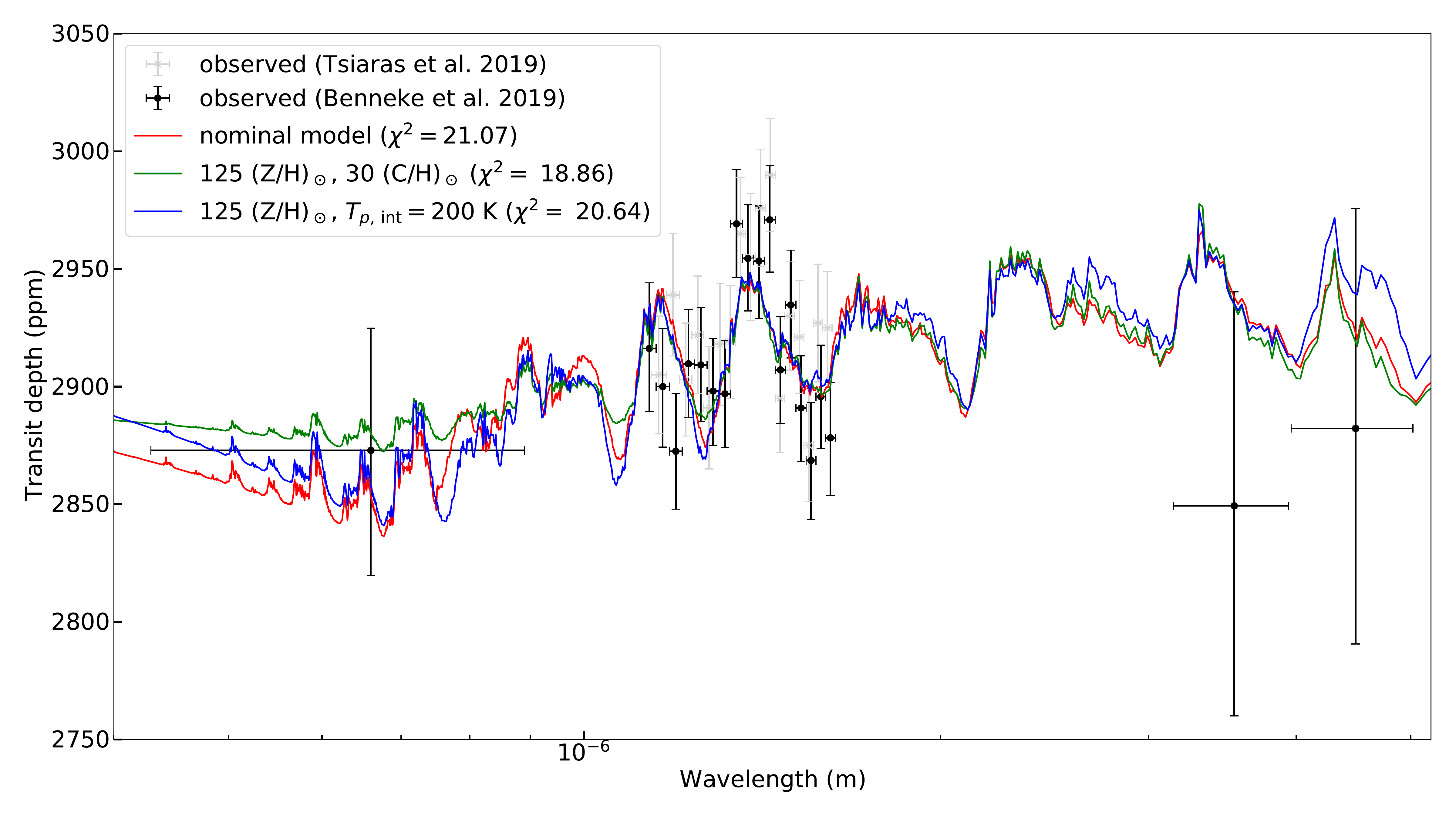}
\caption[Best fit]{Comparison of K2-18b transmission spectrum best-fits for three different scenarios. Red: 175 (Z/H)$_\odot$, 175 (C/H)$_\odot$, $K_{zz} = 10^6$ cm$^2\cdot$s$^{-1}$, $T_{p,\,\text{int}} = 80$ K. Green: 125 (Z/H)$_\odot$, 30 (C/H)$_\odot$, $K_{zz} = 10^6$ cm$^2\cdot$s$^{-1}$, $T_{p,\,\text{int}} = 80$ K. Blue: 125 (Z/H)$_\odot$, 125 (C/H)$_\odot$, $K_{zz} = 10^{10}$ cm$^2\cdot$s$^{-1}$, $T_{p,\,\text{int}} = 200$ K. The $\chi^2$ against \citet{Benneke2019} data is indicated in parentheses.}
\label{fig:best_fit_3_scenarios}
\end{figure*}

We analysed the transmission spectrum of K2-18b between 0.43 and 5.02 $\mu$m using a combination of K2, HST, and Spitzer data retrieved from \citet{Benneke2019} and \citet{Tsiaras2019}. We studied the effect of irradiation, metallicity, clouds, internal temperature, eddy diffusion coefficient, and C/O ratio using our self-consistent model Exo-REM and assuming an atmosphere primarily composed of H$_2$ and He. We also analysed the \citet{Benneke2019} dataset with the retrieval algorithm TauREx 3, and provided a S/N estimation for the detection of CH$_4$ using CRIRES+.

We found that the data are compatible with a highly metal-enriched atmosphere, between 65 and 500 (Z/H)$_\odot$ against \citet{Tsiaras2019} data -- or between 100 and 200 (Z/H)$_\odot$ against \citet{Benneke2019} data -- when assuming a solar C/O ratio, and $\gtrapprox$ 40 (Z/H)$_\odot$ when assuming a sub-solar C/O ratio. According to our results, the atmosphere of K2-18b appears quite similar to that of Neptune or Uranus and seems to follow the C/H-mass relationship of the giant planets of our solar system, as shown in \autoref{fig:mass_metallicity}. Most of the extrasolar planets seem to have an O/H ratio below that expected from the C/H-mass relationship for solar-system giant planets. Part of the explanation could reside in the fact that for most of these planets, the amount of CO and CO$_2$ is unknown, leading to an underestimation of the O/H ratio \citep[as pointed out by e.g.][]{Wakeford2017, Welbanks2019}. A summary of our results and results of other teams is displayed in \autoref{fig:result_comparison}, and the spectra of our best fit for each of the explored scenarios are displayed in \autoref{fig:best_fit_3_scenarios}.

We also show that thick to no H$_2$O-ice clouds are allowed by the data. Liquid H$_2$O clouds are possible on planets similar to K2-18b but receiving at most 80$\%$ of its irradiation. With or without clouds, we found that the Bond albedo of the planet should be around 0.02. The other studied parameters are not well constrained by the data. We note that in most of the cases we studied, CH$_4$ absorption should dominate or be on par with that of H$_2$O in the HST spectral window, as first outlined by \citet{Bezard2020}.

Combining the work of \citet{Bezard2020} and our own retrievals using TauREx 3, we show that the discrepancy between our self-consistent results and the results from retrieval algorithms can be explained by either the discarding of CH$_4$ absorptions, as in the case of the \citet{Tsiaras2019} dataset, or a strong overfitting of the data, as in the case of the \citet{Benneke2019} dataset. 

Accordingly, in addition to our nominal scenario (i.e. Z/H = 65--500 (Z/H)$_\odot$ and a solar C/O), scenarios with a CH$_4$-depleted atmosphere could satisfactorily fit the observed spectrum. These could be obtained with a high internal temperature ($\gtrapprox 200$ K) or a low C/O ratio ($\gtrapprox$ 0.01), with H$_2$O being the dominant absorber in HST/WFC3 spectral band if C/O $\lessapprox$ 0.1. However, the high internal temperature scenario seems very unlikely, and the C-depleted scenario requires the existence of an unknown process. Therefore, it seems that a CH$_4$-depleted atmosphere, a scenario favoured by all other teams who have analysed the same data so far, is more difficult to defend than a nominal scenario with a standard abundance of CH$_4$. Moreover, a spectrum dominated by H$_2$O absorptions seems even more unlikely given the relatively small range of self-consistent solutions allowing for this scenario and the fact that it is favoured by retrieval algorithms only because of overfitting. Another possibility is that our thermochemical model does not accurately describe the chemical conversion between CO and CH$_4$ in the deep atmosphere, as suggested by \citet{Benneke2019b} in their atmospheric analysis of GJ3470b (noting that GJ3470b is a much warmer planet with an effective temperature of $\approx$ 800 K). It should be possible to discriminate between these scenarios by acquiring data in different spectral ranges and/or at higher resolution using the incoming JWST and ARIEL space telescopes. In complement, ground-based instruments such as CRIRES+ could be used to detect individual lines of species. We estimate that with our nominal scenario, we could detect CH$_4$ with CRIRES+ in the H band with an 'optimistic' S/N of 4.7 after three hours of transit observation.
 
This study highlights the critical need for more precise spectroscopic data on sub-Neptunes. This will allow us to better characterise these intriguing bodies, which have no analogue in our Solar System.

\begin{acknowledgements}
D. B. acknowledges financial support from the ANR project "e-PYTHEAS" (ANR-16-CE31-0005-01). We acknowledge support from the Programme National de Planétologie of INSU/CNRS co-funded by CNES. We thank M. Rey for providing the TheoReTS CH$_4$ line lists up to 13400 cm$^{-1}$ over the whole Exo-REM temperature grid. We thank \href{https://orcid.org/0000-0003-2854-765X}{O. Venot} for providing us with the results from \citet{Venot2020}'s thermochemical model for K2-18b. We thank \href{https://orcid.org/0000-0002-3555-480X}{J. Leconte} for his valuable insight on tidal heating. 
\end{acknowledgements}

\bibliographystyle{aa} % style aa.bst
\bibliography{bibliography} % your references Yourfile.bib

\onecolumn
\appendix

\section{Note on the effect of the \texorpdfstring{H$_2$O}{H2O} CIA}
\label{sec:effect_of_the_H2O_CIA}
\begin{figure*}[pt]
\begin{subfigure}{0.5\textwidth}
\centering
\includegraphics[width=1.0\linewidth]{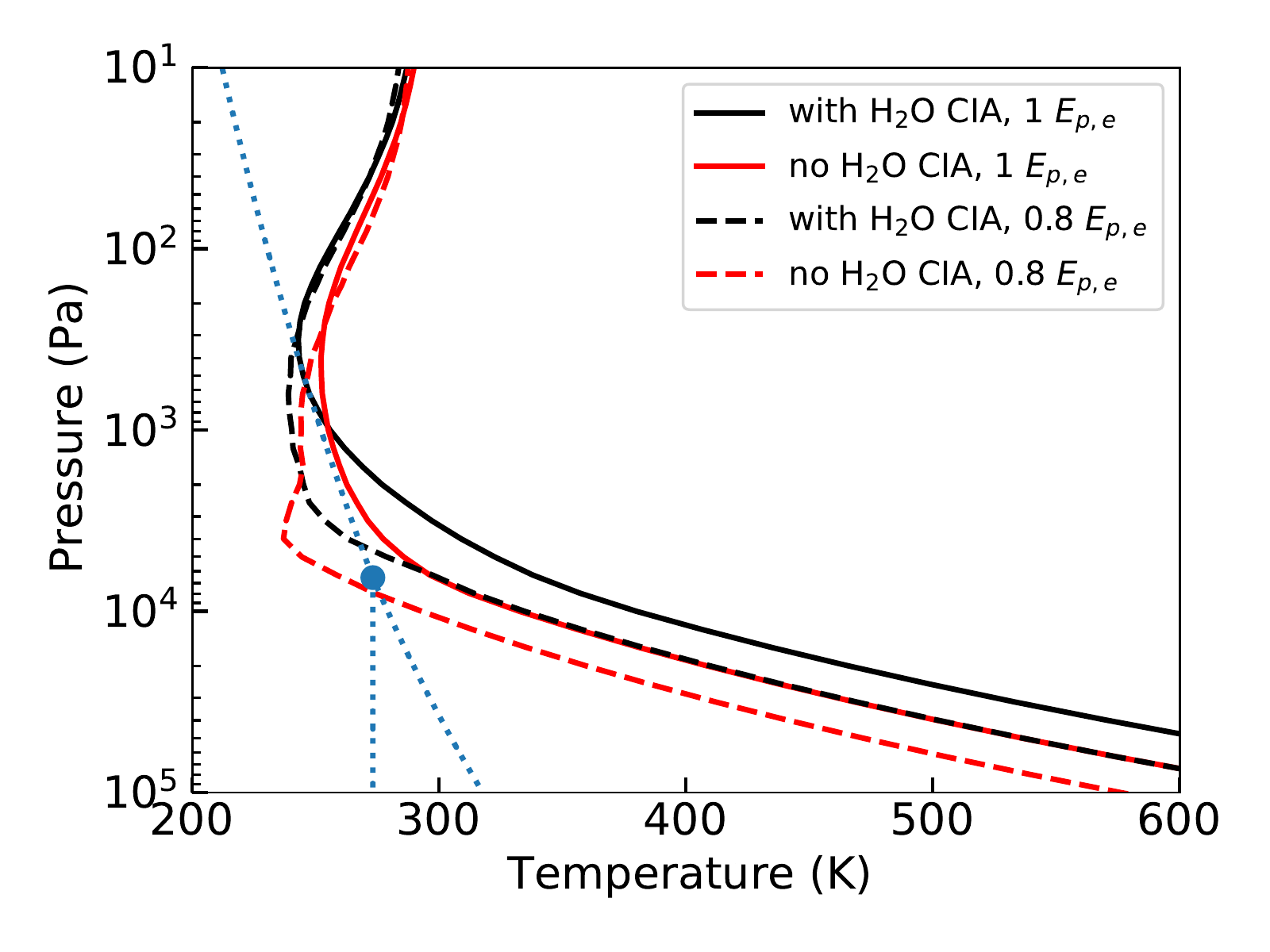}
\end{subfigure}
\begin{subfigure}{0.5\textwidth}
\centering
\includegraphics[width=1.0\linewidth]{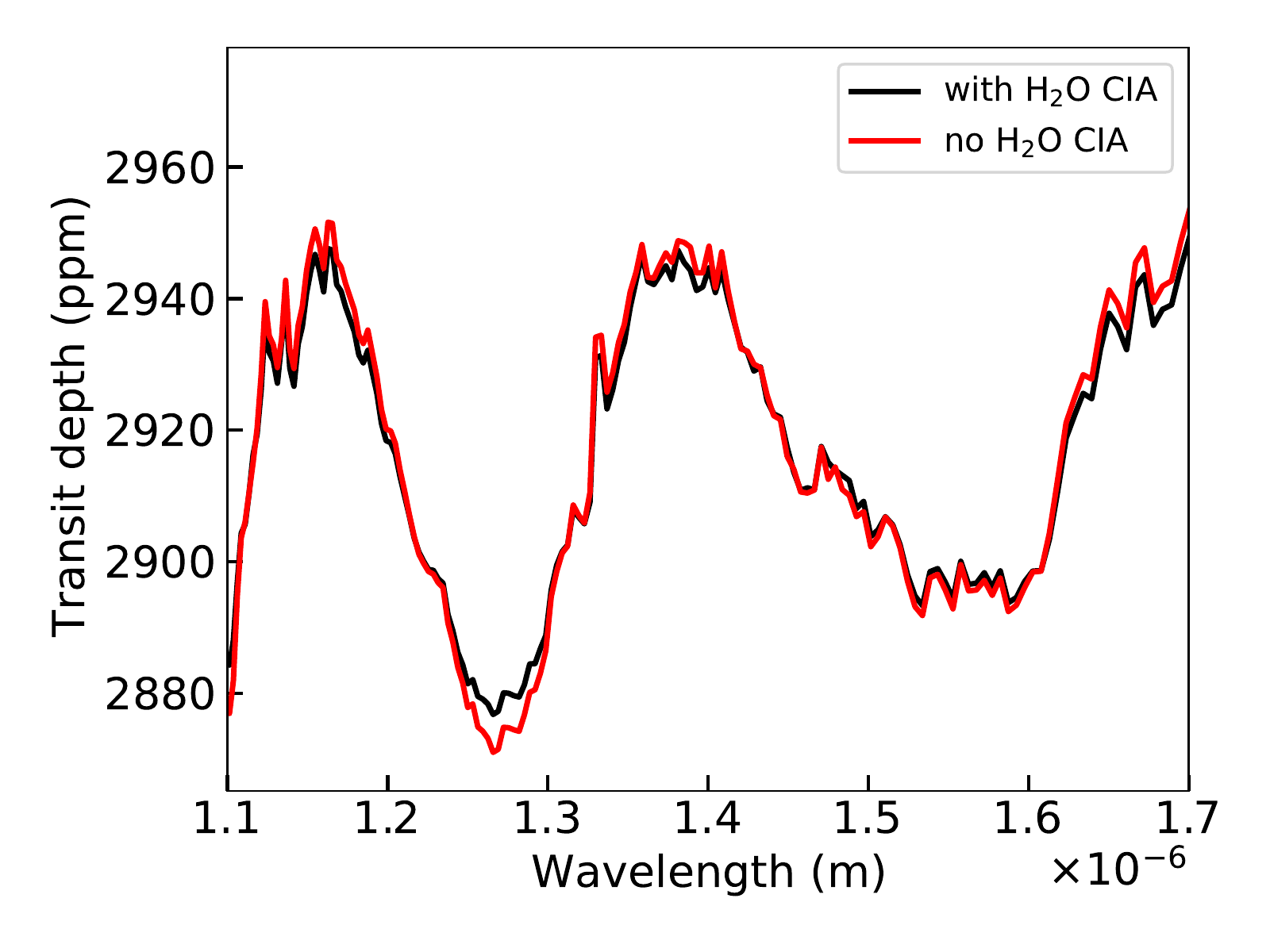}
\end{subfigure}
\caption[H2O CIA effect]{Effect of the H$_2$O CIA at 175 (Z/H)$_\odot$ on the simulated transmission spectrum and temperature profile of K2-18b. Left: Temperature profiles. Solid black: Nominal model. Dotted-dashed black: Nominal model at 0.8 time the nominal irradiance. Red: Nominal model without H$_2$O CIA. Dotted-dashed red: Nominal model at 0.8 times the nominal irradiance and without H$_2$O CIA. The phase diagram of H$_2$O is represented as dotted blue lines, the dot corresponding to the H$_2$O triple point. Right: Transmission spectra at 0.8 time the nominal irradiation, with (black) and without (red) H$_2$O CIA.}
\label{fig:h2o_cia_effect}
\end{figure*}

The effect of H$_2$O CIA on transmission spectra and temperature profiles is shown in \autoref{fig:h2o_cia_effect}. In our K2-18b simulations at 175 (Z/H)$_\odot$, we found that neglecting the H$_2$O CIA leads to an underestimation of the atmospheric temperature by $\approx$ 40 K at $10$ kPa. This can have an effect on the nature of H$_2$O condensation as well. Taking a model with 0.8 time the nominal irradiance of K2-18b, if we include H$_2$O CIA, we find that H$_2$O condenses into its solid phase, whereas if we do not, it condenses into its liquid phase at slightly higher pressures ($\approx$ 8 kPa instead of 3 kPa). Moreover, because the cloud is forming lower without the CIA, it has less effect on the transmission spectrum. As a consequence, the amplitude of the absorption features of the transmission spectrum is increased. However, the effect on the temperature profile is smaller at low metallicities, with a temperature difference at $10$ kPa reaching 20 K at 30 (Z/H)$_\odot$, and 2 K at 3 (Z/H)$_\odot$,  at $\approx$ 300 kPa for both cases.

CIA are induced by molecules colliding with themselves or other species. On planets with low metallicity, H$_2$--H$_2$ and H$_2$--He collisions are the only significant contributors to CIA because species other than H$_2$ and He are present in low quantities. However, this no longer holds true at higher metallicity. As the abundance of other species increases, their CIA increase which has an overall warming effect on the atmosphere.

\section{TauREx 3 retrievals}
\label{sec:app_retrievals}
From our TauREx 3 retrievals, we found an Atmospheric Detectability Index \citep[ADI, see][]{Tsiaras2018} of $\geq 1.17,$ including only CH$_4$ (2.12$\sigma$ , or 3:1 relative odds), and $\geq 6.37$ (3.99$\sigma$, or 584:1 relative odds), including at least H$_2$O, indicating 'weak' to 'strong' detections of atmospheric absorptions. Similarly to \citet{Madhusudhan2020}, we find only a 'weak' detection of clouds by comparing similar models with and without clouds: the difference in log-evidence is at most 1.04 ($2.05\sigma$ or 2.8:1 relative odds) with our CH$_4$-only models and at least 0.29 ($1.48\sigma$ or 1.3:1 relative odds) with our H$_2$O-only models. Comparing our 'all absorbers with clouds' model with our 'no H$_2$O' and 'no CH$_4$' models, we also find that H$_2$O is detected at 273:1 (3.79$\sigma$, that is, a 'strong' detection), while CH$_4$ is detected at 1.30:1 (1.44$\sigma$, meaning a 'not significant' detection).

Regarding our retrieval with all absorbers and clouds (see \autoref{fig:posteriors}), our retrieved VMR of H$_2$O (0.072--5.3$\%$) is similar to what \citet{Benneke2019} found (0.033--8.9$\%$). Our 2$\sigma$ (95.4$\%$) upper limits for CO, CO$_2$, NH$_3$, and N$_2$ are respectively 4.39$\%$, 0.789$\%$, 0.002$\%,$ and 2.73$\%$, all lower than those from \citet{Benneke2019} (7.45$\%$, 2.4$\%$, 13.5$\%,$ and 10.9$\%$), especially for NH$_3$. The cloud top pressure we retrieve (1.8--66.1 kPa) is slightly higher than theirs (0.77--13.9 kPa). From our Exo-REM simulations, this level corresponds more to NH$_4$Cl condensation than to H$_2$O condensation, but we showed that including NH$_4$Cl clouds does not significantly enhance our fits (see Section~\ref{sec:clouds}).

\end{document}